\documentclass[pdflatex,sn-mathphys-num]{sn-jnl}
\usepackage{graphicx}%
\usepackage{multirow}%
\usepackage{amsmath,amssymb,amsfonts}%
\usepackage{amsthm}%
\usepackage{mathrsfs}%
\usepackage[title]{appendix}%
\usepackage{xcolor}%
\usepackage{textcomp}%
\usepackage{manyfoot}%
\usepackage{booktabs}%
\usepackage{algorithm}%
\usepackage{algorithmicx}%
\usepackage{algpseudocode}%
\usepackage{listings}%
\usepackage{chngcntr}

\counterwithin{figure}{section}

\newcommand{\ket}[1]{| #1 \rangle}
\def\eps{\varepsilon}
\def\Hnl{\mathcal{H}_{\rm nl}}

\def\PN{{politician network}}
\def\SN{{netscience network}}

\begin{document}

\title{Wealth Thermalization Hypothesis  {\vskip 0.01cm  and Social Networks}}

\author*[1,2]{\fnm{Klaus M.} \sur{Frahm}}\email{klaus.frahm@utoulouse.fr}
\equalcont{These authors contributed equally to this work}
\author[1,2]{\fnm{Dima L.} \sur{Shepelyansky}}
\equalcont{These authors contributed equally to this work.\\
 received: September 21, 2025}

\affil*[1]{\orgname{Lab. Physique Th\'eorique}, \orgdiv{Universit\'e Toulouse III - Paul Sabatier, UPS, CNRS}, \city{Toulouse}, \country{France}}
%\affil*[1]{Lab. Physique Th\'eorique, Universit\'e Toulouse III - Paul Sabatier, UPS, CNRS}, Toulouse, France}
\affil[2]{\orgdiv{Univ Toulouse}, \orgname{CNRS}, 
\orgname{LPT},
  \city{Toulouse}, \country{France}}

\abstract{
  In 1955 Fermi, Pasta, Ulam and Tsingou performed
  first numerical studies with the aim to obtain
  the thermalization in a chain of
  nonlinear oscillators from dynamical equations of motion.
  This model happend to have several specific features
  and the dynamical thermalization was established only
  later in other studies. In this work we study
  more generic models based on Random Matrix Theory
  and social networks with a nonlinear perturbation
  leading to dynamical thermalization above a certain chaos border.
  These systems have two integrals of motion
  being  total energy and norm so that the theoretical 
  Rayleigh-Jeans thermal distribution depends on temperature and
  chemical potential. We introduce the wealth thermalization hypothesis
  according to which the society wealth is associated with
  energy in the  Rayleigh-Jeans distribution.
  At relatively small values of total wealth or energy
  there is a formation of the Rayleigh-Jeans 
  condensate, well studied in physical systems such 
  as multimode optical fibers.
  This  condensation leads to a huge fraction of
  poor households at low wealth and a small oligarchic fraction which
  monopolizes a dominant fraction of total wealth thus generating
  a strong inequality in human society. We show that this thermalization
  gives a good description of real data of Lorenz curves of US, UK, the whole world
  and capitalization of companies at Stock Exchange of New York SE (NYSE),
  London and Hong Kong. It is also shown that above a chaos border the
  dynamical Rayleigh-Jeans thermalization takes place also in
  social networks with the Lorenz curves being similar to those
  of wealth distribution in world countries.
  Possible actions for inequality reduction are briefly discussed.
}

\maketitle

%% the following command removes the global section 
%% number and applies a small shift to the left 
%% of the section title to compensate the white space
\renewcommand\thesection{\hspace{-0.27cm}}

\section{Introduction}

In 1872 Ludwig Boltzmann published the fundamental work  \cite{boltzmann1}
that became the foundation of the  theory of statistical physics
and thermalization emerging from the dynamical laws of 
time reversible classical motion of many-body systems.
Of course, certain approximations had been used there,
thus considering only pair collisions of gas particles.
But this was only the first step in the demonstration that
the statistical laws follow from the dynamical equations.
One of the important results obtained in \cite{boltzmann1}
was the Boltzmann H-theorem that a system entropy is
monotonically growing with time  or remains constant at the steady-state.

The first numerical experiment
with the aim to  directly obtain
statistical  thermalization
from dynamical equations of motion
was done by Fermi, Pasta, Ulam and Tsingou
in 1955 studying the dynamics of a chain of nonlinear oscillators
on the most powerful MANIAC I computer (at this time) 
with an expectation to obtain the thermal
energy equipartition between oscillator modes \cite{fpu1955}.
However, opposite to expectations the conclusion of authors 
was that ``The results show very little, if any, tendency toward
equipartition of energy between the degrees of freedom.'' \cite{fpu1955}.

Several explanations had been proposed to explain this result.
Zabusky argued  in \cite{zabusky} that in the continuum
limit the Fermi-Pasta-Ulam (FPU) problem
is close to the Korteweg-de Vries equation
with stable solitons 
shown to be completely integrable \cite{greene},
as well as the nonlinear Schr\"odinger equation \cite{zakharov}.
Also, at weak nonlinearity
the FPU $\alpha$-model is close to the completely
integrable Toda lattice \cite{toda,benettin}.
Another explanation of  absence of thermalization 
in the FPU problem
was given in \cite{chirikovfpu1,chirikovfpu2,livi}
showing that below a certain chaos border, determined by the 
strength of the nonlinear
interactions between oscillators,   the system
is located in the regime of Kolmogorov-Arnold-Moser (KAM)
integrability and only above this border
an overlap of nonlinear resonances takes place with
emergence of chaos and thermalization.
Indeed, above a chaos border numerical simulations demonstrated an
emergence of dynamical thermalization
with energy equipartition as reported in \cite{chirikovfpu2,livi}.
Possibilities of low energy chaos in the FPU model were discussed in 
\cite{dls1997}, 

It should be pointed out that while impressive mathematical
results and theorems were obtained by
mathematicians (see e.g. \cite{arnold,sinai} and Refs. therein)
they remained usually not applicable to
thermalization in physical nonlinear systems
which usually have a divided phase space
(see \cite{chirikov1979,lichtenberg})
where integrable islands of stable motion are
often embedded in a chaotic component.
Thus the KAM theorem is valid for unrealistically weak nonlinear
perturbations \cite{arnold,sinai}
and it is more appropriate to use
the Chirikov criterion of overlapping resonances
to estimate more realistic parameters
for a chaos border \cite{chirikov1979,lichtenberg}
(even if this criterion is not working
for completely integrable systems like the Toda lattice 
for example). 

An overview of the full richness of the FPU model
and various regimes of its nonlinear dynamics
has been presented 
50 years after \cite{fpu1955} in the book  \cite{fpu50}.
The variety of studies presented there clearly demonstrates that this
model has an important role
in the investigations of nonlinear dynamics.
However, at the same time the variety of different features of
FPU dynamics 
indicates that the FPU model does not belong to
a class of generic oscillator systems with nonlinear interactions.

With the aim to capture the generic features of
dynamical thermalization the nonlinear random matrix model (NLIRM)
was proposed in the work \cite{rmtprl},
submitted 150 years after the Boltzmann article \cite{boltzmann1}.
In this model, the linear unperturbed Hamiltonian
is described by a random matrix
that can be also viewed as a system of linear oscillators
with complex linear couplings.
The chaos in this system is induced
only by a nonlinear perturbation
that can be local or can have a certain interaction range.
Thus in this model the unperturbed properties
of eigenmodes and eigenenergies are described by
the generic  Random Matrix Theory invented by 
Wigner for a description of
the spectra of complex nuclei, atoms, and molecules
in many-body quantum mechanics \cite{wigner}.
Indeed,  RMT finds a variety of applications in
multiple areas of physics \cite{mehta,guhr}
including systems of quantum chaos where the dynamics is
chaotic in the classical limit \cite{bohigas,haake}.

In \cite{rmtprl} it was shown that above a certain chaos border
the dynamical thermalization takes place leading to the
Rayleigh-Jeans thermal distribution (see Eq. (\ref{eqrj} below)
well known in classical thermodynamics \cite{landau,mayer}.
However, the dynamics of the NLIRM system has two integrals of motion
being the total energy and norm (or probability that is
very natural for quantum evolution). Due to
this the Rayleigh-Jeans distribution is characterized by
temperature and chemical potential.
Such a situation appears in various classical systems
including dynamics of nonlinear waves (see e.g. \cite{zakharovbook}).
In fact this type of thermal distribution was described and
experimentally observed for light propagation in multimode optical
fibers (see e.g. review \cite{picozziphrep} and Refs. below).
The important feature of the Rayleigh-Jeans thermalization
is the condensation of a main fraction of system norm or probability
at the lowest energy modes of the system.
However, in \cite{picozziphrep} the emergence of such thermalization
and condensation was attributed to the turbulence like energy
flows similar to those of the Kolmogorov-Zakharov turbulence spectra of
nonlinear waves \cite{zakharovbook}
(even if it is stated \cite{picozziphrep} that the dynamics is Hamiltonian).
In contrast, it is argued in \cite{rmtprl,ourfiber} that such  Rayleigh-Jeans
thermalization and condensation appear due to dynamical chaos
emerging above a chaos border while below this border
the  thermalization is absent and the system is located in the integrable
KAM regime. Certain similarities of this condensation
with the Fr\"ohlich condensate proposed for molecular systems
at room temperature 
\cite{frohlich1,frohlich2} are discussed in \cite{ourfiber}.
The striking applications of  Rayleigh-Jeans
thermalization are discussed in this work.

This work is composed of two parts.
In the first Part I, we push forward the Wealth Thermalization Hypothesis
according to which the wealth is associated with system energy and the wealth distribution
in the human society is described by the 
Rayleigh-Jeans thermalization and condensation
which are at the origin of strong inequality in the world.
The comparison of the obtained thermalization results
with the real data of wealth inequality
in countries and stock exchange markets confirms the validity
of the  Rayleigh-Jeans thermal description.
In the second Part II, we provide certain additional arguments and
justifications for this hypothesis.
In particular, we argue that social networks,
actively investigated in the network science and society
(see e.g. books \cite{dorogovtsev10,newmanbook}),
provide a reliable model of social relations in the society.
While the previous studies of social networks
describe the social relations and links
only in the frame work of linear matrix algebra we
introduce a nonlinear interaction in such social networks. 
Similarly to the NLIRM results \cite{rmtprl}
our studies show that above a certain chaos border for 
the strength of the nonlinearity 
dynamical thermalization takes place in social networks
being well described by the steady-state Rayleigh-Jeans distribution.
This gives an additional support to the Wealth Thermalization Hypothesis.
In this way the problem of emergence of statistical laws
from dynamical equations of motion finds
new application perspectives. 

The main results of this research are presented in Part I
for links between wealth inequality and  Rayleigh-Jeans thermalization
and in Part II for dynamical thermalization in social networks
supporting the Wealth Thermalization Hypothesis 
discussed in Part I. Figures in there are marked as Fig.IXX and
Fig.IIXX respectively.
Additional material and Figures are
presented in Appendix A related to Part I
and Appendix B related to Part II, 
in the Appendix A and B Figures are marked as Fig.~AXX and Fig.BXX
omitting the word Appendix A or B.

%%%%% BEGINNING of PART I

\section{Part I: Wealth Thermalization Hypothesis}
\setcounter{equation}{0}
\renewcommand{\theequation}{I.\arabic{equation}}
\setcounter{subsection}{0}
\renewcommand\thefigure{I.\arabic{figure}}
\renewcommand\thesubsection{I.\arabic{subsection}}

\subsection{Prologue I}

The wealth distribution in the human society is characterized by
a striking inequality (see e.g. \cite{piketty1,piketty2,boston}).
Thus for the whole world 50\% of the population owns only 2\%
of total wealth,
while 10\% of population owns 75\% of total wealth
and 1\% of population owns 38\% of total wealth \cite{piketty2}.

The distribution of wealth  is usually described by the Lorenz curve
\cite{lorenz,boston} which gives the dependence of accumulated
normalized wealth $0\leq w \leq 1$ on the cumulated normalized fraction of
population or households $0 \leq h \leq 1$.
Thus the equipartition of wealth corresponds to
the diagonal $w=h$ and the doubled area between diagonal
and the Lorenz curve $w(h)$ determines the Gini coefficient
$0 \leq G \leq 1$  \cite{gini,boston}.
Values of $G$ can be found in \cite{wikigini}
for world countries in 2021 being in the range
$0.59 < G < 0.90$; for the whole world $G = 0.889$.

The sharing of wealth
varies from country to country but the global features
remain rather similar with a big fraction of very poor population
with scanty wealth and a very small fraction of rich people
having a  significant fraction of a country's total wealth.
This gives an insight that some fundamental underground reasons
can be at the origin of this inequality.

Diverse methods of statistical mechanics 
and physical kinetics \cite{mayer,landau,lifshitz}
have been proposed and used by different research groups
\cite{angle,redner,yakovenko1,bouchaud,yakovenko2,chakraborti,boghosian1,boghosian2,perotti}.
Various models of interacting agents are investigated
including Random Asset Exchange models
\cite{angle,redner,yakovenko1,bouchaud,yakovenko2,chakraborti,boghosian1,boghosian2,perotti}.
In several of these models
there is appearance of some kind of oligarchic phase with  
a significant wealth accumulation by a group of agents
\cite{bouchaud,boghosian1,boghosian2,perotti}.
The specific arguments are presented
in a favor of the Boltzmann-Gibbs type description of
distribution of money, wealth and income \cite{yakovenko1,yakovenko2}.
Also a nonlinear Fokker–Planck description of asset exchange
is proposed \cite{boghosian1,boghosian2} with emergence of oligarchic phase.
A few important elements are stressed in \cite{boghosian1,boghosian2}:
the conservation of two integrals of system evolution
being the total wealth and total norm (or number of agents),
the argument in favor of consideration of wealth instead of money
based on the small-transaction approximation.
The conservation of two integrals is rather natural assumption
since a Gross domestic product and population of a country or the whole
world are only weakly changed on a typical time scale of one year.

The above models give interesting insights for understanding of
certain features of wealth distribution in the world countries
but they remain model specific and their universality remains
questionable. The universality of the Boltzmann-Gibbs
thermal distribution is the ground element of the approach
developed in \cite{yakovenko1,yakovenko2}
but it does not capture emergence of a huge
condensate of poverty in various countries.

Our studies here are based on the Wealth Thermalization Hypothesis (WTH)
according to which the wealth of a country is
described by the Rayleigh-Jeans (RJ) thermal distribution:
\begin{equation}
\rho_m = \frac{T}{E_m-\mu} \; ({\rm RJ}) .
\label{eqrj}
\end{equation}
Here we assume that the system wealth
has certain states $0 \leq m < N$ with energies $E_m$
and the population probabilities in these
states are $\rho_m$.
Thus a systen wealth 
is associated with a system energy.
Also in (\ref{eqrj}) the parameters $T$ and  $\mu(T)$
are the system temperature and
its  chemical potential dependent on $T$.
As in \cite{boghosian1} there are two conserved
integrals of motion being the total norm of population,
fixed to be unity for convenience, 
$\sum_m \rho_m =1$, 
and the system average wealth
being its total energy
$ \sum_m E_m \rho_m =E$.
For a given system energy $E$ and unity norm
these two integrals of motion determine
the system temperature $T$ and its chemical potential $\mu(T)$. 
The entropy $S$ of the system is determined by
the usual relation \cite{mayer,landau}:
$S= - \sum_m \rho_m \ln \rho_m$
with the implicit  theoretical dependencies on temperature
$E(T)$, $S(T)$, $\mu(T)$.

The RJ thermalization (\ref{eqrj})  is universal
and describes a variety of classical systems \cite{mayer,landau}
including nonlinear waves \cite{zakharovbook},
light propagation in multimode optical fibers with a nonlinear media
\cite{wabnitz,picozzi1,picozzi2,babin,chris,picozzi3},
dynamical thermalization for nonlinear perturbation
of the Random Matrix Theory (RMT) \cite{rmtprl} and 
the nonlinear Schr\"odinger equation (NSE) in quantum chaos billiards \cite{ourfiber}. 
It is pointed out in \cite{wigner,mehta,guhr} that RMT finds a variety of
applications in multiple areas
of physics including nuclei, complex atoms and systems of quantum chaos
whose dynamics is chaotic in the classical limit.
Thus almost any physical  nonlinear interaction  above a chaos border \cite{rmtprl}
leads to dynamical RJ thermalization (\ref{eqrj}).
An example of such a system can be an ensemble of $N$
nonlinear RMT oscillators with random frequencies $\omega_m \propto E_m$
of an ensemble of $N$ agents with nonlinear interactions 
leading to the RJ  thermalization (\ref{eqrj}). The thermalization can have a dynamical origin
when chaotic nonlinear dynamics leads to   (\ref{eqrj})
or it can appear due to an external thermal bath.
We suppose that for WTH a dynamical origin is more
adequate since in a first approximation on a scale of one year a country or the whole world
can be considered to be quasi-isolated from slow external processes.

Due to the presence of two integrals of motion, energy and norm,
RJ thermalization has  the phase of RJ condensate
emerging at relatively low total energy $E$ or low temperature $T$
\cite{picozzi1,picozzi2,ourfiber}. Thus at low energy and a big number of oscillators,
as in \cite{rmtprl}, or a big number of interacting agents,
the fraction of RJ condensate is approaching unity
being  concentrated at a vicinity
of the ground state energy $E_0$ being zero or very
close to zero \cite{ourfiber}. Thus  the RJ condensate (\ref{eqrj})
very naturally has a huge fraction of very poor agents
that naturally describes the huge world wealth inequality
where 50\% of population owns only 2\% of the total wealth \cite{piketty2}.
Below we describe in detail various consequences of WTH (\ref{eqrj})
and compare the results of this theory with real Lorenz curves
of certain countries and the whole world.

\begin{figure}[htbp]
\begin{center}
%\isPreprints{\centering}{} % Only used for preprints
  %\includegraphics[width=0.7\textwidth]{fig_K2.pdf}\\
\includegraphics[width=0.7\textwidth]{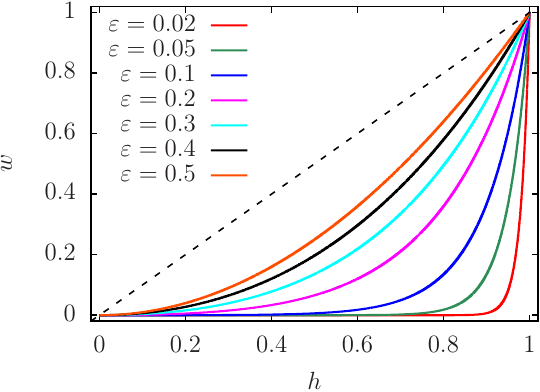}  
\caption{\label{figI_1}
Lorenz curves for the RJS model with the linear spectrum 
$E_m=m/N$ (for $N=10000$) 
for different values of the rescaled energy 
$\varepsilon=E/B$. 
%$\varepsilon=(E-E_0)/(E_{N-1}-E_0)$. 
The $x$-axis corresponds the 
cumulated fraction of households ($h$) and the $y$-axis to
the cumulated fraction of wealth ($w$). 
The dashed line is the line of perfect equipartition $w=h$. 
The Gini coefficients $G$ for all curves are 
$G=0.9600,\,0.9000,\,0.8006,\,0.6250,\,0.4990,\,0.4066,\,0.3333$ (bottom 
to top).
}
\end{center}
\end{figure}

\begin{figure}[htbp]
\begin{center}
%\isPreprints{\centering}{} % Only used for preprints
  %\includegraphics[width=0.7\textwidth]{fig_K2.pdf}\\
\includegraphics[width=0.7\textwidth]{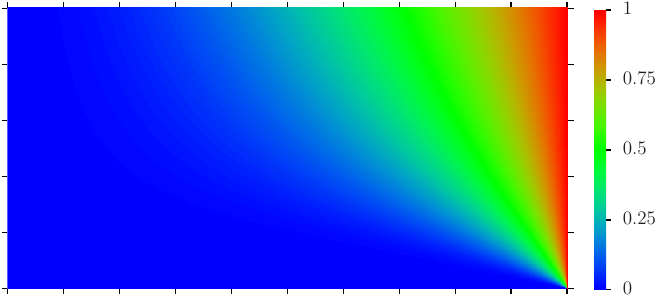}  
\caption{\label{figI_2}
Color plot of wealth $w$ from Lorenz curves of the RJS model 
($N=10000$). The $x$-axis corresponds to 
the fraction of households $h\in[0, 1]$ 
and the $y$-axis to the rescaled 
energy $\varepsilon=E/B\in[0, 0.5]$.
The ticks mark integer multiples of 0.1 
for $h$ and $\varepsilon$. 
}
\end{center}
\end{figure}

\begin{figure}[htbp]
\begin{center}
\includegraphics[width=0.7\textwidth]{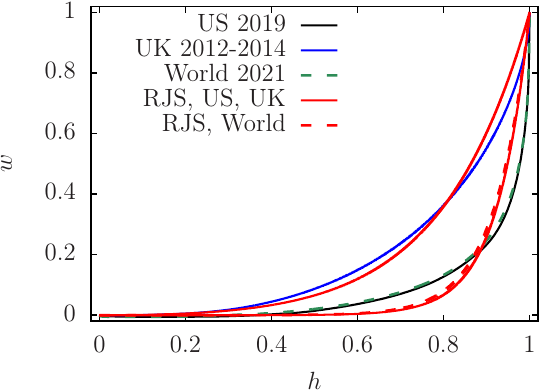}  
\caption{\label{figI_3}
Comparison of the Lorenz curves for US 2019 (black), 
UK 2012-2014 (blue), World 2021 (dashed green) 
with those of RJS model (red curves; $N=10000$);
US and World curves are rather close.
For the three referenced curves 
Gini coefficients are $G=0.852$, $0.626$,
$G=0.842$ respectively and the 
rescaled energies $\varepsilon=E/B$ of RJS model 
are respectively fixed as $\varepsilon=0.07420$, $\varepsilon=0.1996$,
$\varepsilon=0.07911$ so
that the corresponding Gini coefficients match the referenced data.
}
\end{center}
\end{figure}

\begin{figure}[htbp]
\begin{center}
\includegraphics[width=0.8\textwidth]{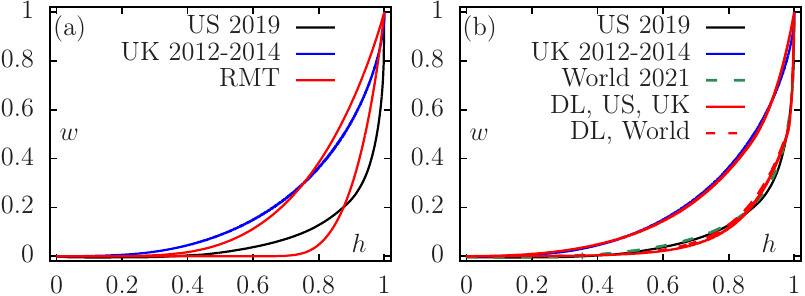}  
\caption{\label{figI_4}
Both panels compare the Lorenz curves for different data sets 
(black for US 2019, blue for UK 2012-2014 and green dashed for World 2021) 
with those of the RMT model (a) and the DL model (b). As in  
Fig.~\ref{figI_3} the Gini coefficients $G$ of the reference curves are
used to fix the rescaled energy $\varepsilon=E/B$ of the corresponding 
model such that the model curves (red) have the same $G$. 
For the RMT model (a) only two data sets are shown
$\varepsilon=0.07996$ (US) and $\varepsilon=0.2027$ (UK).
For the DL model (b) the parameter values are  $a=16$ (US and World) and $a=3$ (UK).
These values are fixed 
to have a best possible fit of the model data with those 
of the reference curves. The chosen values 
$\varepsilon=0.01434$ (US), $\varepsilon=0.1355$ (UK), 
$\varepsilon=0.01535$ (World) match the $G$ values of the reference 
data.
In (b) the curves for US and World are rather close 
and a zoomed view is given in Appendix  Figure~\ref{figA4}.
In (a) the RMT Lorenz curves are shown for one realization of 
a random matrix, other realizations give practically the same curves.
}
\end{center}
\end{figure}

\subsection{RJ thermalization and condensation}
%\sec2
We start from a model with $N$ equidistant energy levels $0 \leq E_m =m/N \leq B$
located in the energy band of total width $B$. This corresponds to certain levels
of wealth for $N$ agents with a fraction of agents on level $m$ being $\rho_m$.
The conserved average system energy is $E = \sum_m E_m \rho_m$ and the dimensionless
parameter  $\varepsilon = E/B$ determines its fraction with respect to the maximal system energy $B$.
We call this model the RJ standard (RJS) model.
On the basis of WTH with RJ distribution (\ref{eqrj}) the local 
(normalized) wealth on level $i$
is $(E_i/E)\rho_i$ and the cumulated wealth on levels $[0,m]$ is
$w = \sum^{m}_{i=0} (E_i/E) \rho_i$ with the cumulated fraction of 
population or households $h = \sum^{m}_{i=0} \rho_i$. Computing both 
sums for all values of $m=0,1,\ldots$ provides the 
Lorenz curve $w(h)$. 
Since the Lorenz curve describes the normalized distribution
of cumulated fractions of wealth $w\in[0,1]$ and households 
$h\in[0,1]$ we use the ratio $E_i/E$ (since $E=\sum_i E_i\rho_i$)
in the definition of wealth ensuring that $w=1$ at $h=1$ for the 
total population. 
At given $\varepsilon$ the relation (\ref{eqrj}) and two integrals of energy and norm
determine the physical parameters $T, \mu, S$. In our numerical studies
we use $N=10000$ which practically corresponds  to the continuous limit
with results independent of $N$. 
The dependencies of $T$ and $\mu$ on $\varepsilon$ in the RJS model are
shown in Appendix Figure~{\ref{figA1}. 
As discussed in \cite{rmtprl,ourfiber}
for $\varepsilon >1/2$ the temperature
$T$ becomes negative and at $\varepsilon$ close to unity there is a formation
of an RJ condensate on highest energy levels with $E_m \rightarrow B$
(see Appendix Figure~\ref{figA2}).
Many unusual properties of RJ thermalization
have been discussed in \cite{rmtprl,ourfiber} but for convenience
we provide some details in Appendix and 
Figures~\ref{figA2},~\ref{figA3}  show the dependence of $\rho_m$ on $E_m/B$ for 
certain values of $\varepsilon$ with a clear formation of an RJ condensate 
at small $\varepsilon$ or $1-\varepsilon$. 
Even if the regime with negative temperatures
has been realized in fiber experiments \cite{wabnitz,picozzi3}
we consider that such a regime is not applicable to human society
and hence we consider only the range with $0 \leq \varepsilon \leq 1/2$.

\begin{figure}[htbp]
\begin{center}
\includegraphics[width=0.7\textwidth]{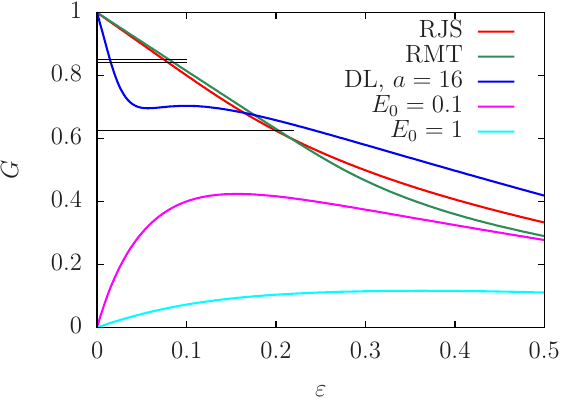}  
\caption{\label{figI_5}
Gini coefficient versus rescaled energy 
$\varepsilon=(E-E_0)/(E_{N-1}-E_0)$ for 
the RJS model (red), RMT model 
(green), DL model (blue; only for the case $a=16$), 
and the EQI model (pink for the offset $E_0=0.1$ and cyan 
for $E_0=1$; same values of $E_0$ are used in Appendix Figure~\ref{figA6}).
The thin black  lines show the values 
of $G=0.852$, $G=0.842$ and $0.626$ for the data of US 2019, World 2021 
and UK 2014. 
The intersection of these lines 
with the red and green  curves correspond to  $\varepsilon$ values
used in  Figs.~\ref{figI_3},~\ref{figI_4}.
}
\end{center}
\end{figure}

The Lorenz curves for the RJS model at several $\varepsilon$ values
are shown in Fig.~\ref{figI_1}. Due to RJ condensate there is a very high fraction
of poor households $f_p$ (with $w \leq 0.02$)
and a small fraction of rich ones  $f_r$ (with $w \geq 0.75$)
who owns a huge fraction of total wealth.
Thus the RJS model naturally describes the big phase of  poor households  
and the oligarchic phase of small fraction of households capturing
the big fraction of total wealth.
At maximal $\varepsilon=0.5$ with ($\mu\to-\infty$) 
all $\rho_m$ are equal and hence the Lorenz curve is $w=h^2$ with
the limiting minimal Gini coefficient $G=1/3$ for the RJS model. 
The dependence of
cumulated wealth $w$ on $h$ and $\varepsilon$ is shown in Fig.~\ref{figI_2}.
It clearly shows the phase of poor households (blue), corresponding to the RJ condensate,
and the oligarchic phase of very rich households (red).
Thus we see that the RJ thermal distribution (\ref{eqrj})
describes the main qualitative features of
wealth inequality of  human society \cite{piketty2}.

\begin{figure}[htbp]
\begin{center}
\includegraphics[width=0.7\textwidth]{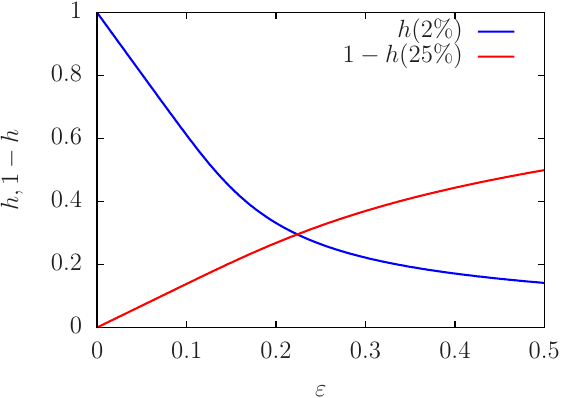}  
\caption{\label{figI_6}
Dependence of fraction of poor households $f_p=h(2\%)$ 
(owning 2\% of wealth)
and fraction of rich oligarchic  households $f_r=1-h(25\%)$
(owning 75\% of wealth) on the rescaled energy 
$\varepsilon = E/B$ for the RJS model.
}
\end{center}
\end{figure}

From Figs.~\ref{figI_1},~\ref{figI_2} we see that for the RJS model
the WTH based on (\ref{eqrj})
captures main elements of wealth inequality
but it is important to see if it can reproduce the real Lorenz curves
for the whole world and specific countries.
For this comparison of WTH theory we choose three cases with
the Lorenz curves for: the whole world from \cite{piketty2}
(integration of front page data gives cumulative $w,h$ values);
USA 2019 case from \cite{usa2019}
and UK 2012-2014 case from \cite{uk2014}. 
These real Lorenz curves are compared with
those obtained from the RJS model (\ref{eqrj}) in Fig.~\ref{figI_3}.
For the comparison  values of $\varepsilon$ are fixed in such a way
that Gini coefficient is the same for
theory and real data curves. The comparison for UK case shows
that there is a good agreement of real and theoretical Lorenz curves
even if there is a certain difference for the range $0.9 \leq h \leq 1$.
The difference is more visible for USA case and the
whole world (Lorenz curves are very similar for these two cases). 
In Appendix Figure~\ref{figA4}, we also show that the
the RJS Lorenz curves have a satisfactory 
agreement with the Lorenz curves for France and Germany
(data are obtained for the year 2010 from \cite{defr}). 

In view of certain differences between real Lorenz curves and 
those obtained from RJS model (see Fig.~\ref{figI_3}) we also
study the case of RJ distribution (\ref{eqrj}) with level energies
$E_m$ taken from a random matrix of size $N=1000$
as it was discussed in \cite{rmtprl}. For this RJ RMT model
the density of states is $\nu=dm/dE_m = \frac{2N}{\pi}\sqrt{1-E^2}$
with typical eigenvalues in the interval $E_m\in[-1,1]$ 
and we shift all $E_m$ to $E_m -E_0$ to have
nonnegative values $E_m \geq 0$ in (\ref{eqrj}).
The comparison of Lorenz curves for US and UK cases
with the results of the RJ RMT model is shown in Fig.~\ref{figI_4}a.
The similarity between real and RMT model data is 
a bit less good then those in Fig.~\ref{figI_3} for the RJS model.
This shows that the density of states $\nu$ can
affect the Lorenz curves. Indeed, we have $\nu=$\ const. for the RJS model
being different from the semi-circle law of RMT model.

To reproduce the real Lorentz curves from
\cite{piketty2,usa2019,uk2014} in a better way we also analyze a 
double-linear (DL) model with energies 
$E_m=m/N$ for $m<N/2$ and 
$E_m=E_{N/2}+a(m-N/2)/N$ for $m\ge N/2$ at $N=10000$
with  $a=16$ ($B=8.5$) for  US and World data,
and $a=3$ $(B=2)$ for UK data.
In this type of model
the density of states takes not one but two values being
$\nu =1 $ and $\nu =1/a$. The existence of two $\nu$ values
can correspond to a society where high wealth energy $E_m$ values
are only accessible to very rich people 
whose  density is lower compared to common people.
The comparison of real Lorenz curves with those of the DL model is shown in
Fig.~\ref{figI_4}b (and its zoomed version Appendix Figure~\ref{figA5})
demonstrating a better proximity
between real Lorenz curves and  those from the DL model
as compared to the results of the RJS model in Fig.~\ref{figI_3}.
However, the DL model has two fit parameters $a, \varepsilon $
while the RJS model has only one $\varepsilon$.

We also remind that for the RJS model the minimal Gini value is $G=1/3$
that is reached at maximal physical value of $\varepsilon=1/2$.
Thus to have $G < 1/3$, we need to significantly modify
the density of states $\nu$. Indeed, we can obtain
a perfect complete wealth and energy equipartition with $w=h$ and $G=0$
for the case when all $E_m = E_0$ values are the same.
In this case, the integrals of energy and norm
give only one conservation law and all states have the same energy and
same population. A small spectrum modification
to $E_m=E_0 +m/N$ with a constant energy offset $E_0$
leads to Lorenz curves being closer to the diagonal
with small Gini values $G < 1/3$ and a finite 
slope $w(h)\approx [E_0/(\varepsilon+E_0)]\, h$ 
at small $h$ (see Appendix with additional discussion of this model 
and related Figure~\ref{figA6}). 
We call this model the equipartition (EQI) model. 

We also show color Figures analogous to Fig.~\ref{figI_2}
for various models discussed above (see Appendix Figure~\ref{figA7}).

\begin{figure}[htbp]
\begin{center}
\includegraphics[width=0.9\textwidth]{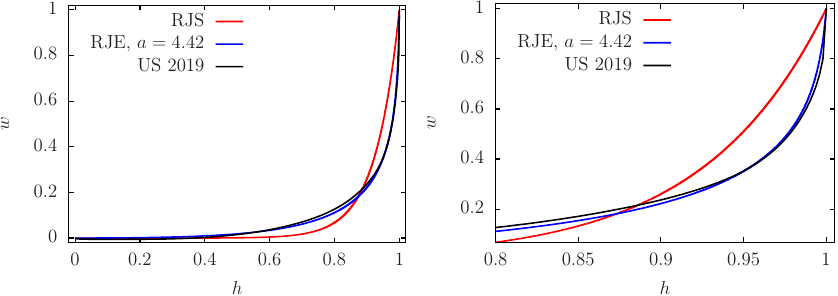}%
\end{center}
\caption{\label{figI_7}
\label{fig_exp_US_2019}
Comparison of the Lorenz curve for the data of US 2019 
(black) with the corresponding curves for the RJS model (red curve; $N=10000$) 
and the RJE model with $a=4.42$ (blue curve; $N=10000$). 
The rescaled energy values $\eps=0.01233$ (RJE) and 
$\eps=0.07420$ (RJS) are obtained by matching the 
Gini coefficient $G=0.8515$.
The value of $a$ is obtained by a fit from the reconstructed spectrum. 
%for the interval $x=m/N\in[0,0.7]$. 
The left (right) panel shows the full range $h\in[0,1]$ (zoomed range 
$h\in[0.8,1]$). }
\end{figure}

The dependence of the Gini coefficient $G$ on $\varepsilon$ is given in 
Fig.~\ref{figI_5} for the different models. 
In global the results show that an increase of $\epsilon$ leads to
a reduction of $G$. Also in Fig.~\ref{figI_6}, we show the dependence of
fractions of poor $f_p$ and rich oligarchic $f_r$ households on 
$\varepsilon$ for the RJS model. 
Thus at $\varepsilon=0.07$ we have $f_p = 0.73$ and $f_r = 0.097$
that is close to the real values $f_p = 0.53$(US), $0.5$(World) and
$f_r =0.09$(US), $0.1$ (World) while for UK we have $f_p=0.32$, $f_r= 0.28$
corresponding to a higher $\varepsilon \approx 0.21$.
Furthermore Fig.~\ref{figI_6} shows
that the fraction of poor households can be significantly reduced and
the fraction of rich households can be increased by increasing 
parameter $\varepsilon$, thus diluting the oligarchic phase.

Finally, we mention that for the RJS model it is possible to work out 
analytic expressions (at $N\to\infty$; see  Appendix Section 3) 
for the Lorenz curve and other quantities that accurately match the numerical data 
(see Appendix Figure~\ref{figA8}). 
These expressions depend on $\mu$ and at small $\varepsilon\le 0.2$ (with $\mu\approx 0$), 
we have $w(h)\approx e^{(h-1)/\varepsilon}$ and 
$G\approx 1-2\varepsilon$, matching 3 values of $G$ in Fig.~\ref{figI_1}. 

\subsection{RJ thermalization  and universality}

Above we presented the comparison of real Lorenz curves of countries and the whole world
with the theoretical results of the RJS model based on the physical phenomenon
of RJ thermalization and condensation.
Since this thermalization is universal for classical systems
when the norm and energy are conserved we expect that other systems
will be also describe by the RJS model and its extensions.

To check this expectation we analyze the capitalization data for
S\&P500 companies at the New York Stock Exchange (NYSE),
companies of London stock exchange and Hong Kong stock exchange.
The data are obtained from the open public sources \cite{sp500},
\cite{london} and \cite{hk} respectively. From these sources we construct
the real Lorenz curves and compare them with those given by
the RJS model for these three cases (see Appendix Section 4
and Figures~\ref{figA8},~\ref{figA9},~\ref{figA10},~\ref{figA11}). The comparison shows that
the RJS model qualitatively describes the real Lorenz curves behavior
approximately with the same level of agreement as it was for countries and the whole world
cases considered before.  It is interesting to note that
for the case of Dow Jones companies with $N=30$ companies the Lorenz curve is very close to the
case of perfect equipartion with $w = h^2$
(see Appendix Figure~\ref{figA12}) and $\epsilon = 0.5, T \rightarrow \infty$
in the RJS model. However, for this case we cannot consider that
these companies form an isolated system.

To obtain an RJ extended (RJE) model giving a better agreement with
the real Lorenz curves for countries and stocks exchange we
make the following extension of the RJS model.

\begin{figure}[htbp]
\begin{center}
\includegraphics[width=0.9\textwidth]{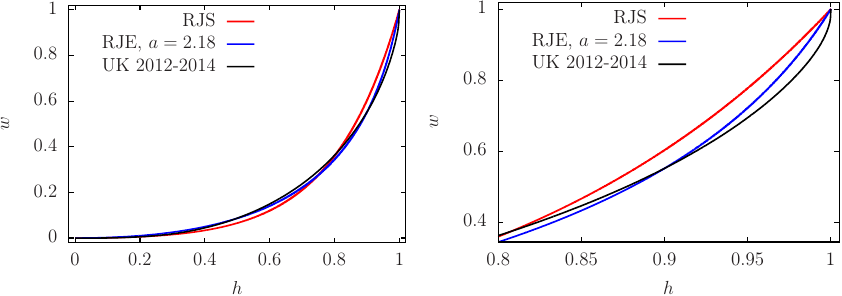}%
\end{center}
\caption{\label{figI_8}
\label{fig_exp_UK_2012-2014}
Comparison of the Lorenz curve for the data of UK 2012-2014 
(black) with the corresponding curves for the RJS model (red curve; $N=10000$) 
and the RJE model with $a=2.18$ (blue curve; $N=10000$). 
The rescaled energy values $\eps=0.1332$ (RJE) and 
$\eps=0.1996$ (RJS) are obtained by matching the 
Gini coefficient $G=0.6255$. 
The value of $a$ is obtained by a fit from the reconstructed spectrum. 
%for the interval $x=m/N\in[0,0.9]$. 
The left (right) panel shows the full range $h\in[0,1]$ (zoomed range 
$h\in[0.8,1]$). }
\end{figure}

\begin{figure}[htbp]
\begin{center}
\includegraphics[width=0.9\textwidth]{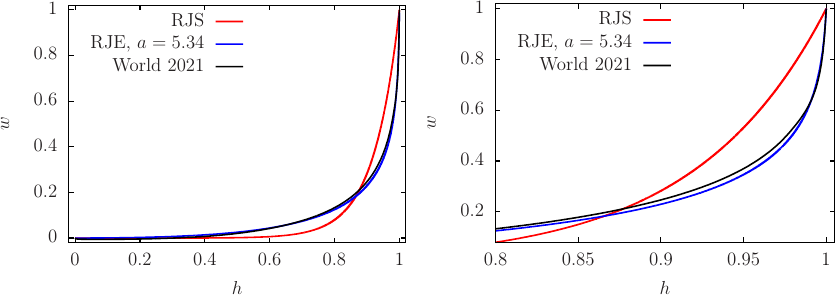}%
\end{center}
\caption{\label{figI_9}
\label{fig_exp_World_2021}
Comparison of the Lorenz curve for the data of World 2021 from \cite{piketty2}
(black) with the corresponding curves for the RJS model (red curve; $N=10000$) 
and the RJE model (\ref{RJEdef}) with $a=5.34$ (blue curve; $N=10000$). 
The rescaled energy values $\eps=0.008553$ (RJE) and 
$\eps=0.07911$ (RJS) are obtained by matching the 
Gini coefficient $G=0.8420$.
The value of $a$ is obtained by a fit from the reconstructed spectrum. 
%for the interval $x=m/N\in[0,0.7]$. 
The left (right) panel shows the full range $h\in[0,1]$ (zoomed range 
$h\in[0.8,1]$). }
\end{figure}

The comparison of the different data with the RJS model shows that typically 
the curves of the RJS model have a slower (final) growth rate. 
Since the latter is 
proportional to the energy $E_m$ one could try an extended model where 
the energy values grow stronger with $m$. One step in this direction 
is the DL model which allowed for a considerable improvement as can be seen 
in the right panel of Fig.~\ref{figI_4} and also its zoomed version
Fig.~\ref{figA5} in Appendix. 
Another possibility is to choose an exponential growth of $E_m$ but in such 
a way that still $E_m\sim m$ for small $m$. This can be achieved by 
the formula 
\begin{align}
\label{RJEdef}
E_m=\frac{\exp({a\,(m/N))}-1}{a}
\end{align}
which we call RJE model 
(RJ extended or RJ exponential model). Here $a$ is an additional 
parameter of the model in addition to the value of $\eps$ which 
is now $\eps=E/B$ with bandwidth $B=E_{N-1}\approx{(e^a-1)/a}$. 
In the limit $a\to 0$, we recover simply the RJS model while with increasing 
values of $a$ the exponential growth of $E_m$ becomes more dominant. 

\begin{figure}[htbp]
\begin{center}
\includegraphics[width=0.9\textwidth]{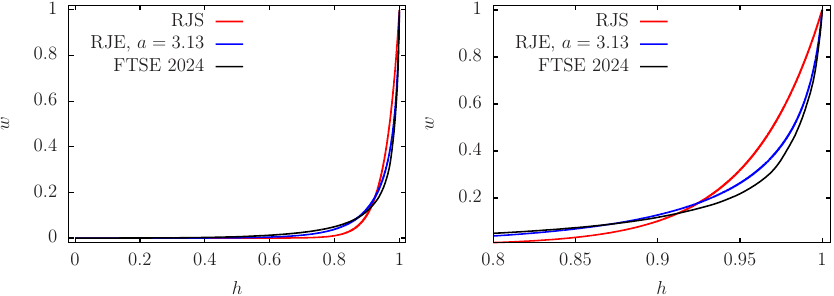}%
\end{center}
\caption{\label{figI_10}
\label{fig_exp_FTSE_2024}
Comparison of the Lorenz curve for the data of the London 
stock exchange FTSE at 31 December 2024 
(black; data from Ref.~\cite{london}) 
with the corresponding curves for the RJS model (red curve; $N=10000$) 
and the RJE model with $a=3.13$ (blue curve; $N=10000$). 
The rescaled energy values $\eps=0.01346$ (RJE) and 
$\eps=0.04376$ (RJS) are obtained by matching the 
Gini coefficient $G=0.9126$.
The value of $a$ is obtained by a fit from the reconstructed spectrum. 
%for the interval $x=m/N\in[0,0.9]$. 
The left (right) panel shows the full range $h\in[0,1]$ (zoomed range 
$h\in[0.8,1]$). }
\end{figure}

The energy spectrum (\ref{RJEdef}) corresponds to a density of states:
\begin{align}
\label{RJEDOS}
\nu(E_m)=\frac{dm}{dE_m}=\frac{d}{dE_m}\left(\frac{N\ln(1+aE_m)}{a}\right)
=\frac{N}{1+aE_m}
\end{align}
which interpolates between a constant density of states 
$\nu(E_m)\approx N$ for $E_m\ll a^{-1}$ (as in the RJS model) 
and a power law decay
$\nu(E_m)\approx N/aE_m\sim 1/E_m$ for $E_m\gg a^{-1}$. 

\begin{figure}[htbp]
\begin{center}
\includegraphics[width=0.9\textwidth]{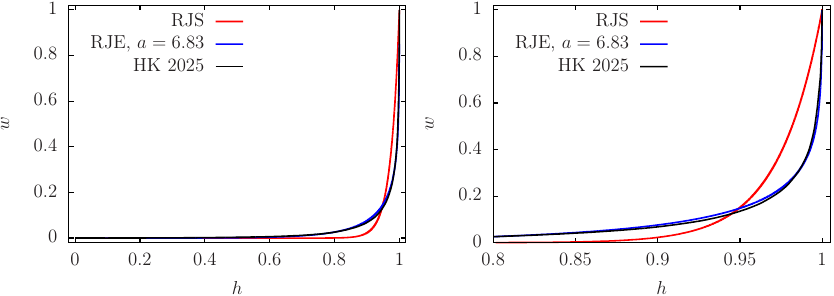}%
\end{center}
\caption{\label{figI_11}
\label{fig_exp_HK_2025}
Comparison of the Lorenz curve for the data of the Hong Kong 
stock exchange at 19 June 2025 (black; data from Ref.~\cite{hk}) 
with the corresponding curves for the RJS model (red curve; $N=10000$) 
and the RJE model with $a=6.83$ (blue curve; $N=10000$). 
The rescaled energy values $\eps=0.0008381$ (RJE) and 
$\eps=0.02648$ (RJS) are obtained by matching the 
Gini coefficient $G=0.9471$.
The value of $a$ is obtained by a fit from the reconstructed spectrum. 
%for the interval $x=m/N\in[0,0.9]$. 
The left (right) panel shows the full range $h\in[0,1]$ (zoomed range 
$h\in[0.8,1]$). }
\end{figure}

\begin{figure}[htbp]
\begin{center}
\includegraphics[width=0.7\textwidth]{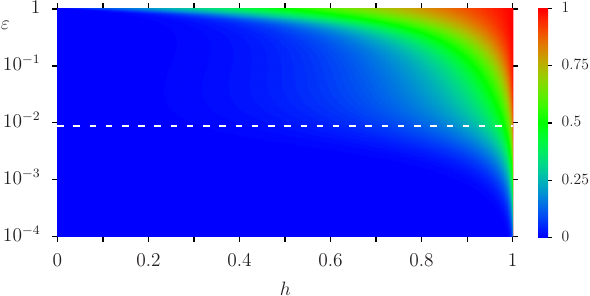}  
\caption{\label{figI_12}
Color plot of cumulated wealth $w$ from the Lorenz curves of the RJE model (\ref{RJEdef})
($N=10000$) with the parameter $a=5.34$. The $x$-axis corresponds to 
the fraction of households $h\in[0, 1]$ 
and the $y$-axis to the rescaled 
energy $\varepsilon=E/B\in[10^{-4}, 1[$ in logarithmic representation. 
The white dashed line corresponds to the value $\eps=0.008553$ obtained by 
matching the Gini coefficient of the RJE model (at $a=5.34, N=10000$) 
with the data of World 2021 from \cite{piketty2} (see Fig.~\ref{figI_9}). Note that 
the color values along the dashed line correspond to the blue RJE 
curve in Fig.~\ref{figI_9}. 
}
\end{center}
\end{figure}

To determine optimal values for the parameter $a$, we compute 
a reconstructed spectrum from a given Lorenz curve of some given data set 
(see Appendix Section 5 for a description and more detailed discussion of 
this spectral reconstruction with Figure~\ref{figA13}) and fit  the reconstructed spectrum 
to the function $E_m\approx C(e^{a\,(m/N)}-1)/a$ with two parameters $C$ and 
$a$. The 2nd parameter $C$ has no importance since one could apply an 
arbitrary fixed factor on (\ref{RJEdef}) without changing the resulting 
Lorenz curve of the RJE model. This is because the construction procedure 
of Lorenz curve involves only the ratio $E_m/E$ 
(with $E$ being the average energy) so that the global energy scale 
(or bandwidth $B$) drops out. 

To fix some procedure, we perform the fit of the reconstructed spectrum 
for two fit intervals for the rescaled level number $x=m/N$ being 
either $x\in[0,0.7]$ or $x\in[0,0.9]$ and select the resulting value of $a$ 
for which the RJE model provides a closer Lorenz curve to the given data set. 
In certain cases, the shorter fit interval provides a better fit value of 
$a$ (cases where the global fit is of reduced quality for small $x$) 
and in other 
cases the longer fit interval is more accurate (cases where the global 
fit is also of rather good quality for small $x$).

The results for US 2019, UK 2012-2014, World 2010, FTSE 2024 (London stock 
exchange) and the Hong Kong 2025 stock exchange are shown in 
Figs.~\ref{figI_7}---\ref{figI_11}, in each case with two panels, top 
for the full range of $h\in[0,1]$ and bottom for the zoomed range 
$h\in[0.8,1]$. Here, we  choose for simplicity 
the value of $N=10000$ for the curves of both RJE and RJS models 
(the RJS curves are also shown for comparison). 
Other values such as $N=1000$ or the given size of the data set, 
give the same Lorenz curves at graphical precision.

In all cases, the agreement of the RJE model
with the data is significantly better than the RJS model. In particular 
for HK 2025, the agreement is close to perfect and even in the zoomed panel 
it is difficult to distinguish the theoretical RJE curve (blue)
from the data (black). For the case UK 2012-2014 the original simpler 
RJS model was already quite good, but also here the RJE model provides 
a significant improvement. The RJE curves for US 2019 and World 2021 are 
also very good, nearly as good as the curve for HK 2025. 
For FTSE 2024 the agreement of the data with the RJE model is a bit 
less perfect (since S\&P500 captures only about 80\% of NYSE)
but still clearly better than the RJS model.

We  also verified that for three other cases  DE 2010, FR 2010 
and NYSE 2025 the RJE model gives a good description with 
values $a=4.2$, $a=3.82$ and $a=2.66$ respectively.
Here the results have also 
a strongly improved agreement of the RJE model with the real data. 
For NYSE 2025 the agreement is slightly less good compared
to other cases
(since S\&P sector captures only about 80 percent of total NYSE)
but even here the RJE model is significantly better 
than the simple RJS model. 

On the basis of presented results we conclude that
the RJ thermalization gives a universal description
of inequality described by the Lorenz curves for countries
and company capitalization at stock exchange.

\subsection{Overview of  wealth thermalization results}

In this Part I   we use the WTH approach (\ref{eqrj}) to describe the wealth distribution
in a closed system that may be a country or the whole world or a stock exchange. 
Our main argument is that in such a system interaction of agents
is described by nonlinear equations with the conservation of two integrals of motion being
total number of agents (norm or total probability analogous to number of system particles)
and total wealth (analogous to total system energy).  Under these conditions
the wealth sharing is described by the universal RJ thermal distribution (\ref{eqrj})
as it is the case for various physical systems
\cite{mayer,landau,zakharovbook,picozzi1,picozzi2,babin,chris,picozzi3,rmtprl,ourfiber}.
The striking feature of RJ thermalization (\ref{eqrj}) is that at low
system  energy  (low $\varepsilon$) there is the physical phenomenon of RJ condensation
when a high fraction of total probability 
is located at lowest energy states that corresponds to the
high fraction of poor households with very low wealth
and also other small fraction of oligarchic households
that monopolizes a big fraction of total wealth.
Thus according to the WTH
 phenomenon a huge wealth inequality in the world \cite{piketty1,piketty2}
  finds a natural thermodynamic explication. We show that 
  the WTH theory gives a good description of
  the Lorenz curves of US, UK and the whole world. 
  It also gives a very good description of capitalization
  of companies at stock exchange of New York, London, Hong Kong
  demonstrating the universality
  of RJ thermalization description.

    On the basis of WTH theory we see that a reduction
  of wealth inequality can be realized by an increase of
  rescaled system energy ($\varepsilon=E/B$). 
This point is illustrated in Fig.~\ref{figI_12} which shows a color plot of 
the Lorenz curves of the RJE model at $a=5.34$ for different 
values $\eps$. For this case there is very good matching with the World 2021 
data at $\eps=0.008553$ as can be seen in Fig.~\ref{figI_9} and 
at larger values of $\eps$ the green/red domain with moderate/high 
wealth increases. The simplest
  way to reach this is to reduce the global dispersion of wealth (given by $B$)
  that can be realized by a high taxation of high wealth revenues.

  In the next Part II, we give more justifications
  for the WTH approach  showing that a nonlinear
  perturbation of social networks
  leads to the RJ thermalization and condensation.
  
%%%%% BEGINNING of PART II

\section{Part II: Dynamical thermalization in social networks}
\setcounter{equation}{0}
\renewcommand{\theequation}{II.\arabic{equation}}
\setcounter{subsection}{0}
\renewcommand\thefigure{II.\arabic{figure}}
\renewcommand\thesubsection{II.\arabic{subsection}}

\subsection{Prologue II}

During last years social networks
gained a significant importance for
communications, opinion formation and relations analysis
in a human society (see e.g. \cite{dorogovtsev10,newmanbook}).
Many fundamental properties of such networks have been studied
with a variety of their applications established for multiple fields of science.
However, all these studies are based on a linear matrix algebra
of links between network nodes provided by their adjacency matrix.
At the same time it may be interesting and important
to analyze the effects of nonlinear
interactions between network nodes (agents or users).
Indeed, it looks rather natural to assume
that in real relations between network agents
nonlinear effects should play an important role.
With this aim we consider a
nonlinear perturbation of two examples of nondirected
social networks. These two networks  
are taken from the database compiled by Newman
\cite{newmannets}. The first one represents a 
collaboration network of scientists studying networks
created by Newman  \cite{newman2001,newman2006,newman2006ref84}
and the second one is a network of politicians generated from Facebook
in \cite{rozemb2018} with network data of \cite{rozemb2018} taken 
from \cite{newmannets}.

Both networks are nondirectional so that their adjacency matrix
is symmetric and can be viewed as a certain Hamiltonian
of a quantum system or a system of coupled linear oscillators.
As in \cite{rmtprl} a nonlinear interaction
is included as a nonlinear frequency shift
on a network site (node, agent).
The dynamical evolution of the obtained system
of nonlinear oscillators has two integrals of motion
being the total energy and total norm
(probability or number of agents).
This corresponds to two integrals of motion
in the evolution of wealth of agents considered
in \cite{boghosian1,boghosian2}
assuming that the wealth is associated to the system energy.

We show that above a certain chaos border of
nonlinear interactions a dynamical thermalization
takes place in the social networks nonlinear dynamical models 
leading to RJ thermal distribution (\ref{eqrj}).
At low values of the total system energy, or total wealth,
RJ condensation emerges in the considered social networks
leading to an enormous phase of poor households
and a small oligarchic fraction capturing a main part of total wealth.
Since social networks can be considered
as realistic models of relations in a human society
the obtained results for dynamical RJ thermalization
provide an additional support and justification
for the WTH related to the origins of wealth inequality
considered in Part I.

In a certain sense the presented studies of dynamical thermalization
in social networks can be considered as an extension of
the studies of the FPU problem \cite{fpu1955}
but with a Hamiltonian part of linear oscillators
based on a typical structure of social network links.
This linear part of the Hamiltonian is
similar to that of the random matrix model NLIRM \cite{rmtprl}.
We attribute this similarity between two systems to the fact that
the nodes in social networks are well connected with each
other and only a few link hoppings are required
to pass from any node to any other node
(see e.g. \cite{dorogovtsev10,newmanbook}).
Indeed, only 4-5 such link transitions, called the Erd\"os number,
are required to connect any node of
the entire Facebook with $8 \times 10^8$ users
to any other node \cite{vigna}.
Due to this the so called ``six degrees of separation'' \cite{dorogovtsev10,newmanbook},
the nonlinear interactions lead to an efficient chaos transition with 
dynamical thermalization and RJ condensation.

\subsection{Model description and numerical methods}

In this work, we consider mainly a nondirected network of $N=379$ known scientists 
with $N_\ell=1828$ links 
from \cite{newman2001,newman2006}, called the 
{\em \SN}, where certain weights are attributed to the links (see 
Eq. (2) of \cite{newman2001}). 
In addition, we also  present a few results for a larger anonymous 
nondirected 
network of $N=5908$ of politicians and $N_\ell=83412$ links, 
called the {\em \PN}, 
obtained from Facebook \cite{rozemb2018}.
In this network all links have the same weight being unity. 
For both cases, links $i\to j$ and $j\to i$ for different nodes 
$j\ne i$ are counted twice in the definition 
of $N_\ell$ and self links $i\to i$ are either absent (\SN)
or taken out (\PN). 

For these networks, we define the adjacency matrix $A_{ij}$ 
by $A_{ij}=w_{ij}$ if there is a link from node $j$ to $i$ and 
where $w_{ij}$ is the weight of this link (which is 1 for the 
\PN) and $A_{ij}=0$ if there is no link $j\to i$. 
For nondirected networks this matrix is symmetric and has 
real eigenvalues. 
We briefly mention that using this matrix $A$ one can define 
a stochastic matrix $S$ by normalizing the columns of $A$ where 
eventual empty columns of $A$ for {\em dangling nodes} 
are replaced by $1/N$ entries in $S$ but this does not happen 
for the two networks above. Then the Google matrix $G$ with 
elements $G_{ij}$ is defined as $G_{ij}=\alpha S_{ij}+(1-\alpha)/N$
with the damping factor $\alpha$ and its typical value $\alpha=0.85$. 
The reason for this is to 
obtain a unique leading eigenvector of $G$ with eigenvalue $\lambda=1$, 
called the {\em PageRank}, which can be computed efficiently by 
the power method. However, in the two networks here also for $\alpha=1$ 
there is a unique PageRank with a gap between $\lambda=1$
and other eigenvalues with $|\lambda|<1$. We refer to Ref. \cite{rmp2015} 
for a review on the network matrices $A$, $S$ and $G$ and the PageRank
(for the more general case of directed networks) . 

We associate to each of these networks a ``quantum Hamiltonian'' 
by 
\begin{align}
\label{eqHdef}
H=A+\kappa H^{\rm GOE}
\end{align}
where $\kappa$ is a small parameter and 
$H^{\rm GOE}$ is a random GOE-matrix \cite{wigner,mehta,guhr} 
with a semicircle density of states of radius unity (at 
large $N\gg 1$) which corresponds to 
random gaussian matrix elements with zero mean and 
variance $\langle (H^{\rm GOE}_{n,n'})^2\rangle = (1+\delta_{n,n'})/(4(N+1))$.
The contribution of $\kappa H^{\rm GOE}$ corresponds to a small 
static perturbation of the network links which we also expect 
in real life. 
In this work, we mainly used one specific random realization of $H^{\rm GOE}$ 
for a given network of size $N$ but we verified with  different 
realizations that the results do not depend on this choice. 

Using the matrix $H$, we consider the nonlinear oscillator system
\begin{align}
\label{eqNLeq1}
i\frac{\partial\psi_n(t)}{\partial t}&=\sum_{n'=1}^N H_{n,n'} \psi_{n'}(t) 
+   \beta \vert\psi_n(t)\vert^2\psi_n(t) 
\end{align}
with complex oscillator amplitudes $\psi_n(t)$ for nodes $n$ 
where $\beta$ is the parameter of the nonlinear perturbation. 
The dynamical system (\ref{eqNLeq1}) has two integrals of motion being
 the {\em norm} $\mathcal{N}$ and the 
{\em classical energy} $\mathcal{H}$:
\begin{align}
\label{eqIntegrals}
\mathcal{N}&=\sum_n |\psi_n|^2\quad,\quad
\mathcal{H}=\sum_{n,n'} \psi_n^* H_{n,n'}\psi_n+\frac{\beta}{2}
\sum_n |\psi_n|^4  \; .
\end{align}
 In fact, the system (\ref{eqNLeq1}) is actually 
a classical Hamiltonian system with Hamilton function $\mathcal{H}$ 
if we write $\psi_n=(q_n+ip_n)/\sqrt{2}$ with canonical coordinates 
$q_n$ and $p_n$.
In this work we fix the norm by $\mathcal{N}=1$. 
The case $\mathcal{N}\neq 1$ can be transformed to the 
case $\mathcal{N}=1$ by a suitable rescaling of 
$\psi$ and $\beta$. 

For $\beta=0$ this system is simply the time dependent Schr\"odinger 
equation for a quantum system with the state 
$\ket{\psi}=\sum_n \psi_n\ket{n}$ (with $\hbar=1$). 
It is useful to diagonalize $H$ by 
$H\phi^{(m)}=E_m\phi^{(m)}$ with eigenvectors $\phi^{(m)}$ 
(and components $\phi^{(m)}_n$ ) and to define 
amplitudes $C_m$ in eigenmode space by
\begin{align}
\label{eqCmdef}
C_m=\sum_n\phi^{(m)*}_n \psi_n\ .
\end{align}
Here we write more general formulas with the complex conjugate of 
$\phi^{(m)}_n$ which are also valid for the more general case 
of complex hermitian matrices $H$ even though in our case $H$ 
is real symmetric where it is possible to choose real eigenvectors 
$\phi^{(m)}_n\in\mathbb{R}$.
Also we assume that the eigenvectors are orthogonal
\begin{align}
\label{eqCmorth}
\sum_n \phi^{(\tilde m)*}_n\phi^{(m)}_n=\delta_{\tilde m,m}
\end{align}
so that the matrix $U_{nm}=\phi^{(m)}_n$, containing 
the eigenvectors in its columns, is orthogonal (or unitary for 
complex hermitian $H$) with the diagonalization identity 
$H=U\hat EU^\dagger$ where $\hat E_{\tilde m,m}=E_m\delta_{\tilde m,m}$ 
and the inverse transformation of (\ref{eqCmdef}) being 
\begin{align}
\label{eqCminverse}
\psi_n=\sum_m\phi^{(m)}_n C_m\ .
\end{align}
Using (\ref{eqCmdef})-(\ref{eqCminverse}) one can show that 
the nonlinear system (\ref{eqNLeq1}) can be rewritten in the 
eigenmode amplitudes $C_m$ as
\begin{align}
\label{eqNLeq2}
i\frac{\partial C_m}{\partial t}&
=E_m C_m+\beta\sum_{m_1,m_2,m_3} Q_{mm_1m_2m_3}C_{m_2}^* 
C_{m_3}^{\phantom *} C_{m_1}^{\phantom *}
\end{align}
with nonlinear transition coefficients 
\begin{align}
\label{eqNLcoeffs}
Q_{mm_1m_2m_3}&=\sum_n \phi_n^{(m)*}\phi_n^{(m_1)}
\phi_n^{(m_2)*}\phi_n^{(m_3)}\ .
\end{align}
At $\beta=0$ the solution of this system is 
$C_m(t)=e^{-iE_mt}C_m(0)$ and for small values of $\beta$ 
there is typically a complicated KAM scenario with a transition to a 
chaotic region in a large part of the phase space for sufficiently 
large $\beta$. 

The two integrals of motion (\ref{eqIntegrals}) can be written in the 
eigenmode amplitudes $C_m$ as:
\begin{align}
\label{eqIntegralsCm}
\mathcal{N}&=\sum_m |C_m|^2\quad,\quad
\mathcal{H}=\sum_{m} E_m\,|C_m|^2 + \Hnl\quad,\\
\label{eqHnonlin}
\Hnl&=\frac{\beta}{2}\sum_n |\psi_n|^4
=\frac{\beta}{2}
\sum_{m_0,\ldots,m_3} Q_{m_0m_1m_2m_3}C_{m_0}^*C_{m_2}^* 
C_{m_3}^{\phantom *} C_{m_1}^{\phantom *}\ .
\end{align}

As in \cite{rmtprl},  we solve the nonlinear system by a symplectic fourth 
order integrator \cite{forest} also known as one of the splitting methods 
\cite{integrator1,integrator2}.
More details about our implementation of this method 
can be found in the supplementary material of \cite{rmtprl}. 
This method has the advantage that it respects the symplectic 
symmetry of the problem. In particular the first integral $\mathcal{N}$
is exactly conserved (up to usual numerical rounding errors) 
while the second integral $\mathcal{H}$ varies only slightly in time 
with a small error $\sim dt^4$ 
and can be used to verify if the choice of $dt$ is appropriate. 

As initial condition, we usually  start with an eigenstate located 
at an initial energy mode $m_0$ with $C_m(t=0)=\delta_{m,m_0}$. 
For the \SN, we also consider a few cases where the 
initial state is localized 
on some specific node $n_0$ with $\psi_n(t=0)=\delta_{n,n_0}$ 
(i.e. $C_m(t=0)=\phi_{n_0}^{(m)*}$). 
As in \cite{rmtprl} an integration time step $dt$ is chosen
in such a way that the second integral of motion 
$\mathcal{H}$ is conserved with a high relative precision
being below $10^{-4}$ for most initial modes (or $\sim 10^{-2}$ 
for very few boundary modes; note that due to the method 
the first integral $\mathcal{N}$ is always conserved ``exactly'' 
with numerical precision $\sim 10^{-15}$).

\subsection{Theoretical elements of RJ thermalization}

At sufficient strong values of $\beta$ and long times $t$, 
we expect the nonlinear system to be chaotic and the amplitudes 
to be somehow ergodic. Assuming a simple behavior $\psi_n(t)\sim 1/\sqrt{N}$ 
the typical value of the nonlinear energy contribution in $\mathcal{H}$ is 
$\Hnl\sim \beta/N$ which can be neglected at $N\gg 1$ and then 
we have:
\begin{align}
\label{eqHlin1}
E=\mathcal{H}\approx \sum_m E_m |C_m(t)|^2
\end{align}
where $E$ is the specific energy value of the integral $\mathcal{H}$. 
However, in real systems, such as the network generated matrices $H$ 
considered here, the assumption $\psi_n(t)\sim 1/\sqrt{N}$ may 
not be realistic, especially at initial times. 
More generally, the nonlinear energy 
contribution is 
$\Hnl=\beta/\xi_{\rm IPR}$ where 
\begin{align}
\label{eqIPR1}
\xi_{\rm IPR}=\left(\sum_n |\psi_n|^4\right)^{-1}
\end{align}
is the {\em inverse participation ratio} (IPR) on the state $\psi_n$.
The IPR corresponds roughly to the number of populated nodes and is
broadly used in the problems of disordered solids (see e.g. \cite{mirlin}).
The identity (\ref{eqHlin1}) is still valid if $\beta/\xi_{\rm IPR}$ 
can be neglected for sufficiently large values of $\xi_{\rm IPR}\gg \beta$. 
For the case of the eigenmode initial condition with 
$\psi_n(t)\approx \phi^{(m_0)}_n$ at small times it is the IPR 
of the eigenstate $\phi^{(m)}$ which fixes the value 
of $\Hnl$ at initial times. We discuss the IPR 
for the eigenstates of both 
models in the next sections pointing out that 
 certain eigenmodes may have relatively small IPR values (depending on 
$E_m$ and $\kappa$). 

In any case, even if the initial value of $\Hnl$ 
is not very small, we expect that it decays with time $t$ and 
that (\ref{eqHlin1}) holds at large times (assuming a chaotic behavior, 
i.e. no KAM localized state for very small $\beta$). 
This situation corresponds to a microcanonical ensemble 
with energy conservation (\ref{eqHlin1}) and an additional second 
constraint 
\begin{align}
\label{eqNorm11}
1=\mathcal{N}=\sum_m |C_m(t)|^2\ .
\end{align}
This special microcanonical 
ensemble can be treated analytically in a simple way only for small 
energies (temperatures) with $E$ in the lower part of the spectrum $E_m$ 
and it is more convenient to replace it with a grand canonical ensemble 
with a probability density 
\begin{align}
\label{eqProbCm}
P(\{C_m\})&=\frac{1}{Z}\exp\left(-\sum_m\frac{E_m-\mu}{T}|C_m|^2\right)\quad,\\
\nonumber
Z&=\int \prod_m d^2 C_m \exp\left(-\sum_n\frac{E_m-\mu}{T}|C_m|^2\right)\\
&=\pi^N\prod_m \frac{T}{E_m-\mu}
=\pi^N\prod_m \rho_m\quad\mbox{with}\quad
\rho_m=\langle |C_m|^2\rangle=\frac{T}{E_m-\mu}
\label{eqZpart1}
\end{align}
being the (thermalized) average {\em occupation probability of the 
mode} $m$. Here the temperature 
$T$ and the chemical potential $\mu$ are determined such
that both constraints are verified in average:
\begin{align}
\label{eqConstraints1}
1=\sum_m\rho_m\quad,\quad E=\sum_m E_m\rho_m\ .
\end{align}
(See also \cite{rmtprl} and Appendix A.1 below for more details on this.) 
The probability density (\ref{eqProbCm}) corresponds to the RJ thermalization 
for classical fields.  The condition $\rho_m>0$ (for all $m$) ensures 
that there is only a unique physically valid solution of (\ref{eqConstraints1}) 
with either $T>0,\,\mu< E_1$ or $T<0,\,\mu>E_N$.

For the numerical system evolution 
we compute the time average $\rho_m(t)=\langle|C_m(\tilde t)|^2\rangle$ 
over time intervals $t/2<\tilde t\le t$ for successive discrete time 
values $t=2^l\le t_{\rm max}$ with $l=2,3,\ldots,l_{\rm max}$ and
$t_{\rm max}$ being typically 
$2^{22}-2^{24}$ for the \SN\ and $2^{19}-2^{20}$ for the \PN. These 
values for $\rho_m(t)$ can be compared 
to the thermalized theoretical values $\rho_{m,RJ}=T/(E_m-\mu)$ 
where $T$ and $\mu$ are determined from the constraints (\ref{eqConstraints1})
using the value $E=\langle E\rangle=\sum_m E_m\rho_m(t)$ to fix the 
energy from the numerically obtained values of $\rho_m(t)$. 

Note that for the eigenmode initial condition 
with $C_m(t=0)=\delta_{m,m_0}$, we typically have $E\approx E_{m_0}$ 
if $\Hnl(t=0)$ can be neglected. 
However, in case of relatively large $\beta$ values 
and small IPR, with a significant 
initial value of $\Hnl(t=0)$, we have the more precise 
relation $E=E_{m_0}+\Hnl(t=0)$
which may give a significant energy shift in the linear part at 
larger times scales 
$\langle E\rangle=\mathcal{H}-\Hnl(t)=E-\Hnl(t)\approx E
=E_{m_0}+\Hnl(t=0)$ 
assuming that $\Hnl(t)$ becomes small for large $t$ (for 
``ergodic states'' in a thermalized regime). However, the initial 
value $\Hnl(t=0)$ may be rather large such that $E_{m_0}$ and 
$\langle E\rangle$ are rather different. Therefore, 
it is more appropriate to use $\langle E\rangle=\sum_m E_m\rho_m(t)$ 
(with numerical values of $\rho_m(t)$ at large $t$) rather than $E_{m_0}$ 
to estimate the value of $E$ 
to determine $T$ and $\mu$ from (\ref{eqConstraints1}) and to compute 
the theoretical RJ values which are to be compared with the numerical results.
We see in the next sections that the numerical data $\rho_m(t)$ 
indeed approach the thermalized theoretical values for sufficiently 
large $t$ and the \SN\ while for the \PN\ the situation is more difficult 
due to a limited available integration time. 

Another quantity of thermalization 
is the entropy of the 
system as a function of either $\langle E\rangle$ or $t$. 
There are two points of view to define the entropy. The first one is to use 
the discrete occupation probabilities $\rho_m$ 
and  define the quantum von Neumann entropy by
\begin{align}
\label{eqSq}
S_q=-\sum_m \rho_m\,\ln(\rho_m)\ .
\end{align}
The latter can also be viewed as the entropy of the associated quantum 
system of the $N$ levels of the Hamiltonian $H$ with 
neglected nonlinear term. 

The second point of view is based on the underlying classical nonlinear 
oscillator system with the classical Boltzmann 
entropy:
\begin{align}
\label{eqSBDef}
S_B&=-\int \prod_m d^2 C_m\ P(\{C_m\})\ln\left(P(\{C_m\})\,h_B^N\right)
\end{align}
where $P(\{C_m\})$ is the classical probability density (in some 
statistical ensemble) of 
the oscillator amplitudes $C_m$ and $h_B$ is a small 
constant which compensates the physical dimension 
in the logarithm. The interpretation of this constant is that we 
use the discrete probabilities $p(\{C_m\})=P(\{C_m\})\,h_B^N$ 
of finding a micro-state in 
a given elementary cell of volume $h_B^N$ at phase space point 
$\{C_m\}$ to define the entropy 
by a sum over a grid of such elementary cells: 
$-\sum p(\{C_m\})\ln(p(\{C_m\})$ 
which gives exactly the above expression 
(\ref{eqSBDef}). Below, we give an explicit numerical choice for 
the parameter $h_B$ which corresponds to a certain constant offset 
in the definition of $S_B$. 

The formula (\ref{eqSBDef}) is numerically not convenient since the classical 
probability density 
in phase space is not easily available from the trajectory $C_m(t)$ 
and the integral itself (over many 
variables) is also difficult to evaluate. 
In thermal equilibrium, we can replace $P(\{C_m\})$ by 
(\ref{eqProbCm}) which gives
\begin{align}
S_B&=\ln(Z/h_B^N)+\sum_m\frac{E_m-\mu}{T}\langle|C_m|^2\rangle
=\ln(Z/h_B^N)+N\\
\label{eqSB}
&=\sum_m\ln\left(\rho_m/h_B\right)+N\left(1+\ln\pi\right)\ .
\end{align}
The expression (\ref{eqSB}) is also more generally valid (outside 
thermal equilibrium) if we assume independently gaussian 
distributed amplitudes:
\begin{align}
\label{eqPGauss}
P(\{C_m\})\sim \exp\left(-\sum_m a_m|C_m|^2\right)
\end{align}
with arbitrary coefficients $a_m$ related to 
$\rho_m=\langle|C_m|^2\rangle=1/a_m$. In thermal equilibrium we have 
$a_m=(E_m-\mu)/T$ but we may assume that (\ref{eqPGauss}) is 
also valid at sufficiently large finite iteration times if $\rho_m=1/a_m$ is 
computed as some suitable time average over the trajectory. 
However, we insist that this assumption is not necessarily very exact 
especially at small $t$ and that outside thermal equilibrium 
(\ref{eqSB}) is only a convenient approximation of (\ref{eqSBDef}) 
in terms of parameters $\rho_m$ obtained from the numerical procedure. 

To have reasonable numerical values of $S_B$ which are (mostly) positive, 
we choose the numerical value $h_B=1/N^2$ for the data 
and figures presented in this work (the numerical choice 
of $h_B$ defines only a certain constant offset in the definition 
of $S_B$). Since typical values of 
$\rho_m$ are $\sim 1/N$ (at rather larger $T$) this gives indeed 
$\rho_m\gg h$. In particular for uniform $\rho_m=1/N$ 
we have $S_B/N=\ln(N)+1+\ln\pi$ which is comparable to $S_q=\ln(N)$. 
However, at very small times and/or small values of $\beta$ it 
is still possible that many values of $\rho_m$ are below $1/N^2$ which 
gives potentially negative values of $S_B$. This is an artificial effect 
of the classical oscillator model and we remind that the entropy of classical 
systems (oscillators, ideal gas etc.) typically 
behaves as $S\sim\ln(T)$ for small 
$T$ with a logarithmic singularity at $T\to 0$. 
For practical reasons and since $S_B$ is extensive, we consider 
typically the entropy per mode $S_B/N$ which has comparable numerical values 
to $S_q$ assuming $\rho_m\sim 1/N$.

We note that in our model the 2nd law of thermodynamics applies to the 
2nd entropy $S_B$ and technically it does not apply to $S_q$. In particular, 
it is possible that $S_q(t)$ may temporarily decrease with $t$ for certain 
specific situations (see below) while $S_B(t)$ always increases with $t$ 
(except for a few cases with a very minimal decrease at very short times 
and/or small values of $\beta$ below the chaos border). 
Furthermore, the usual thermodynamic relation $dS/dE=1/T$ only holds for 
$S=S_B$ (and not for $S_q$), that can be verified by a rather simple calculation 
from (\ref{eqSB}) using $\rho_m=T(E)/(E_m-\mu(E))$. 
In particular the terms $\sim d\mu(E)/dT$ cancel exactly due to 
the two constraints (\ref{eqConstraints1}).

\subsection{Netscience network model}

In this section, we present the results for the \SN. Note that 
additional results and figures related to this section 
(and also the subsequent sections) are presented (and sometimes discussed 
in more detail) in Appendix B. 

The issue of thermalization depends in a sensitive way 
on the ``ergodic'' structure of the eigenvector components 
$\phi^{(m)}_n$ of the matrix $H$ given in (\ref{eqHdef}). 
We  compute the eigenvalues and eigenvectors of 
this matrix for the \SN\ with $N=379$ 
for different values of the parameter $\kappa$. 
For certain eigenvectors of modes $m$ with minimal 
energies $E_m$, $m=1,2,3,\ldots$, we also determine the ranking index 
$K_m$ such $|\phi^{(m)}_n|$ is ordered in $n$ in this index, i.e. 
$|\phi^{(m)}_n|\ge |\phi^{(m)}_{n'}|$ if $K_m(n)<K_m(n')$. 
The PageRank eigenvector is computed at damping factor $\alpha=1$ and 
$\alpha=0.85$. This  is the leading eigenvector of $G(\alpha)$; see 
text above (\ref{eqHdef}) and \cite{rmp2015}.
Both cases have a similar ranking index $K$.

\begin{table}
\caption{\label{tab1}
Table of node names with largest eigenvector components. 
The first three columns show names of 10 top nodes in $K_m$-rank order 
($K_m=1,\ldots,10$) 
obtained by ordering the eigenvectors components of 
modes $m=1,2,3$ of lowest energies $E_m$ 
with decreasing values of $|\phi_n^{(m)}|$ (in $n$ for given value of 
$m$). The 
4th (6th) column corresponds to the $K$-rank order for the PageRank 
at $\alpha=1$ ($\alpha=0.85$) and the 5th column contains 
the PageRank probabilities (at $\alpha=1$) for the node list 
in the 4th column. }
\begin{tabular}{@{}llllll@{}}
\toprule
$K_1$ & $K_2$ & $K_3$ & $K(\alpha=1)$ & $P(\alpha=1)$ & $K(\alpha=0.85)$ \\
\midrule
 Barabasi &  Pastorsatorras & Newman &  Barabasi & 0.03064 &  Barabasi \\
 Jeong &  Sole & Pastorsatorras &  Newman & 0.02349 &  Newman \\
 Albert &  Vespignani & Vespignani &  Jeong & 0.01839 &  Sole \\
 Oltvai &  Newman & Watts &  Pastorsatorras & 0.01736 &  Jeong \\
 Ravasz &  Valverde & Girvan &  Moreno & 0.01532 &  Pastorsatorras \\
 Bianconi &  Watts & Moore &  Vespignani & 0.01532 &  Boccaletti \\
 Demenezes &  Ferrericancho & Stauffer &  Sole & 0.01532 &  Vespignani \\
 Dezso &  Girvan & Sneppen &  Boccaletti & 0.01226 &  Moreno \\
 Vicsek &  Montoya & Park &  Kurths & 0.01124 &  Kurths \\
 Yook &  Moore & Lusseau &  Vazquez & 0.01124 &  Stauffer \\
\botrule
\end{tabular}

\end{table}

In Tab.~\ref{tab1}, we present the names of the top 10 nodes 
for the eigenmodes with $m=1,2,3$ and also both PageRank vectors 
(at $\alpha=1$ and $\alpha=0.85$) ordered by their respective 
ranking index $K_m$ or $K(\alpha)$. The data in this table was 
computed for $\kappa=0$ but the top rankings of the eigenmodes at 
$\kappa=0.5$ are actually identical to those of $\kappa=0$. 
The top 4 PageRank names at $\alpha=0.85$ being
{\em Barabasi}, {\em Newman}, {\em Sole} and {\em Jeong} also appear 
as nodes with strong ``community centrality'' in \cite{newman2006} 
(see additional data at Ref. [84] therein). They also play an 
important role in the PageRank at $\alpha=1$ with rank values 
$K=1,2,7,3$ respectively. {\em Barabasi} and {\em Jeong} occupy 
the positions $K_1=1$ and $2$ in the first eigenmode ranking for $m=1$ 
while {\em Newman} and {\em Sole} appear with $K_2=4$ and $2$ in the 2nd 
eigenvector. {\em Newman} also holds the first ranking position 
$K_3=1$ in the 3rd eigenvector. We also mention 
that each of the three eigenvectors with modes a $N+1-m$ ($m=1,2,3$, 
$N=379$) at largest energies (not shown in the table) has 
actually a rather strong overlap 
with top nodes with the lowest modes $m$ (e.g. $\phi^{(1)}$ and $\phi^{(379)}$ 
have some common node names in their top ranking index and similarly 
for the 2nd and 3rd eigenvectors) while there is no overlap between $m=1$ and 
$m=2$. 

\begin{figure}[htbp]
\begin{center}
\includegraphics[width=0.95\columnwidth]{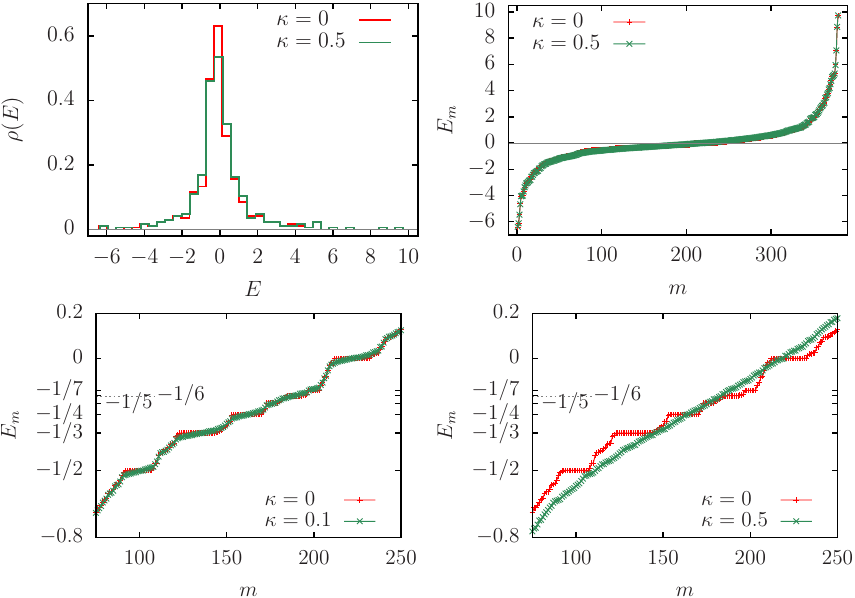}%
\end{center}
\caption{\label{figII_1}
\label{fig_specNet}
Spectral properties of the eigenvalue spectrum of the matrix 
$H$ for the \SN\ with $N=379$ eigenvalues $E_m$ ($E_1<E_2<\ldots <E_N$). 
The top left panel shows the density of states 
for $\kappa=0$ and $\kappa=0.5$ normalized by $\int_{E_1}^{E_N} dE\,\rho(E)=1$ 
with histogramm bin width $dE=10(E_N-E_1)/N\approx 0.436$. 
The top right panel shows the eigenvalue $E_m$ versus index $m$ for 
$\kappa=0$ and $\kappa=0.5$. 
Note that $m/N\approx \int_{E_1}^{E_m} dE\,\rho(E)$. 
Bottom panels show $E_m$ versus $m$ in a zoomed range
for either $\kappa=0$ and $\kappa=0.1$ (left) or 
$\kappa=0$ and $\kappa=0.5$ (right). The plateau values 
for $\kappa=0$ (red data points) at $E_m=0$ and $E_m=1/p$ for $p=2,\ldots,7$
correspond to degenerate energy levels. 
Many eigenvectors of these energies (and other energies with 
nice rational values) have a small support length $l(m)\ll N$ 
where $l(m)$ is the number of non-zero values of eigenvector components 
(with numerical precision $10^{-12}$). These degeneracies are 
lifted by small GOE perturbations at $\kappa=0.1$ or $\kappa=0.5$. 
Bottom and top eigenvalues (at $\kappa=0.5$) are 
$-6.38,-6.12,-5.46,-4.67$ and $5.94, 7.06,8.86,9.73$. 
(Bottom/top eigenvalues at $\kappa=0$ are very close).}
\end{figure}

Fig.~\ref{fig_specNet} shows in different panels the density 
of states of $H$ at $\kappa=0$ and $\kappa=0.5$ and also 
the dependence of $E_m$ on $m$. The global spectral energy band
is in the range $E_1\approx -6.4$ and $E_N\approx 9.7$ (see caption 
of this figure for 4 bottom and top eigenvalues) with 
strong gaps of boundary eigenvalues while the density of states 
has peaked structure around $E=0$, with a slightly stronger peak 
for $\kappa=0$. The global dependence of $E_m$ on $m$ seems rather 
similar between $\kappa=0$ and $\kappa=0.5$ but the zoomed bottom 
panels show that at $\kappa=0$ there are several plateau values 
of degenerate levels at $E=0$ and $E=-1/p$ for $p=2,\ldots,7$ which 
are lifted by small GOE perturbations. At $\kappa=0.1$ the degeneracies 
are only weakly lifted and one can still see the effect of them in 
the (zoomed) $E_m$ vers $m$ curve while at $\kappa=0.5$ this 
curve is essentially a straight line in the shown interval 
$-0.8<E_m<0.2$. 

In order to understand these degeneracies we have analyzed 
the eigenvector structure in more detail by computing for each 
eigenvector $m$ (at $\kappa=0$) the {\em support length} $l(m)$ 
which we define as the number of nodes $n$ with $\phi^{(m)}_n\neq 0$ 
(or more precisely with $|\phi^{(m)}_n|>10^{-12}$ due to the 
limited numerical precision) and also the IPR (for $\kappa=0$ and 
$\kappa=0.5$). 

It turns out that the eigenvectors of the degenerate energies 
(at $\kappa=0$) visible in Fig.~\ref{fig_specNet} have small 
values of the support length in the interval $8\le l(m)\le 48$
($l(m)=48$ for $E_m=0$). There are also other eigenmodes (non  
degenerate or only with a double degeneracy ) with very small support 
length in the range $2\le l(m)\le 8$. In total there are 104 out 
of 379 eigenmodes with $l(m)\le 48$. These modes are all characterized 
by energies $E_m=p/q$ with nice rational values (maximal $q=420$ 
and other $q\le 12$). These points are also illustrated in 
In Fig.~\ref{figB1} in Appendix B.1 which provides also a more 
detailed discussion on this. 

The IPR values of the eigenvectors at $\kappa=0$ (in the 
interval $1.77\le \xi_{IPR}\le 45.98$) are actually not strongly 
correlated to the support length (see Fig.~\ref{figB2}). 
Globally the IPR is rather small, also for modes with maximal 
$l(m)=N$, and for some modes with small $l(m)$ the IPR value may be close 
to $\xi_{\rm IPR}\approx 20$ (about 50\% of the possible maximal value).

Globally, the \SN\ and its related adjacency matrix has some specific 
algebraic structure explaining these modes. Even though the \SN\ 
has only one single component of maximal size $N=379$  there is some hidden 
subblock structure in some other base obtained by linear combinations 
of certain states. 

We mention, without going into much details, that for $\kappa=0$ and 
even strong interaction values such as $\beta=10$, typical states 
with initial modes $m_0$ in the band center do not thermalize well to all modes 
$E_m$ according to the RJ-values of $\rho_m$. For many values of 
$m$ (corresponding to the degenerate modes) the values of $\rho_m$ stay 
very small even for long times such as $t=2^{24}$. For this reason, 
we  focus in the following 
on the case $\kappa=0.5$ with a rather significative GOE-perturbation which 
clearly lifts the degeneracies, where all eigenvectors have the 
maximal value $l(m)=N$ and where the IPR values are roughly a factor 10 
larger than for the case $\kappa=0$ (see also Fig.~\ref{figB2}). 
Only a few number of boundary modes, which are clearly in the perturbative 
regime due to the large gaps, have roughly the same IPR values 
between $\kappa=0.5$ and $\kappa=0$.
We assume that it is natural to have in human society
a presence of such small random links between society members
that can appear due to global information
sources (e.g. radio, TV). Thus we present results mainly for
the case at relatively small $\kappa =0.5$
where the RMT perturbation takes out the degeneracies present at $\kappa=0$.

The values of typical Lyapunov exponents (obtained for
initial mode initial conditions), show that at $\beta=10$ 
and $\kappa=0.5$ all modes are in the chaotic regime (see Fig.~\ref{figB3}).
Even for a very small value of $\beta=0.2$ the values of Lyapunov exponent 
is not very small for modes with $-2<E_{m_0}<2$ while boundary modes 
are in the KAM regime. 
The basic properties of the theoretical thermalized values of $\rho_m$ 
for the energy spectrum of the \SN\ and the energy dependence of 
temperature $T$ and chemical $\mu$ are illustrated in 
Figs.~\ref{figB3} and \ref{figB4}.

\begin{figure}[htbp]
\begin{center}
\includegraphics[width=0.95\columnwidth]{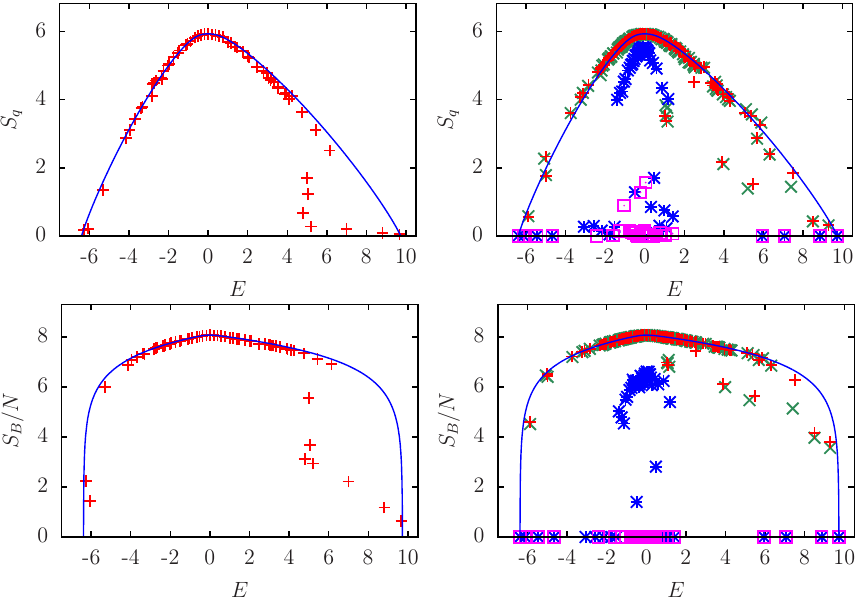}%
\end{center}
\caption{\label{figII_2}
\label{fig_SENet}
Entropy $S_q$ ($S_B/N$) versus energy $E$ in top (bottom) panels 
for certain cases of the \SN\ with $\kappa=0.5$, $N=379$. 
Left panels correspond to 64 selected modes at $\beta=4$ and 
$t=2^{22}$ (red $+$ symbols) and right panels 
correspond to 128 selected modes at $\beta=10$, $t=2^{22}$
(red $+$ symbols), $\beta=10$, $t=2^{20}$ (green $\times$ symbols), 
64 modes at $\beta=0.2$, $t=2^{22}$ (blue $*$ symbols) 
and 32 modes at $\beta=0.05$, $t=2^{22}$ (pink $\square$ symbols) 
In the right bottom panel the data points with $S_B<0$ (certain 
points for $\beta=0.2$ and all points for $\beta=0.05$) 
have been shifted up to $S_B=0$. 
The blue line shows the energy dependence of the theoretical 
thermalized entropy for both entropy quantities. 
$S_q$ ($S_B$) has been computed by Equation~(\ref{eqSq}) 
(Equation~\ref{eqSB} with $h_B=1/N^2$) using $\rho_m$ values 
obtained as the time average $\rho_m=\langle |C_m(\tilde t)|^2\rangle$ 
for $t/2<\tilde t\le t$ (for $t=2^{22}$ or $t=2^{20}$ according 
to the selected data in this Figure). }
\end{figure}

We now turn to the discussion of how well the numerical results for the
nonlinear system (\ref{eqNLeq1}) are in agreement with 
the RJ-theory. As already explained above, we solve (\ref{eqNLeq1}) 
numerically for the \SN\ using initial conditions localized on 
one energy mode $m_0$ with $C_m(t=0)=\delta_{m,m_0}$. Due to the nonlinear 
term we expect, for sufficiently large values of the parameter $\beta$, 
that the probability  states to diffuse approaching to the RJ-distribution.
To verify this point,  we compute 
long time averages $\rho_m(t)=\langle|C_m(\tilde t)|^2\rangle$ 
over time intervals $t/2<\tilde t\le t$ for successive discrete time 
values $t=2^l$, $l=1,2,\ldots$ with values up to $t=2^{25}$. 
Using these numerical averages, we compute $S_q$ and $S_B/N$ 
using the formulas (\ref{eqSq}) and (\ref{eqSB}) in terms of $\rho_m$ and 
using $h_B=1/N^2$ in (\ref{eqSB}). Then we can compare with 
the theoretical values of $S_q$ and $S_B/N$ using the 
RJ-thermalized occupation probabilities $\rho_{m,RJ}=T/(E_m-\mu)$. 
To obtain the values of $T$ and $\mu$, we need to solve the implicit equations 
(\ref{eqConstraints1}) with a given energy value $E$ as parameter. 
One possible choice for the comparison is $E=E_{m_0}$ where $m_0$ is 
the used initial mode of the numerical data. However, it turns out 
that the agreement between numerical and theoretical values is better 
if we choose $E=\langle E\rangle =\sum_m E_m \rho_m$ (using the numerical 
values of $\rho_m$ at a given time $t$) 
to solve (\ref{eqConstraints1}) to obtain 
$T$, $\mu$ and $\rho_{m,RJ}$. Typically, we have 
$\langle E\rangle \approx E_{m_0}$ but for boundary modes with 
small initial IPR, and larger initial nonlinear energy contribution, 
there may be a significant energy shift between $E_{m_0}$ and 
the final $\langle E\rangle$ value (see also the discussion at the beginning 
of the last section around Eqs. (\ref{eqHlin1})-(\ref{eqNorm11})).

Thus Fig.~\ref{figII_2} compares the energy dependence of the numerical data 
of $S_q$ and $S_B/N$ for a selected number of initial modes 
with the theoretical values (for $E=\langle E\rangle$) 
for the \SN\ at $\kappa=0.5$ and different values of $\beta$. For 
$\beta=4$ and $\beta=10$ at $t=2^{22}$ 
the numerical entropy values agree very well 
with the theoretical curves for $E\le 4$ (with 2-3 exceptions at $\beta=10$). 
Some of the boundary modes at $E>4$ have very low entropy values which 
can be explained by the large energy gaps for theses modes which are 
very stable with respect to the nonlinear perturbation. They have also 
slightly smaller Lyapunov exponents compared to the modes with $E_m$ in the band 
center. Furthermore, the effect of the energy shift due to the 
initial nonlinear energy contribution pushes these modes more to the 
right boundary 
for $\beta>0$ while boundary modes at the lower part of the 
spectrum are more pushed to the band center with a reduced chaos border 
by this effect. For $\beta=10$ the second set of data 
with $t=2^{20}$ is very close to the first set if data at $t=2^{22}$ (with 
a few exceptions at $E>0$) showing that most entropy values are already 
quite stable for $t\ge 2^{20}$. 
These results are confirmed for other values $1\le \beta\le 20$ 
(data not shown in this or other figures here) with the 
same kind of exceptions for $E>4$ 
and for $1\le \beta\le 3$ a few boundary modes at the lower energy border for 
$E\le -5$ are not well thermalized as well. 

The entropy data for very small coupling constants $\beta=0.2$ and 
$\beta=0.05$ (blue and pink data points in the right panels) 
are clearly below the theoretical curves. Note that for $S_B/N$ 
negative values have been artificially shifted up to zero values for 
a better visibility. This vertical 
shift concerns some boundary modes (all modes) 
for $\beta=0.2$ ($\beta=0.05$) with rather strong negative values 
of $S_B$ (using the parameter $h_B=1/N^2$). However, for $\beta=0.2$ 
the entropy values of the center energy modes in the interval 
$-1.5<E<1.5$ are not 
very far from the theoretical curves and they still continue to increase 
in time (at largest available time values). Also their Lyapunov exponents 
are somewhat stronger than those of the boundary modes at same value 
of $\beta$. These results indicate that center modes at $\beta=0.2$ 
are already in a ``weak'' chaotic regime, but with reduced Lyapunov exponent 
and much larger thermalization time scales, while boundary modes are still in 
a pertubative KAM regime. For $\beta=0.05$ the entropy values are very low 
for all modes, most $S_q$ values are close to zero and only three values 
are between $1$ and $2$ which is 1 to 3 times smaller than the theoretical 
$S_q$ value. We know that in the energy band center there are many
oscillators with very close frequencies $E_m$ and in such cases
the KAM theory is not valid and even a very small nonlinearity
for e.g. 3 oscillators with equal frequencies
have about 50\% of the phase space being chaotic
(see e.g. \cite{chirikovyadfiz,mulansky2}).

For the case $\kappa=0$ and $\beta=10$ we show in Appendix Fig.~\ref{figB6} 
the energy dependence of $S_q$. Globally there a is similar agreement 
as for the case with $\kappa=0.5$ but for this longer iteration 
time scales $t=2^{24}$ are required. 

\begin{figure}[htbp]
\begin{center}
\includegraphics[width=0.95\columnwidth]{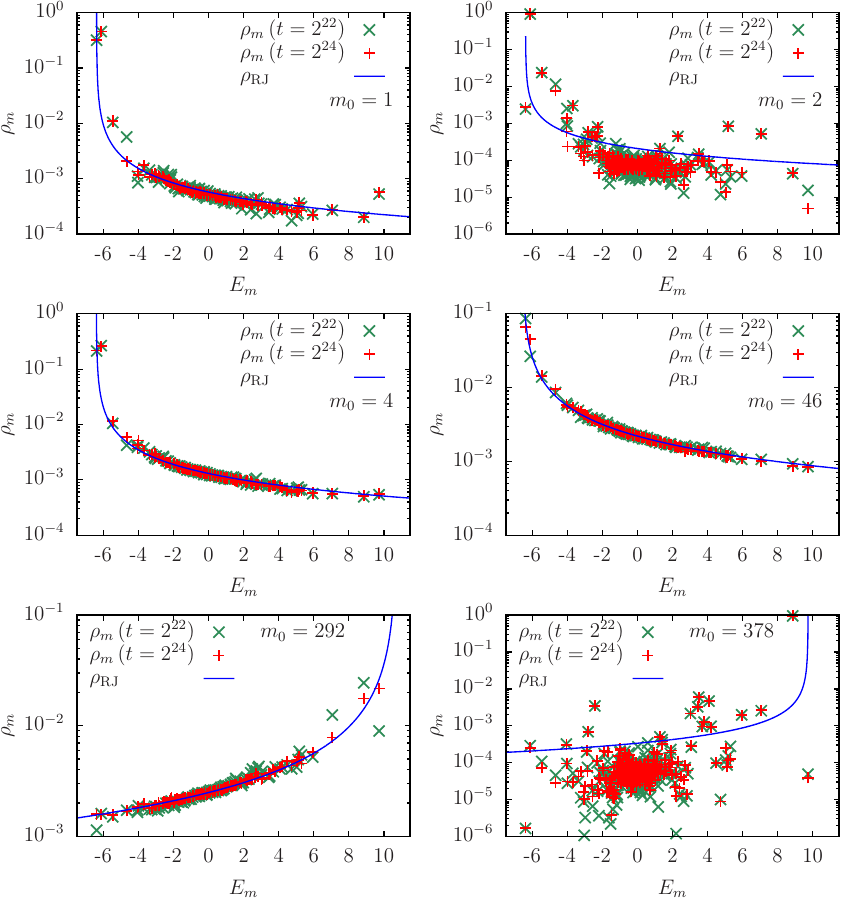}%
\end{center}
\caption{\label{figII_3}
\label{fig_stateNET}
Dependence of $\rho_m(E_m)$ on $E_m$ for the \SN\ with 
$\kappa=0.5$, $\beta=10$, $N=379$, 
the initial condition $C_m(t=0)=\delta_{m,m_0}$
where the initial modes are 
$m_0=1,2,4,46,292,378$. $\rho_m$ has been 
obtained as the time average $\rho_m=\langle |C_m(\tilde t)|^2\rangle$ 
for $t/2<\tilde t\le t$ for $t=2^{22}$ (green $\times$ symbol) 
and $t=2^{24}$ (red $+$ symbol). 
The blue curve shows the RJ theoretical curve 
$\rho_{\rm RJ}(E_m)=T/(E_m-\mu)$ with $T$ and $\mu$ determined 
from the implicit equations (\ref{eqConstraints1})
and using the mean linear energy of 
the state $\langle E\rangle =\sum_m E_m\rho_m(t=2^{24})$ for the value of 
$E$. The values of $T$, $\mu$ and $\langle E\rangle$ for the 
6 initial modes $m_0=1,2,4,46,292,378$ are 
$T=0.003689, 0.001347, 0.008348, 0.01443, -0.02676, -0.003266$, 
$\mu=-6.388, -6.384, -6.402, -6.555, 10.74, 9.729$ and 
$\langle E\rangle =-4.99, -5.874, -3.238, -1.088, 0.6025, 8.492$.
Note that the case of the initial mode $m_0=1$ has finally at large 
times higher values of $T$, $-\mu$ and $\langle E\rangle$ than 
the case for $m_0=2$ which is due to a stronger energy shift from 
the nonlinear energy contributions for $m_0=1$.}
\end{figure}

In Fig.~\ref{figII_3}, we show the dependence of numerical values $\rho_m(t)$ 
on $E_m$ for the \SN, $\beta=10$, $\kappa=0.5$, several initial modes 
$m_0$ and two time values $t=2^{22}$ and $t=2^{24}$ (note that 
$\rho_m(t)=\langle |C_m(\tilde t)|^2\rangle$ for $t/2<\tilde t\le t$). 
The blue curve corresponds to the thermalized expression 
$\rho_{RJ}(E_m)=T/(E_m-\mu)$ with $T$ and $\mu$ obtained 
by solving the implicit equations (\ref{eqConstraints1}) with 
$E=\langle E\rangle =\sum_m E_m \rho_m(t=2^{24})$. 
The states with initial modes $m_0=4,46,292$ are very well thermalized 
with values of $\rho_m$ that are in good agreement with the theoretical curve.
There are somewhat stronger fluctuations for $m_0=292$ and the shorter time 
value $t=2^{22}$. Even the first (left) boundary mode $m_0=1$ 
is quite well thermalized while the second mode $m_0=2$ is not 
well thermalized with most $\rho_m$ values below the theoretical curve 
and a few data points strongly above it. The reason for this strange behavior 
is that the first mode $m_0=1$ has a very strong energy shift effect 
($E_1=-6.38$ while $\langle E\rangle=-4.99$) due to its particularly small 
value of the initial IPR ($\approx 2.8$, see also first left data point in 
right panel of Fig~\ref{figB2} in Appendix B).

The right boundary mode $m_0=378$
has a significant nonlinear energy shift
being close to maximal possible energy values.
Thus the  energy integral of motion (energy constraint) does not allow to diffuse 
to a more ergodic state and the system remains in the integrable KAM regime.

More generally, for modes with $T<0$ and $E_{m_0}>0$ the energy shift effect 
has a tendency to increase the energy to a region with a stronger condensation 
and due to the energy constraint it is more difficult to thermalize while 
at $E_{m_0}<0$ the energy shift effect facilitates thermalization (to 
a certain modest degree). This can be seen at the modes $m_0=1$ and $m_0=4$ 
and also in Fig.~\ref{figII_2} where many modes with $E>4$ do not 
thermalize and their entropy values are clearly below the theoretical 
curves (for the cases $\beta=4$ and $\beta=10$ with good 
thermalization at $E<4$). 
Further two examples of well thermalized states at the smaller value $\beta=4$ 
are shown in Fig.~\ref{figB7} for initial modes $m_0=8$ and $m_0=54$. 

\begin{figure}[htbp]
\begin{center}
\includegraphics[width=0.95\columnwidth]{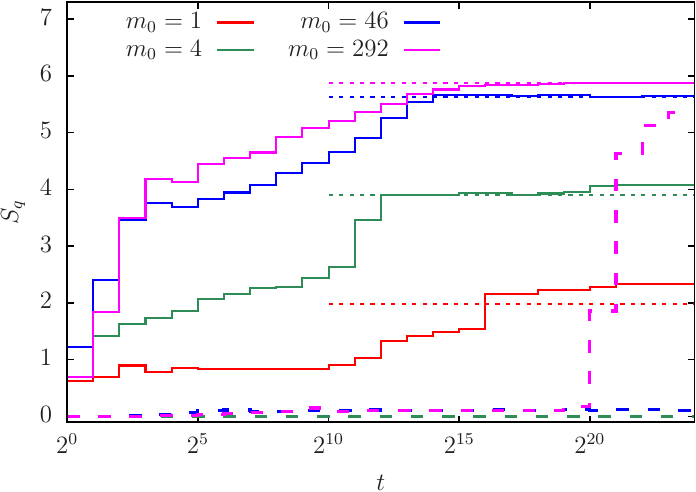}%
\end{center}
\caption{\label{figII_4}
\label{fig_Stime_q_Raw}
Time dependence of $S_q$ for the \SN\ with $\kappa=0.5$, $N=379$. 
The full lines with colors red, green, blue, pink correspond to $\beta=10$ 
for modes $m_0=1,4,46,292$ respectively, the dashed lines correspond 
to $\beta=0.2$ (for $m_0=4,46,292$ with colors green, blue, pink) and the 
dotted lines indicate the theoretical thermalized RJ values. 
$S_q$ has been computed by 
(\ref{eqSq}) using $\rho_m(t)$ values obtained as the time 
average $\rho_m=\langle |C_m(\tilde t)|^2\rangle$ 
for $t/2<\tilde t\le t$. }
\end{figure}

Figs.~\ref{figII_4} and \ref{figII_5}, show the entropy time dependence of 
of $S_q(t)$ and $S_B(t)/N$ for the 4 well thermalized modes 
$m_0=1,4,46,292$ at $\beta=10$ (full lines) and also 
for $m_0=4,46,292$ at $\beta=0.2$ (dashed lines of same color for 
corresponding modes). 
The plateau values correspond to the used intervals for the 
time average in the computation of $\rho_m(t)$ between $t/2$ and 
$t$ for $t=2^l$, $l=1,2,\ldots 24$. 

For $\beta=10$ both entropy quantities increase with time and 
saturate at values close the theoretical thermalized values 
(dotted lines). The initial values between $t=1$ and $t=2$ 
are already 
$\sim 1$ (for $S_q$) or $\sim 4$ (for $S_B$). Note this 
figure does not show any data for the very initial time interval
$t\in[0,1[$ with at least 10 (or more) basic integration steps with $dt=0.1$
(or less) at which there is already some initial diffusion from 
$S_q=0$ (or $S_B=-\infty$) of the mode localized initial condition 
to some finite values. 
For $S_q$ (at $\beta=10$) and the modes $m_0=1,4$ the latest values 
of $S_q(t)$ are even a bit above the 
thermalized values. A similar effect for a well thermalized boundary mode 
was also observed in \cite{rmtprl} (for intermediate time scales) 
and such a behavior is indeed possible 
since $S_q$ is different from the thermodynamical entropy $S_B/N$. 
Furthermore,  $S_q$ is not maximal at the thermalized $\rho_{m,RJ}=T/(E_m-\mu)$ values but 
$S_B/N$ is of course maximal 
for $\rho_{m,RJ}$ 
which can be verified by a standard textbook calculation 
by maximizing (\ref{eqSq}) and (\ref{eqSB}).
Mathematically, $S_q$ from (\ref{eqSq}) is maximal at the Gibbs 
values $\rho_{m,G}=e^{-(E_m-\mu_G)/T_G}$ 
where the Gibbs temperature $T_G$ and chemical potential $\mu_G$
are determined from the implicit equations 
(\ref{eqConstraints1}) using $\rho_m=\rho_{m,G}$.

\begin{figure}[htbp]
\begin{center}
\includegraphics[width=0.95\columnwidth]{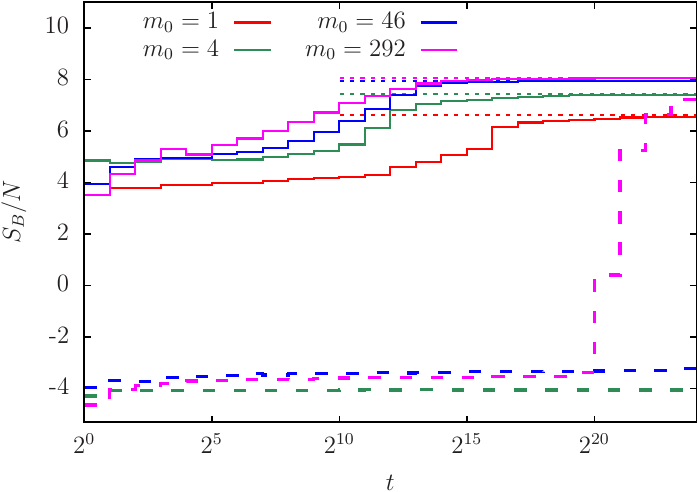}%
\end{center}
\caption{\label{figII_5}
\label{fig_Stime_B_Raw}
Time dependence of $S_B$ for the \SN\ with $\kappa=0.5$, $N=379$. 
The full lines with colors red, green, blue, pink correspond to $\beta=10$ 
for modes $m_0=1,4,46,292$ respectively, the dashed lines correspond 
to $\beta=0.2$ (for $m_0=4,46,292$ with colors green, blue, pink) and the 
dotted lines indicate the theoretical thermalized RJ values. 
$S_B$ has been computed by (\ref{eqSB}) using $h_B=1/N^2$ and 
$\rho_m(t)$ values obtained as the time 
average $\rho_m=\langle |C_m(\tilde t)|^2\rangle$ 
for $t/2<\tilde t\le t$. }
\end{figure}

For $\beta=0.2$ the modes $m_0=4,46$ stay localized, even with 
significant negative values of $S_B/N$ (for the parameter choice 
$h_B=1/N^2$) and $S_q\approx 0$. It is likely that there are in the KAM regime.
The mode $m=292$ at $\beta=0.2$ is very interesting with a very 
late onset of thermalization at $t\approx 2^{20}$, with a ``final'' 
value at $t=2^{24}$ only slightly below the theoretical value. 

We also consider (for $\kappa=0.5,\,\beta=1,4,10$ and the \SN) 
two example states where the initial condition 
is localized on one specific node $n_0$ (instead of 
some eigenmode $m_0$) with $\psi_n(t=0)=\delta_{n,n_0}$ 
(i.e. $C_m(t=0)=\phi_{n_0}^{(m)*}$) being either {\em Barabasi} or {\em Newman} 
which are the two top PageRank nodes (see also Table~\ref{tab1}). 
For this case, we show some results in Fig.~\ref{figB8} of Appendix B. 
For example the time evolution of both entropy quantities is similar 
to Figs.~\ref{figII_4} and \ref{figII_5} with a good convergence to 
the theoretical thermalized entropy, now in a regime of negative temperature 
$T<0$ since the conserved energy is $E\approx \mathcal{H}\approx\beta/2$. 
Also the data of $\rho_m(t)$ at large times match very nicely the theoretical 
thermalized curves.
For more details on this case see Fig.~\ref{figB8} and its 
related discussion in Appendix B. 

Finally, the results for dynamical thermalization
in the netscience network with nonlinear interactions
show that the time evolution
is converging to the theoretical RJ distribution 
for a majority of initial conditions if the system is above
a certain chaos border with $\beta > \beta_c$
We estimate that $\beta_c \sim 1$ even if a small chaotic component
can survive even below $\beta_c$ as it as the case in \cite{mulansky2}.
In the regime of dynamical thermalization
the Boltzmann entropy is growing monotonically with time
reaching its maximal value in the thermal state
in agreement with
the Boltzmann H-theorem \cite{boltzmann1}.

\subsection{Politician network model}

In this section and related Appendix B subsection, we present a few 
results on the \PN\ for $N=5908$ politicians via Facebook using 
data from \cite{rozemb2018}. For this case, we do not know the 
names associated to each node and each of the $N_\ell=83412$ links 
has the same unit weight $1$ so that $A_{ij}=1$ for the non-vanishing 
matrix elements of the adjacency matrix. 

\begin{figure}[htbp]
\begin{center}
\includegraphics[width=0.95\columnwidth]{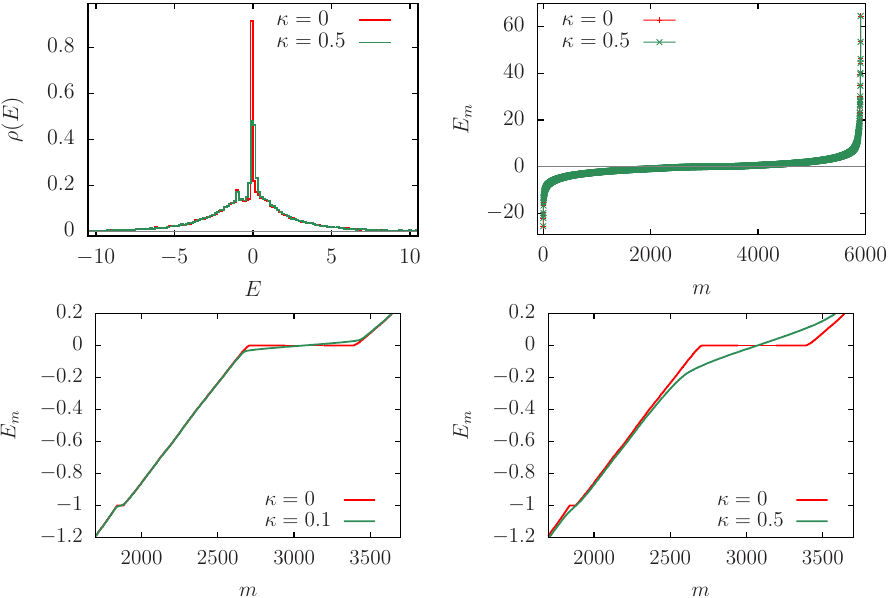}%
\end{center}
\caption{\label{figII_6}
\label{fig_specPolit}
Spectral properties of the eigenvalue spectrum of the matrix 
$H$ for the \PN\ with $N=5908$ eigenvalues $E_m$.
The top left panel shows the density of states 
for $\kappa=0$ and $\kappa=0.5$ normalized by $\int_{E_1}^{E_N} dE\,\rho(E)=1$ 
with histogram bin width $dE=10(E_N-E_1)/N\approx 0.153$. 
The top right panel shows the eigenvalue $E_m$ versus index $m$ for 
$\kappa=0$ and $\kappa=0.5$. 
Note that $m/N\approx \int_{E_1}^{E_m} dE\,\rho(E)$. 
Bottom panels show $E_m$ versus $m$ in a zoomed representation 
for either $\kappa=0$ and $\kappa=0.1$ (left) or 
$\kappa=0$ and $\kappa=0.5$ (right). The plateau values 
for $\kappa=0$ (red data points) at $E_m=0$ and $E_m=-1$
 correspond to degenerate energy levels 
lifted by small GOE perturbations at $\kappa=0.1$ or $\kappa=0.5$ 
but the larger (smaller) degeneracy at $E_m=0$ ($E_m=-1$) 
has still a strong (modest) 
effect on the density of states with a strong (small) peak at $E=0$ and a 
(slightly) deformed $E_m$ vers $m$ curve. 
Bottom and top eigenvalues (at $\kappa=0.5$) are 
$-25.48,-22.02, -20.76,-19.80$ and $44.46, 46.23, 53.50, 64.58$. 
(Bottom/top eigenvalues at $\kappa=0$ are very close).}
\end{figure}

Fig.~\ref{figII_6} shows for the cases of this network 
and different values of $\kappa$ the density of states $\rho(E)$ of the 
eigenvalues $E_m$ of the matrix $H=A+\kappa H_{\rm GOE}$ and also the 
dependence of $E_m$ on $m$. 
The global interval for the energies is between $E_1\approx-25.5$ and 
$E_N\approx 64.6$, which is significantly larger than for the \SN. 
The overall form for $\rho(E)$ and the $E_m$ dependence on $m$ is 
rather similar to the \SN\ but with a somewhat reduced (relative) 
sub-interval for the bulk of eigenvalues in the center 
(in comparison to the global energy interval). Now, at $\kappa=0$, 
there are two degenerate 
eigenvalues $E_0=0$ (with about 700 modes) and $E_0=-1$ (with about 
40 modes) which produce visible peaks in $\rho(E)$ (a strong one for 
$E_m=0$ and a modest one for $E_m=-1$). 
At $\kappa=0.1$ and $\kappa=0.5$ these degeneracies are lifted but 
their effect on $\rho(E)$ is still visible at $\kappa=0.5$ with 
slightly reduced peaks. 

The IPR values of the eigenmodes are very small for boundary 
modes and have a broad distribution for center modes with a reduced 
probability to have small IPR values at $\kappa=0.5$ in comparison 
to $\kappa=0$. See Fig.~\ref{figB9} in Appendix B and the related 
Appendix discussion for more details.

\begin{figure}[htbp]
\begin{center}
\includegraphics[width=0.95\columnwidth]{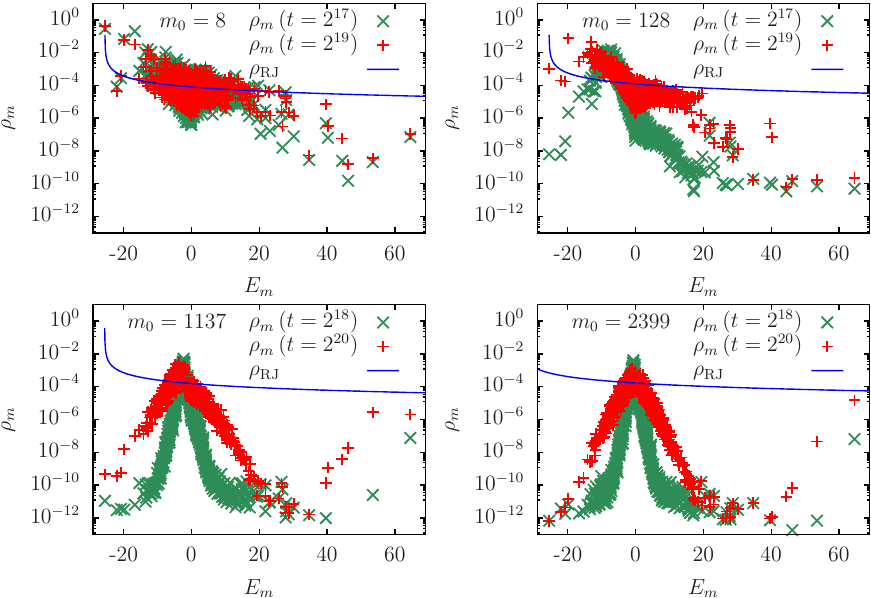}%
\end{center}
\caption{\label{figII_7}
\label{fig_statePolit}
Dependence of $\rho_m(E_m)$ on $E_m$ for the \PN\ with 
$\kappa=0.5$, $\beta=10$, $N=5908$, 
the initial condition $C_m(t=0)=\delta_{m,m_0}$
where the initial modes are 
$m_0=8,128,1137,2399$. $\rho_m$ has been 
obtained as the time average $\rho_m=\langle |C_m(\tilde t)|^2\rangle$ 
for $t/2<\tilde t\le t$ for two values of $t$. 
The blue curve shows the RJ theoretical curve 
$\rho_{\rm RJ}(E_m)=T/(E_m-\mu)$ with $T$ and $\mu$ determined 
in the usual way as described in the text (see also the caption 
of Figure~\ref{fig_stateNET}). 
The values of $T$, $\mu$ and $\langle E\rangle$ for the 
4 initial modes $m_0=8,128,1137,2399$ (and the larger $t$ value) are 
$T=0.002053, 0.003161, 0.003972, 0.005611$, 
$\mu=-25.49, -25.5, -25.55, -33.53$ and 
$\langle E\rangle =-13.36, -6.824, -2.083, -0.3803$. 
For comparison the values of $E_{m_0}$ for these modes are 
$E_{m_0}=-13.53,-6.84,-2.085,-0.382$. 
The cases with $m_0=8, 128$ ($m_0=1137, 2399$) have been computed using the 
time step $dt=1/32$ ($dt=0.1$) for the symplectic integrator up to $t=2^{19}$ 
($t=2^{20}$).}
\end{figure}

Fig.~\ref{figII_7} shows (same style as Fig.~\ref{figII_3}) 
the dependence of $\rho_m$ on $E_m$ for four example states of the \PN\ 
at maximal iterations times $t=2^{19}$ ($m_0=8,128$) or $t=2^{20}$ 
($m_0=1137,2399$). The latter two modes have been chosen because of their 
large IPR $\approx 1250-1300$ values close to the maximal values in the 
hope to optimize the chance to observe a rapid thermalization. 
However, due to the large matrix size $N=5908$ for the \PN\ 
the numerical effort is very high 
and only the limited time scales used in Fig.~\ref{figII_7} 
are available. Indeed, the states visible in Fig.~\ref{figII_7} 
are clearly not yet thermalized at $t=2^{19}$ or $t=2^{20}$. 
However, the time evolution between the two sets of data points at 
$t/100$ and $t$ 
still indicates a tendency for convergence to the thermalized curve at 
much longer time scales numerically not easily accessible. 
In particular, for the modes $m_0=8,128$ the cloud of data points approaches 
the theoretical curve, with a delayed ``convergence'' for the values 
$\rho_m$ with $E_m$ close to $0$ which can be explained by the 
effects of the strong initial degeneracy at $E_m=0$ (for $\kappa=0$) 
and typical reduces IPR values (also for $\kappa=0.5$). 

Furthermore, the entropy values of the states shown in Fig.~\ref{figII_7} 
(and some other states we computed) are already rather close to the 
theoretical value as can be seen in Fig.~\ref{figB10} in Appendix B. 
This figure shows for each mode three data points 
for successive values of $t$ clearly indicating a convergence 
to the theoretical entropy values. In summary, we can say that 
the available numerical data provides  indications for the onset 
of thermalization for the \PN\ but at longer time scales than yet 
accessible by the numerical method.

Thus the results of this Section show that in large size networks
the thermalization time scale can be rather high.

\subsection{Entropy in the RMT model}

In Ref. \cite{rmtprl}, we already studied the thermalization 
problem for a nonlinear perturbation of a  GOE matrix with semicircle radius 1 
(corresponding to $A=0$, $\kappa=1$ and $N=64$ in the notations here). 
However, in \cite{rmtprl} only the quantity $S_q$ was computed (and called 
$S$ there). Therefore, we provide here also a few results for $S_B/N$ 
using the data of \cite{rmtprl}. 
Specifically, Figs.~\ref{figB11} and \ref{figB12} in Appendix B 
show the time dependence of $S_q(t)$ and $S_B(t)/N$ for the 
cases already shown in Fig. S1 of \cite{rmtprl} which is actually 
very similar to Fig.~\ref{figB11} (the latter is provided for convenience
and also shows more theoretical values than Fig. S1 of \cite{rmtprl}). 

Both entropy values converge for the cases with $\beta=1$ clearly 
to the thermalized entropy values. For the initial mode $m_0=3$ at $\beta=1$, 
we observe that $S_q(t)$ takes at intermediate times ($t\sim 2^{14}$) 
even larger values 
(roughly by a factor 1.5) than the thermalized theoretical value 
before the curve drops to its final value at larger times. 
This behavior is indeed possible since $S_q$ is not the thermodynamical 
entropy of the problem. However, $S_B(t)$ increases 
monotonically with $t$ all modes (including this mode) and converges 
in (certain cases) to the thermalized value. 
For the mode $m_0=30$ Figs.~\ref{figB11} and \ref{figB12} 
show each three curves (of either $S_q$ or $S_B/N$) at $\beta=1, 0.1, 0.02$. 
For $\beta=1$ ($\beta=0.1$) there is a rapid (slow/delayed) convergence 
to the (same) thermalized value while for $\beta=0.02$ the mode does not 
thermalize since it is very likely in the integrable KAM regime. 
Thus the time dependence of $S_B(t)$ is well in agreement
with the Boltzmann H-theorem \cite{boltzmann1}.

At the same time we note that a nonmonotonic
time dependence of $S_B(t)$ was found in numerical simulations
of dynamical thermalization in
quantum chaos billiard described by the nonlinear Schr\"odinger equation \cite{ourfiber}.
However, in this system the number of linear modes is formally
unlimited and there is a question
how to define a finite Boltzmann entropy in such a case
since the definition (\ref{eqSB}) is diverging in such a case.

\subsection{Wealth inequality and Lorenz curves}

It is interesting to compute Lorenz curves for the specific spectra 
of both networks discussed above. We briefly remind the construction 
procedure which was introduced in Part I. For a given 
energy spectrum $E_m$ (e.g. for the \SN\ at $\kappa=0.5$), we first 
compute a shifted spectrum $\bar E_m=E_m-E_1$ such that 
$\bar E_1=0$ and other $\bar E_m>0$. Then for a specific value 
of the energy $\bar E=E-E_1$ with $\bar E_1<\bar E<\bar E_N$ 
corresponding the rescaled energy 
$\varepsilon=(E-E_1)/(E_{N}-E_1)=\bar E/\bar E_{N}$ 
we compute the RJ thermalized values of $T$, $\mu$ and $\rho_m$ 
in the usual way ($T$ and $\rho_m$ are not modified by the shift 
and for $\mu$ the same shift as for $E_m$ is applied). In particular, 
the relation $\bar E=\sum_m \bar E_m\rho_m$ is verified. 
Using these values of $\rho_m$ we compute (for $0\le m\le N$)
the cumulated household fraction $h(m)=\sum_{i=1}^m \rho_i$ and 
the associated cumulated wealth fraction 
$w(m)=\sum_{i=1}^m (\bar E_i/\bar E) \rho_i$ 
such that $h(0)=w(0)=0$, $h(N)=w(N)=1$ and $h,w\in[0,1]$. 
The set of points $(h(m),w(m))$ for $0\le m\le N$ then provides the 
Lorenz curve. 

To characterize the degree of ``inequality'' one uses the 
{\em Gini coefficient} defined as the area 
between the line $w=h$ (of perfect ``equality'') and the curve 
divided over its maximal possible value if $w=0$ (i.e. $1/2$ 
for the area of the triangle below the line $w=h$). 

\begin{figure}[htbp]
\begin{center}
\includegraphics[width=0.95\columnwidth]{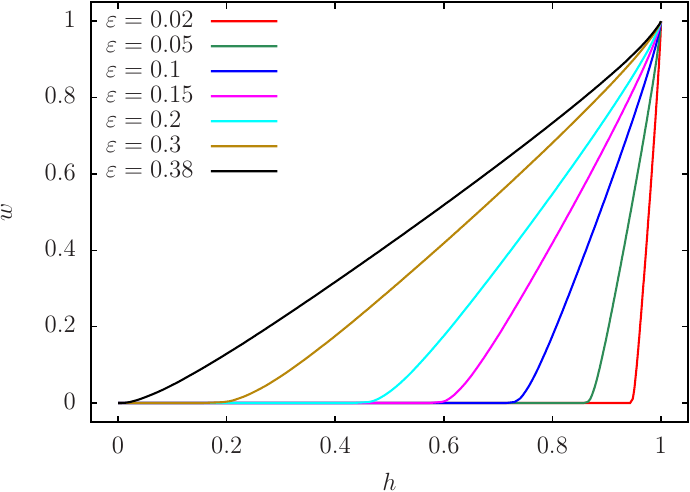}%
\end{center}
\caption{\label{figII_8}
\label{fig_lorenz_379}
Lorenz curves for the \SN\ with $\kappa=0.5$, $\beta=10$, $N=379$ 
for different values of the rescaled energy 
$\varepsilon=(E-E_1)/(E_{N}-E_1)=\bar E/\bar E_{N}$. 
The $x$-axis corresponds the 
cumulated fraction of households ($h$) and the $y$-axis to
the cumulated fraction of wealth ($w$). 
The largest value $\eps=0.38$ is slightly below the critical value 
$\eps_c=0.39622$ at which the transition from $T>0$ to $T<0$ appears. 
The Gini coefficients $G$ for all curves are 
$G=0.9534, 0.8834, 0.7668, 0.5336, 0.6502, 0.301, 0.1321$ (bottom to top).}
\end{figure}

Fig.~\ref{figII_8} shows for the \SN\ a certain number of Lorenz curves 
for different values of the rescaled system energy $\eps$. The largest used value 
$\eps=0.38$ is close to the critical value $\eps_c=0.39622$ at which the 
transition from $T>0$ to $T<0$ appears. As expected at smaller values 
of $\eps$ the curves describe a strong inequality with a value of 
$G$ close to 1. Here for $\eps\approx \eps_c$ we have $G\approx 0$ and 
a curve quite close to the line $w=h$ of perfect equality. This is different 
from the RJS model (of a uniform spectrum) used in Part I where at 
$\eps=\eps_c=1/2$ we have found $w=h^2$. 
Apart from that, for most curves there is a large interval $h\in [0,h_0]$ 
where $w(h)=0$ which is due to the very large energy gaps of the 
first modes at $m=1,2,3,\ldots$ in comparison to the level spacings 
of modes in the bulk. For $h>h_0$ the curves increase to the final 
value $w(h=1)=1$ and there are rather close to the straight line between 
$(h_0,0)$ and $(1,1)$. 
This behavior can be modified a bit (at least for the 
larger values of $\eps$) by choosing an ever stronger value of 
$\kappa$. Fig.~\ref{figB13} in Appendix B illustrates 
this for $\kappa=6$ where the straight lines are a bit more curved. 
However, in this case also the density of states is strongly modified 
and it is rather close to the semicercle law with radius $\kappa=6$ 
with a significant reduction of the initial energy gaps. 

The results of Fig.~\ref{figII_8} show that e.g. ar $\varepsilon = 0.15$
the phase of absolutely poor households is approximately 60\% of all
households while the top 10\% of most rich households own approximately
32\% of total wealth.

\begin{figure}[htbp]
\begin{center}
\includegraphics[width=0.95\columnwidth]{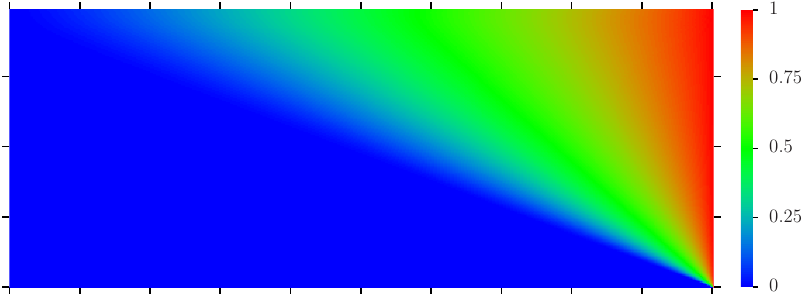}%
\end{center}
\caption{\label{figII_9}
\label{fig_LorcolNet}
Color plot of wealth $w$ from Lorenz curves for the \SN\ with 
$\kappa=0.5$, $\beta=10$, $N=379$. 
The $x$-axis corresponds to 
the fraction of households $h\in[0, 1]$ 
and the $y$-axis to the rescaled energy  
$\varepsilon=(E-E_1)/(E_{N}-E_1)\in[0, \eps_C[$ 
where $\eps_c=0.39622$ is the critical value 
at which the transition from $T>0$ to $T<0$ appears. 
The ticks mark integer multiples of 0.1 
for $h$ and $\varepsilon$. }
\end{figure}

Fig.~\ref{figII_9} shows a color density plot for the Lorenz curves 
for a continuous distribution of $\eps\in[0,\eps_c[$ essentially 
confirming the observations of Fig.~\ref{figII_8} 
the length $h_0$ of the initial interval (with $w(h)=0$) 
behaves roughly as $h_0\approx \eps/\eps_c$.

\begin{figure}[htbp]
\begin{center}
\includegraphics[width=0.95\columnwidth]{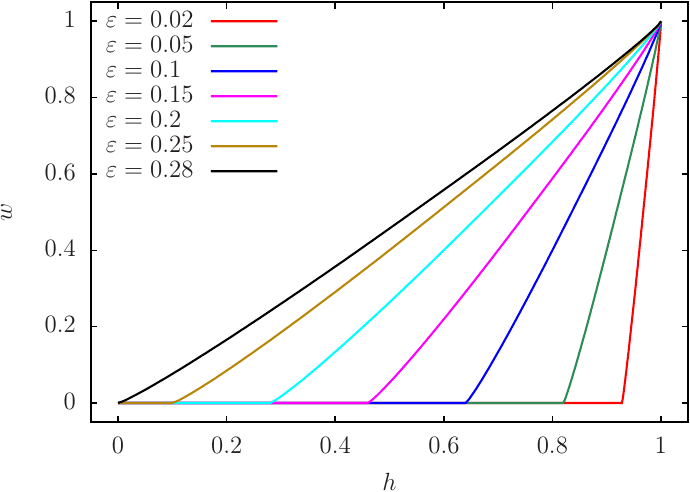}%
\end{center}
\caption{\label{figII_10}
\label{fig_lorenz_5908}
Lorenz curves for the \PN\ with $\kappa=0.5$, $\beta=10$, $N=5908$ 
for different values of the rescaled energy 
$\varepsilon=(E-E_1)/(E_{N}-E_1)$. 
The $x$-axis corresponds the 
cumulated fraction of households ($h$) and the $y$-axis to
the cumulated fraction of wealth ($w$). 
The largest value $\eps=0.28$ is slightly below the critical value 
$\eps_c=0.28297$ at which the transition from $T>0$ to $T<0$ appears. 
The Gini coefficients $G$ for all curves are 
$G=0.9328, 0.8321, 0.6642, 0.4963, 0.3284, 0.1605, 0.06647$ 
(bottom to top).}
\end{figure}

\begin{figure}[htbp]
\begin{center}
\includegraphics[width=0.95\columnwidth]{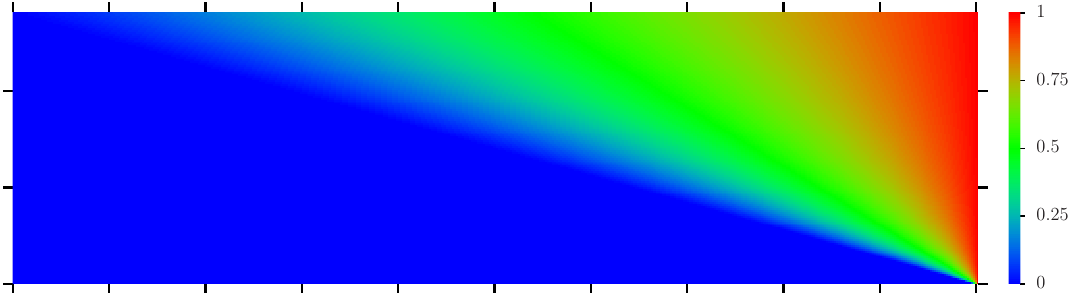}%
\end{center}
\caption{\label{figII_11}
\label{fig_LorcolPol}
Color plot of wealth $w$ from Lorenz curves for the \PN\ with 
$\kappa=0.5$, $\beta=10$, $N=5908$. 
The $x$-axis corresponds to 
the fraction of households $h\in[0, 1]$ 
and the $y$-axis to the rescaled energy  
$\varepsilon=(E-E_1)/(E_{N}-E_1)\in[0, \eps_C[$ 
where $\eps_c=0.28297$ is the critical value 
at which the transition from $T>0$ to $T<0$ appears. 
The ticks mark integer multiples of 0.1 
for $h$ and $\varepsilon$. }
\end{figure}

Figs.~\ref{figII_10} and \ref{figII_11} are as 
Figs.~\ref{figII_8} and \ref{figII_9} respectively but for the 
\PN. Now the critical value for the transition from positive to negative 
$T$ is $\eps_c=0.28297$. 
The structure of the curves and the color plot are rather similar 
as for the \SN\ but the effect of straight lines for $h>h_0$ is even 
a bit stronger which is certainly related to the larger initial energy 
gaps of the politician network. 

The results of this section
show that the RJ thermalization and condensation in social networks
presented in the previous sections leads to the formation of 
an enormous phase
of very poor households and a small fraction of rich households
that owns a significant fraction of total system wealth.
The presence of a significant energy gap in the
energy spectrum of two social networks studied here
enhance the fraction of the poor phase
comparing to the other spectrum models considered in Part I.
The question of how typical such energy gaps are 
for social networks requires further studies.
Another feature of the considered social networks
is a stronglly peaked density of states
approximately at the middle of the energy band.
As a result the energies $E_m$ are very flat in this
energy region. This is rather different from the RMT model
or RJS model of Part I where the density of states is approximately
constant in this energy range. We suppose that the origin of
this difference is related to the fact that in the
considered networks we have links between the members of the same
society layer or class: scientists linked to scientists,
politicians to politicians. Probably this is a general feature of
internet connections where there is little if any distinction
between classes of network members.
In a real human society there is some kind of natural
society statification: factory workers are mainly linked with workers,
peasants with peasants, businessmen with businessmen,
aristocrats with aristocrats. This feature is well present in a 
real human society
with its society classes and wealth gradient between classes
(of course with fluctuations and
relatively weak links between classes).
Thus it is possible that the society networks
should be revised and updated to include
the above feature of human society.

\subsection{Overview of social networks results}

The presented studies of dynamical thermalization in social networks
show that chaos in these systems lead to the RJ thermalized distribution
if nonlinear interactions are above a certain chaos border.
The time scale for onset of this RJ distribution
is determined by the strength of the nonlinearity. 
This time scale can be relatively long.
On a first glance this seems rather surprising
in view of a rather small number of links to hope and
connect any pair of nodes,
with the Erd\"os number
$N_E \approx 4 - 5$ \cite{dorogovtsev10,newmanbook,vigna}.
However, such link transitions are provided only
by the linear part of system Hamiltonian
while only the nonlinear interactions
lead to transitions between eigenmodes (see (\ref{eqNLeq2}))
with eventual thermalization.
Our results show that the RJ thermalization process in the social
networks has close similarities with those
in the NLIRM model \cite{rmtprl}.
Thus we expect that the Lyapunov exponents $\lambda_m$ 
and the thermalization time scale $t_{RJ}$
have a similar to \cite{rmtprl}  dependence 
on nonlinearity $\beta$ and number of oscillators (nodes)
where it was found that the typical Lyapunov exponent values
are $\lambda \sim \beta^{1.5}/N^{1.9}$.
We expect that $t_{RJ} \propto 1/\lambda $ but further studies
are required to confirm these dependencies. 

The emergence of RJ condensate 
leads to a formation of an enormous phase
with high fraction of total norm
located at low energy, or wealth, states.
This leads to a massive fraction of poor households
in the social networks as it is well seen in
the figures of the Lorenz curves in Section~II.7.
At the same time a significant part of total wealth
is captured by a small oligarchic group
of rich households. Thus the obtained results
for dynamical thermalization in social networks
provide a confirmation of WTH origin 
and highlight the problem of wealth inequality in
human society from a new view point.

\section{Discussion and conclusion}

In 1955 Fermi, Pasta, Ulam and Tsingu performed
the first numerical simulations of
a chain of nonlinear oscillators
with the aim to find a dynamical thermalization
and energy equipartition between the degrees of freedom. 
However, this model
happened to have various specific features
so that no tendency to equipartition was found in 1955 \cite{fpu1955}.
To extend these studies a generic model of coupled oscillators was proposed
  in \cite{rmtprl} on the basis of nonlinear perturbation
  of Random Matrix Theory showing that chaos leads
  to dynamical thermalization with the resulting RJ distribution over
  the linear energy eigenmodes. This model has two integrals of motion
  being the total energy and total norm (probability).
  Thus the RJ distribution in this isolated Hamiltonian system
  is characterized by the system
  temperature $T(E)$ and chemical potential $\mu(E)$.

  In fact the emergence of the RJ thermalization
  had been studied earlier numerically and experimentally
  for light propagation in multimode optical fibers
  \cite{picozziphrep,wabnitz,picozzi1,picozzi2,babin,chris,picozzi3},
  even if the origin of this thermalization was
  attributed to the Kolmogorov-Zakharov turbulence \cite{zakharovbook}
  without links to chaos and KAM integrability.
  The emergence of RJ condensation was established numerically \cite{picozzi1}
  and experimentally \cite{wabnitz,picozzi2}.
  The emergence of an RJ condensate and thermalization in
  quantum chaos fibers with the nonlinear Schr\"odinger equation
  was demonstrated in numerical and
  analytical studies reported in \cite{ourfiber}. 
  At the same time it should be pointed out that the Fr\"ohlich condensate
  for molecules at room temperature,
  discussed in \cite{frohlich1,frohlich2}, 
  has also certain similarities with the RJ condensate,
  even if in \cite{frohlich1,frohlich2} the system is
  considered under external pumping and dissipation
  (see discussion in \cite{ourfiber}).

  In Part I we analyzed the consequences of RJ thermalization and condensation
  associating system energy and norm, both conserved by time evolution. 
  These quantities are related to the global wealth and the number of 
  interacting households which are also conserved as justified in 
  \cite{boghosian1,boghosian2}.
  The performed analysis shows that this WTH description depicts very well
  the shape of real Lorenz curves of wealth of households for several countries
  and the whole world. Also the WTH approach well reproduces
  the Lorenz curves for the stock exchange markets of New York, London and 
  Hong Kong.
  To provide more arguments in support of the WTH description
  we study in Part II  the dynamical thermalization
  in social networks induced by a nonlinear perturbation.
  Our results show the emergence of RJ thermalization and condensation
  in social networks with nonlinear interactions between network agents
  with the interactions  being above a certain chaos border.
  As in Part I the RJ condensation leads to the Lorenz curves with an enormous
  fraction of poor households and a small fraction which owns a main part
  of total wealth.
  On the basis of the results of this work we argue that the WTH description
  provides new perspectives for the understanding of
  the nontrivial aspects  of the wealth inequality in the world.

\section*{Declarations}

\begin{itemize}
\item Funding: The authors acknowledge support from the grant
 ANR France NANOX $N^\circ$ ANR-17-EURE-0009 in the framework of 
the Programme Investissements d'Avenir (project MTDINA).

\item Conflict of interest: The authors declare no conflicts of interest.
%Conflict of interest/Competing interests (check journal-specific guidelines for which heading to use)
%\item Ethics approval and consent to participate
%\item Consent for publication
\item Data availability: This work uses specific datasets 
about wealth data for countries, stock market, 
network data etc. available at Refs. 
\cite{newmanbook,piketty1,piketty2,boston,wikigini,uk2014,defr,sp500,
london,hk,newmannets,newman2001,newman2006,newman2006ref84,rozemb2018}. 
%%Data sets generated during the current study are 
%%available from the corresponding author on reasonable request.

%\item Materials availability
%\item Code availability 
%\item Author contribution
\end{itemize}

\bmhead{Acknowledgements}
% {\bf Acknowledgments:}
We thank A. D. Chepelianskii and L. Ermann for useful discussions.
The authors acknowledge support from the grant
 ANR France NANOX $N^\circ$ ANR-17-EURE-0009 in the framework of 
the Programme Investissements d'Avenir (project MTDINA).

%%%%%%%%%%%%%%%%%%%%%%%%%%%%%%%%%%%%%%%%%%
\begin{appendices}
\def\folgt{\quad\Rightarrow\quad}

\section{Additional material for Part I}
\renewcommand{\theequation}{A.\arabic{equation}}
\renewcommand\thefigure{A.\arabic{figure}}
\renewcommand\thesubsection{A.\arabic{subsection}}
\setcounter{equation}{0}

\subsection{General features of the thermalization in the RJS model}

Here we remind a bit more details about thermalization of the RJS model 
(see also Refs.~\cite{rmtprl,ourfiber} for more details). Let us assume that we have 
$N$ linear classical oscillators with individual energies 
$E_m$, $m=0,\ldots,N-1$ which are coupled by some small 
non-linear perturbation (see Ref.~\cite{rmtprl} for an example) 
such that there are two conserved quantities being the global (squared) 
amplitude and total energy:
\begin{align}
\label{eqS1}
1=\sum_{m=0}^{N-1} \rho_m\quad,\quad E=\sum_{m=0}^{N-1} E_m\rho_m
\end{align}
where $\rho_m$ is the time averaged squared amplitude and occupation 
probability of each oscillator. If the non-linear terms are 
sufficiently strong or if there is some weak coupling to an external 
system (which respects both constraints (\ref{eqS1})) one can assume 
that the system thermalizes. 
Applying the framework of the grand canonical ensemble one introduces 
two parameters: temperature $T$ and chemical potential 
$\mu$ to satisfy both constraints (\ref{eqS1}) in average and it can be shown 
(see e.g. Ref.~\cite{rmtprl}) that 
\begin{align}
\label{eqS2}
\rho_m=\frac{T}{E_m-\mu}\quad,\quad T=\frac{E-\mu}{N}
\end{align}
where the expression for the temperature $T$ is obtained 
from $\sum_m (E_m-\mu)\rho_m=(E-\mu)$ which follows directly 
from (\ref{eqS1}). The chemical potential is determined 
(using standard numerical techniques) by solving the implicit
equation:
\begin{align}
\label{eqS3}
1=\frac{E-\mu}{N}\sum_{m=0}^{N-1}\frac{1}{E_m-\mu}
\end{align}
which allows for one physical solution of $\mu$ outside the 
energy interval $[E_{\rm min},E_{\rm max}]$ with either 
$\mu<E_{\rm min}$ ($T>0$) or $\mu>E_{\rm max}$ ($T<0$) 
(depending on the value of $E$ we have either $T<0$ or $T>0$) 
such that $\rho_m>0$. The data presented in this work were obtained 
by this procedure for different model spectra and certain values of $N=10000$ 
(or $N=1000$ for the RMT model). Concerning the RJS model we have also 
considered the cases $N=100$, $N=1000$ and verified that the obtained 
Lorenz curves are very close (in graphical precision).

\begin{figure}[htbp]
\begin{center}
\includegraphics[width=0.8\textwidth]{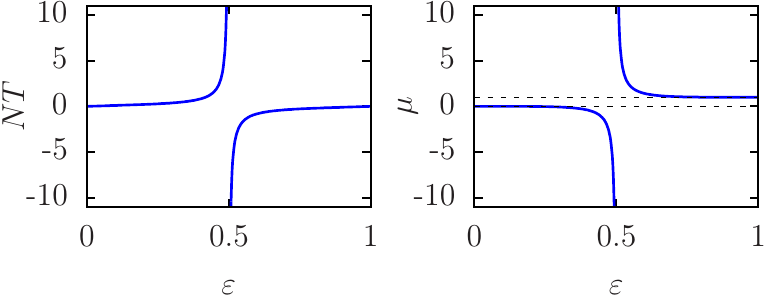}
\caption{\label{figA1}
The left (right) panel shows the (rescaled) temperature $NT$ 
(the chemical potential 
$\mu$) versus the rescaled energy 
$\varepsilon=E/B$ for the RJS model $E_m=m/N$, $N=10000$. 
The dashed black lines in the right panel correspond to the values of 
$E_0=0$ and $B\approx 1$ showing that either $\mu<E_0$ (for $T>0$) 
or $\mu>B$ (for $T<0$). 
}
\end{center}
\end{figure}  

As illustration Fig.~\ref{figA1} shows for the RJS model 
with $E_m=m/N$, $m=0,1,\ldots,N-1$, $N=10000$ both $T$ and $\mu$ 
as a function of $\eps=E/B$ (here $B=E_{\rm max}-E_{\rm min}\approx 1$).
Note that the left panel shows the rescaled temperature $NT$ since 
typical numerical values of $T$ are $\sim 1/N$ due to the finite 
value of $B$ in our particular model. (We note that the construction 
of the Lorenz curve is independent of a global scaling factor one 
could apply to the energy levels.) 
The figure illustrates that $-\mu\to 0$ ($\mu\to -\infty$) for $\eps\to 0$ 
($\eps\to 1/2$).

\begin{figure}[htbp]
\begin{center}
\includegraphics[width=0.65\textwidth]{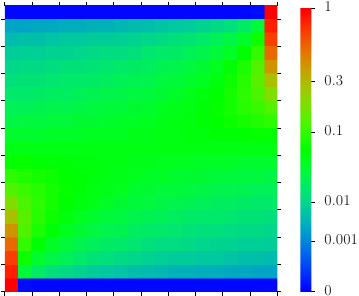}
\caption{\label{figA2}
Color plot of the coarse-grained thermalized occupation 
probabilities $\rho_m=T/(E_m-\mu)=(E-\mu)/[N(E_m-\mu)]$
for the RJS model. The $x$-axis corresponds 
to the fraction $E_m/B\in[0,1]$ (left to right) 
and the $y$-axis to the rescaled 
energy $\varepsilon$ (top to bottom 
for increasing values). The tics indicate integer multiples of 0.1 
for both quantities. 
The color values shown in the color bar 
correspond to the value of $\rho_m$ averaged over intervals of size 
$1/20$ (for $E_m/B$ on the $x$-axis) and computed for 
21 values $\varepsilon=i/20$, 
$i=0,1,\ldots, 20$ (for the $y$-axis; the 
minimal value $\varepsilon=0$ has been 
slightly enhanced and the maximal value $\varepsilon=1$ has been slightly 
reduced to have a stable computation of the thermalized $\mu$-value). 
To increase visibility of small values a non-linear color bar scale 
has been chosen (e.g. green color corresponds to $1/16$). 
}
\end{center}
\end{figure}  

\begin{figure}[htbp]
\begin{center}
\includegraphics[width=0.65\textwidth]{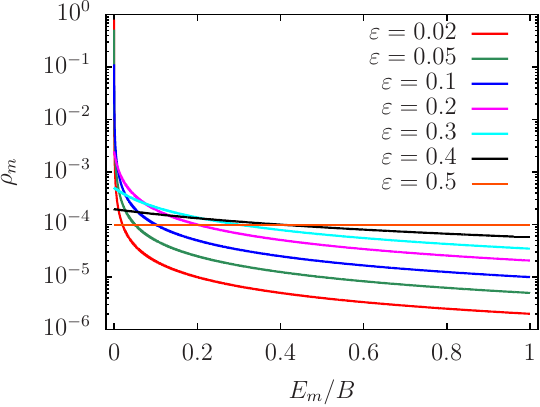}
\caption{\label{figA3}
Dependence of the thermalized occupation 
probabilities $\rho_m=T/(E_m-\mu)=(E-\mu)/[N(E_m-\mu)]$ on 
$E_m/B$ for the RJS model 
$E_m=m/N$, $N=10000$ and the same values of 
$\varepsilon=E/B$ used in Figure~\ref{figI_1} of the main part. 
}
\end{center}
\end{figure}

Using (\ref{eqS3}) one can show that 
$-\mu\approx E/(N-1)\ll E$ for very small energies $0<E\ll 1/N$ 
and in this particular case we have $\rho_0\approx 1$ (strong 
condensation) and other 
$\rho_m\sim E/(NE_m)\ll 1/N$ (for $m>0$). 
With increasing values of $E$ (or $\eps$) 
the values of ``$-\mu$'' increase and more probability is shifted to 
the other $\rho_m$ values for $m>0$. At $\eps\approx 1/2$ 
we have very large values of ``$-\mu$'' (and of $T$) such that all 
$\rho_m\approx 1/N$ are uniformly constant. Further increase of 
$\eps$ enters the regime of negative temperatures (with $\mu>E_{\rm max}$) 
with possible 
condensation at the last oscillator with $\rho_{N-1}\gg 1/N$ 
(in this work we do not insist on the regime of $T<0$). 
These features 
are visible in both figures Figs.~\ref{figA2} and \ref{figA3} showing 
$\rho_m$ versus $E_m/B$ for different values of $\eps$ (as color plot 
or curves in log-scale). The effect of 
condensation for small $\eps$ with a finite 
probability $\rho_0\gg 1/N$ is clearly 
visible in both figures and qualitatively one could even say that it extends 
even up to $\eps=0.2$ with $\rho_0=0.002495$ still being larger than $1/N$. 
However, here also some other values of $\rho_m$ with small $m$ are 
significantly larger than $1/N$ (as can be seen in \ref{figA3}
for the first 5\% of modes with $\rho_m\ge 3/N$). Also the coarse-grained 
average value at the first 5\% of modes at $\eps=0.2$ is roughly 
0.35 times the 
maximal coarse-grained value at $\eps\approx 0$ (according to 
Fig.~\ref{figA2}). This effect corresponds to (modest) condensation 
on several modes or a given small mode interval.

When constructing the Lorenz curve we have $w=0$ for $h<\rho_0$ and 
in the presence of (strong) condensation there is a finite 
interval of households with no wealth at all. Even for modest condensation 
over several modes the wealth value is initially very low. 
This can also be seen in Fig.~\ref{figI_1} where $w(h\le 0.1)\approx 0$ 
for $\eps=0.2$ showing the effect of modest condensation. 

Below, we will present a continuous version of the RJS model with the exact 
limit $N\to\infty$ and some analytic formulas for the key quantities. 

\subsection{Additional data}

\begin{figure}[htbp]
\begin{center}
\includegraphics[width=0.65\textwidth]{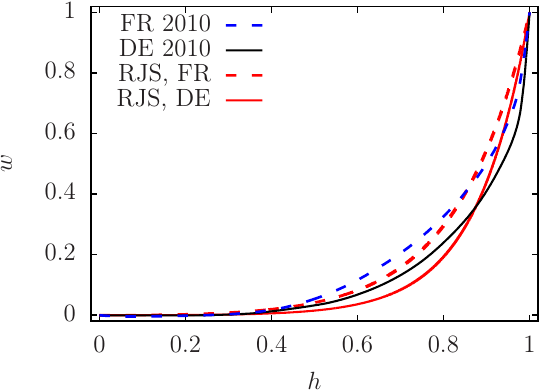}
\caption{\label{figA4}
Comparison of the Lorenz curves for DE 2010 (black) and 
FR 2010 (blue dashed) 
with those of the RJS model (red curves; $N=10000$). The data of 
DE and FR were extracted from Ref.~\cite{defr}. 
As in Fig.~\ref{figI_3} the Gini coefficients $G=0.758$ (DE) and $G=0.679$ (FR) 
were used to determine the $\varepsilon$ values of the RJS model as 
$\varepsilon=0.1220$ (DE) and $\varepsilon=0.1659$ (FR) to match 
the Gini coefficients of the reference data. 
}
\end{center}
\end{figure}

In this section, we present additional data. First Fig.~\ref{figA4} shows 
the Lorenz curves from Germany and France and the corresponding curves 
of RJS model (with matching Gini coefficients). These data were extracted from 
Ref.~\cite{defr} with best possible precision and correspond to the 
period of 2010. The agreement with the RJS is comparable (not perfect 
but still rather close) as with the cases of US and World in Fig.~\ref{figI_3}.
The Gini coefficients of Germany ($G=0.758$) and France ($G=0.679$) are 
both intermediate between UK ($G\approx 0.62$) and US/World 
($G\approx 0.85/0.84$). 

\begin{figure}[htbp]
\begin{center}
\includegraphics[width=0.65\textwidth]{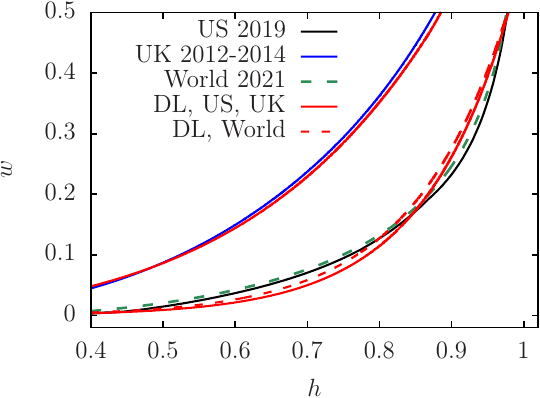}
\caption{\label{figA5}
As panel (b) of Fig.~\ref{figI_4} but with a zoomed representation for 
$h\in[0.4,1]$ and $w\in[0,0.5]$ to increase the visibility between the 
close curves for US 2019 and World 2021.
}
\end{center}
\end{figure}

The next Figure~\ref{figA5} shows a zoomed representation of panel (b) of 
Figure~\ref{figI_4} for 
$h\in[0.4,1]$ and $w\in[0,0.5]$ to increase the visibility between the 
close curves for US 2019 and World 2021 and to also to enhance the small 
differences with respect to the DL model (red lines).

\begin{figure}[htbp]
\begin{center}
\includegraphics[width=0.65\textwidth]{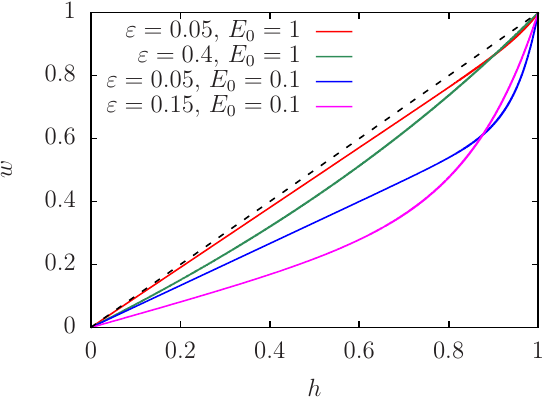}
\caption{\label{figA6}
Lorenz curves of the thermalized EQI model ($N=10000$) with the two 
offset values $E_0=0.1$ and $E_0=1$ and for each case 
for two values 
of the rescaled energy $\varepsilon=(E-E_0)/(E_{N-1}-E_0)$. 
Note that for $E_0=1\Rightarrow E_{N-1}\approx 2E_0$ and  
for $E_0=0.1\Rightarrow E_{N-1}\approx 11E_0$. 
The dashed line corresponds to the line of perfect equipartition $w=h$. 
The Gini coefficients $G$ for all curves are 
$G=0.4239,\,0.3000,\,0.1162,\,0.04286$ (bottom to top). 
These curves show a finite initial slope with 
value $E_0/(\eps+E_0)=0.4,\,0.6667,\,0.7143,\,0.9524$ 
(bottom to top). 
}
\end{center}
\end{figure}

Furthermore, Fig.~\ref{figA6} presents results for the EQI model with $E_0>0$, 
$E_m=E_0+m/N$ and $\varepsilon=(E-E_0)/(E_{N-1}-E_0)\approx E-E_0$. 
In this case, 
the finite value $E_0>0$ induces an initial finite slope 
$E_0/E=E_0/(E_0+\eps)$ in the Lorenz curve. We have verified that for the 
four cases shown in Fig.~\ref{figA6} this formula indeed represents the 
initial slope (see figure caption for the values). In this model, even the 
poorest households own a significant fraction of the wealth which 
is given by this slope. Here the range of possible Gini coefficients is 
quite limited with maximum values of $G_{\rm max}\approx 0.1$ or $0.4$ 
for $E_0=1$ or $E_0=0.1$ respectively. Due this reason it is not possible 
to match the data of US, UK, World etc. (with much larger Gini coefficients) 
to this model (for the cases shown in Figs.~\ref{figI_6},~\ref{figA6}).

\begin{figure}[htbp]
\begin{center}
\includegraphics[width=0.9\textwidth]{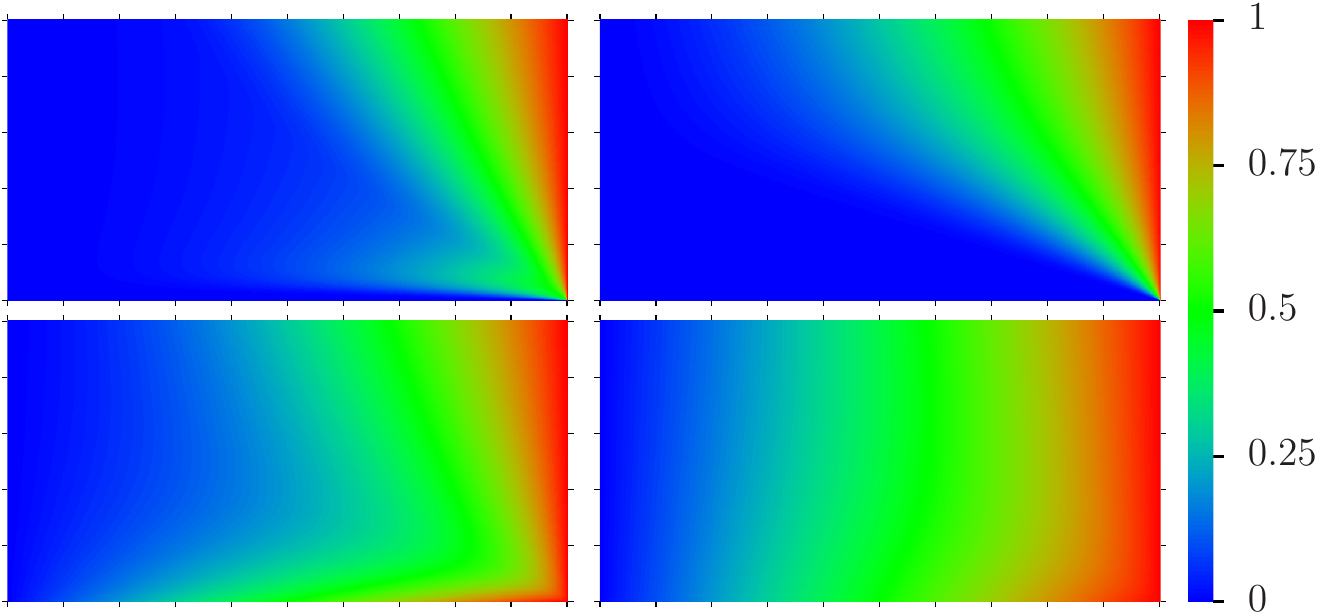}
\caption{\label{figA7}
Color plot of Lorenz curves for different models in the same style 
of Figure~\ref{figI_2}. The panels correspond to 
the DL model with parameter $a=16$ (top left), 
to the shifted RMT semi-circle spectrum (top right), 
to the EQI model with given offset $E_0=0.1$ (bottom left) and 
$E_0=1$ (bottom right). 
For the EQI model the rescaled energy $\varepsilon\in[0,0.5]$ for the vertical 
axis is given by $\varepsilon=(E-E_0)/(E_{N-1}-E_0)$ (same expression for the 
other models but with $E_0=0$). 
All cases correspond to $N=10000$ levels except for RMT with $N=1000$.
}
\end{center}
\end{figure}

Finally, Fig.~\ref{figA7} shows several color plots in the same style as 
Fig.~\ref{figI_2}, i.e. the color value (visible in the color bar) shows $w$ of the Lorenz 
curve as a function of $h$ ($x$-axis) and $\eps$ ($y$-axis). The 2nd 
panel for the RMT model is rather similar to Fig.~\ref{figI_2} for the RJS model, 
with a slight tendency for smaller $G$ values for $\eps>0.2$ 
(at given $\eps$, see also Fig.~\ref{figI_5}). The first panel for the DL model 
with $a=16$ has a stronger condensation effect 
(i.e. more poor or poorer households) at $\eps\approx 0.08$ as compared to 
$\eps\approx 0.03$. Both bottom panels correspond to the EQI model 
with $E_0>0$ (here $\eps=(E-E_0)/(E_{N-1}-E_0)$) with reduced 
Gini coefficients and where poor people own a significant fraction 
of the wealth, even at small values of $\eps$. 

\subsection{Analytical results for  RJS model}

For the RJS model with finite $\varepsilon$ it is  possible to obtain explicit formulas 
in the limit $N\to\infty$ by replacing the sums over $m$ with integrals 
over an energy variable $\tilde E=m/N\in[0,1]$. In the following, we 
also use $\eps=E$ (since $E_0=0$ and $B=(N-1)/N\to 1$ for $N\to\infty$).
In the limit $N\to\infty$ the implicit equation (\ref{eqS3}) becomes:
\begin{align}
\label{eqmuA} 
1&=(\eps-\mu)\int_0^1 \frac{1}{\tilde E-\mu}\,d\tilde E
=(\eps-\mu)\ln\left(\frac{1-\mu}{-\mu}\right)
\end{align}
which can be rewritten in the following form:
\begin{align}
\label{eqmu2} 
\mu=-(1-\mu)e^{-1/(\eps-\mu)}\ .
\end{align}
Both equations determine $\mu$ as a function of $\eps\in]0,1[$. 
In the limit of small $\eps$ one can simply iterate Eq.~(\ref{eqmu2}) by 
inserting $\mu_0=0$ in the RHS which gives $\mu_1=-e^{-1/\eps}$ on the LHS 
which can be inserted in the RHS to obtain a better value $\mu_2$ etc. 
This procedure converges nicely for small $\eps$ and for other values 
of $\eps$ one 
can use standard techniques to solve these 
equations numerically and efficiently. For $\eps \ll 1$, the 
first approximation $(-\mu)\approx e^{-1/\eps}\ll \eps$ is already very good.

To understand the limit of $|\mu|\gg 1$ it is more useful to consider 
$\eps$ as a function of $\mu$ which is determined by (\ref{eqmuA}). 
Expanding the logarithm in (\ref{eqmuA}) up to 3rd order in $1/\mu$ one 
finds that 
\begin{align}
\label{eqmuinfty}
\eps\approx \frac12\left(1+\frac{1}{6\mu}\right)\to \frac12
\end{align}
for $|\mu|\to\infty$ 
which is expected from the curve of $\mu$ in Fig.~\ref{figA1}. The $1/\mu$ 
correction in (\ref{eqmuinfty}) will be useful below. 

As explained in the main part of this work, to compute the Lorenz curve 
we have to compute a partial sum over $\rho_m$ to obtain the household 
fraction $h$ and over $(E_m/\eps)\rho_m$ to obtain the wealth variable. 
Now, we replace these partial sums also by integrals up to some 
arbitrary value $s\in[0,1]$ which provides functions $h(s)$ and $w(s)$ 
allowing to determine the Lorenz curve $w(h)$. These partial integrals are:
\begin{align}
\phantom{\folgt}h(s)&=(\eps-\mu)\int_0^s \frac{1}{\tilde E-\mu}\,d\tilde E
\label{eqh1}
=(\eps-\mu)\ln\left(\frac{s-\mu}{-\mu}\right)\\
\label{sofh}
\folgt s(h)&=(-\mu)\left(e^{h/(\eps-\mu)}-1\right)
\end{align}
and
\begin{align}
\nonumber
w(s)&=\frac{\eps-\mu}{\eps}\int_0^s \frac{\tilde E}{\tilde E-\mu}\,d\tilde E\\
\label{eqwofs}
&=\frac{1}{\eps}\bigg[(\eps-\mu)s+\mu
(\eps-\mu)\ln\left(\frac{s-\mu}{-\mu}\right)\bigg]
\ .
\end{align}
Inserting (\ref{eqh1}) and (\ref{sofh}) in (\ref{eqwofs}) we obtain 
the following analytical expressions for the Lorenz curve:
\begin{align}
\label{eqw1}
w(h)&=\frac{-\mu}{\eps}\left(
(\eps-\mu)\left(e^{h/(\eps-\mu)}-1\right)-h\right)\\
\label{eqlorA}
&=\frac{1-\mu}{\eps}\,e^{-1/(\eps-\mu)}\left(
(\eps-\mu)\left(e^{h/(\eps-\mu)}-1\right)-h\right)\ .
\end{align}
Here (\ref{eqlorA}) has been obtained by replacing the global factor 
$\mu$ with (\ref{eqmu2}) which gives a more convenient expression. 
Using (\ref{eqmu2}), one can verify that (\ref{eqw1}) (and therefore also 
(\ref{eqlorA})) satisfy the conditions $w(0)=0$ and $w(1)=1$. 

The expression (\ref{eqlorA}) allows 
to take the limit $\eps\ll 1$ with $\mu\approx -e^{-1/\eps}\ll \eps$ such 
that for $\eps\ll 1$ we have the simplified Lorenz curve (replacing $\mu=0$ 
in (\ref{eqlorA})): 
\begin{align}
\label{eqlorA0}
w(h)&\approx
e^{-1/\eps}\left(e^{h/\eps}-1-\frac{h}{\eps}\right)\approx e^{(h-1)/\eps}\ .
\end{align}
Here both expression are equivalent approximations for small $\eps$ with 
$e^{-1/\eps}\ll 1$. The first (second) expression does not exactly verify the 
condition for $w(1)$ (or $w(0)$). The second expression is very simple 
and numerically quite sufficient for $\eps\le 0.2$. 

We have verified that both expressions (\ref{eqw1}) and (\ref{eqlorA}) 
coincide with the numerical data shown in Fig.~\ref{figI_1} up to graphical precision 
with an error below $10^{-4}$ and for all values of $\eps$ used in Fig.~\ref{figI_1}. 
The approximate formulas (\ref{eqlorA0}) 
are valid for $\eps\le 2$ with an error $\sim 10^{-2}$ for $\eps=0.2$ (and 
smaller errors for smaller values of $\eps$). 
This can be seen in Fig.~\ref{figA8} comparing the data for $\eps=0.1,0.2,0.3$ 
between the analytic expressions and the data for $N=10000$. Even 
for $\eps=0.3$ only a modest deviation of the approximate curve is visible 
while here and in all other cases the more precise expression (\ref{eqlorA}) 
matches the numerical data very closely. 

Using the analytical expressions for $w(h)$ one can compute several other 
quantities. For example it is interesting to consider the 2nd order expansion 
in $h$ for $|h/(\eps-\mu)|\ll 1$ which gives:
\begin{align}
\label{eqlor2}
w(h)&=\frac{(-\mu)}{2\eps(\eps-\mu)}\,h^2\ .
\end{align}
We know that the limit $|\mu|\to\infty $ corresponds to $\eps\to 1/2$ 
and in this case (\ref{eqlor2}) is valid for all $h\in[0,1]$. This gives 
the very simple formula $w=h^2$ (which is also 
obvious from the fact that $\rho_m=1/N=$ const. for $|\mu|\to\infty$ and 
the way the Lorenz curve is constructed from $\rho_m$). 

\begin{figure}[htbp]
\begin{center}
\includegraphics[width=0.65\textwidth]{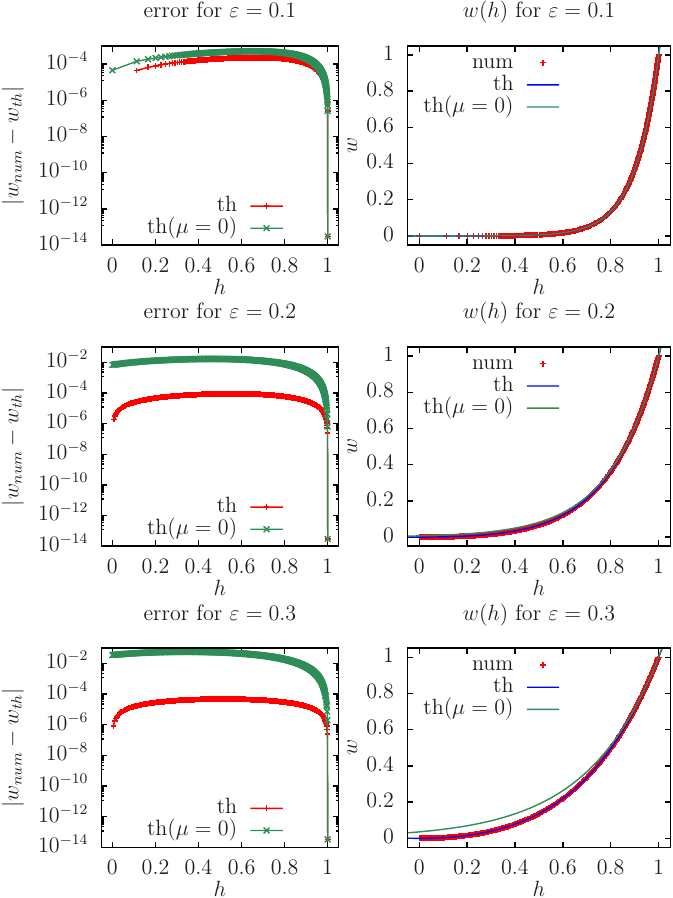}
\caption{\label{figA8}
Comparison of Lorenz curves of wealth fraction $w$ 
versus household fraction $h$ for the analytical model with 
the numerical data of the RJS model for finite $N=10000$ and 
for three key values of the rescaled energy 
$\eps=0.1,0.2,0.3$ (top to bottom). 
Left panels shows the difference between the analytical model 
and numerical data and right panels show directly the curves $w$ versus $h$ 
for the numerical data (red lines and plus symbols) and the analytical model.
Blue lines/data points correspond to the formula (\ref{eqlorA}) valid 
for all values of $\eps$ and using the appropriate value of the 
chemical potential $\mu$ determined by the implicit equation (\ref{eqmuA}). 
Green lines/data points correspond to the (second) approximate formula  
(\ref{eqlorA0}) valid for small $\eps\le 0.2$. 
The discrete points of data in the top right panel for $\eps=0.1$ at 
values close to $w=0$ indicate finite values for $\rho_0=0.1129$, 
$\rho_0+\rho_1=0.1660$, etc. which are due to RJ condensation. 
}
\end{center}
\end{figure}

It is also possible to compute the Gini coefficient:
\begin{align}
\nonumber
G&=1-2\int_0^1 w(h)\,dh\\
\label{eqGini1}
&=1+\frac{2\mu}{\eps}\left[
(\eps-\mu)^2(e^{1/(\eps-\mu)}-1)-(\eps-\mu)-\frac12
\right]\\
\label{eqGini2}
&=1-\frac{\mu}{\eps}-2(\eps-\mu)\ .
\end{align}
Here the second simpler expression (\ref{eqGini2}) has been obtained 
by replacing the exponential term in (\ref{eqGini1}) using the implicit 
equation of $\mu$. The limit $\eps\ll 1$ with $\mu\approx -e^{-1/\eps}$ 
gives $G\approx 1-2\eps$ which matches well the values of $G$ given in the 
caption of Fig.~\ref{figI_1} for $\eps\le 0.1$ (rather close value for $\eps=0.2$). The 
other values are matched exactly by the more precise expression 
(\ref{eqGini2}). 
Furthermore, inserting the expression (\ref{eqmuinfty}) for large $|\mu|$ 
in (\ref{eqGini2}) one finds (confirms) that $G=1/3$ for $\eps=1/2$ 
(here it is necessary to keep the $1/\mu$ correction in (\ref{eqmuinfty}) 
to obtain the correct result for $G$). 

Using the analytical expression (\ref{eqlorA}) for $w(h)$, it is also 
straightforward to compute (with simple numerics) the inverse 
function $h(w)$. Using this and the analytical expression (\ref{eqGini2}) 
for the Gini coefficient, we have also recomputed the curves 
for $G(\eps)$, $h(2\%)$, $1-h(25\%)$ (both as a function of $\eps$) 
and verified that the analytical curves coincide with the numerical curves 
shown in Fig.~\ref{figI_5} and Fig.~\ref{figI_6} (for the RJS model at $N=10000$) up to graphical 
precision (typical error $\sim 10^{-4}$). 

One might be concerned that the integral approximation is not very 
good for small $\mu$ (close to the singularity of the first term in 
(\ref{eqS3})) and some finite but large value of $N$ such as $N=10000$. This 
is true but the integral provides a modified logarithmic singularity which 
allows also to mimic correctly the condensation effect with correct 
probabilities. Therefore even though the values of $\mu$ are modified 
for $\eps\ll 1$ (but still $0<-\mu\ll \eps\ll 1$ for both models !) 
the resulting probabilities (e.g. integrals or sums of $\rho_m$ over 
some interval in $\tilde E=m/N$) are the same. The values of $\mu$ obtained 
by the continuous analytical model match very well the curve shown 
in Fig.~\ref{figA1} but of course this figure does not allow to verify if 
$\mu\approx -e^{-1/\eps}$ (continuous model) or $\mu\approx -\eps/(N-1)$ 
(for the finite $N$ model with discrete sums) which are both close to zero 
in the figure. 
In any case, we find that the analytical expressions given here 
(if $\mu$ is properly evaluated by its implicit equation (\ref{eqmuA}) 
and if properly evaluated by avoiding numerical instabilities of some 
formulas in some special cases) match the numerical data with an 
error that scales with $1/N$.

Without going into details, we mention that 
we have also considered a more refined version 
of the continuous model using a finite value of $N$ and keeping the 
first singular term separate from 
the integral (which starts at $s=1/N$ and not $S=0$). In this case, we obtain 
a modified implicit equation of $\mu$ which results in 
values of $\mu$ closer to the model of finite $N$ but the resulting physical 
quantities ($w(h)$ curves, Gini coefficients etc.) are (numerically 
with an error $<10^{-4}$) the same as both 
the numerical data and the simple model. The resulting 
analytical expressions of the refined model 
are slightly modified (essentially replacing $h$ by 
$h-\rho_0$ for $h\ge \rho_0$ 
in the formula of the Lorenz curve and using $w(h)=0$ for $h<\rho_0$ 
where $\rho_0$ may now have a finite value). Note that the 
initial interval 
$h\in[0,\rho_0[$ with exactly $w(h)=0$ for the refined and also the 
discrete model translates to exponentially small values 
$w(h)\approx h^2\,e^{-1/\eps}/(2\eps)$ 
for the simple analytical model (replacing $\mu\approx -e^{-1/\eps}$ in 
(\ref{eqlor2})). 

\subsection{Data for companies of stock exchange at New York, London, Hong Kong}

We present here the Lorenz curves for the capitalization of companies
at stock exchanges of New York, London, Hong Kong.
They are obtained respectively from Refs.~\cite{sp500,london,hk}.

First, we present in Fig.~\ref{figA9} the Lorenz curve 
for the data of 504 S\&P500 companies of the New York Stock Exchange (NYSE)
of June 16, 2025 (see Ref.~\cite{sp500}). This Fig.~\ref{figA9}  
shows the direct comparison of the Lorenz curve of NYSE
and the corresponding RJ thermal distribution of the RJS model 
(at same Gini value). Here, we use the standard value $N=10000$ for 
the RJS curve but using a reduced value $N=504$ (as the number of companies) 
gives the same RJS curve within graphical precision. 
The quality of agreement with the RJS model 
is comparable to the cases of US or World in Fig.~\ref{figI_3}. 
We also note the characteristic values:
$h=0.191$ at wealth $w=0.02$;
$w=0.092$ at $h=0.5$;
the wealth of top 10 percent of $h$ companies
is $1-w(0.9)=0.602$
and the wealth of top 1 percent of companies is
$1-w(0.99)=0.267$. Thus we see that there is a small
fraction of oligarchic companies that monopolize
a big fraction of total wealth.
The fraction of poor companies, corresponding to the RJ condensate,
is smaller than the fraction of poor households in the US or World cases.
We attribute this to the fact that these 504 companies of S\&P500
represent only about 80 percent of the 
total capitalization of US companies.

\begin{figure}[htbp]
\begin{center}
\includegraphics[width=0.65\textwidth]{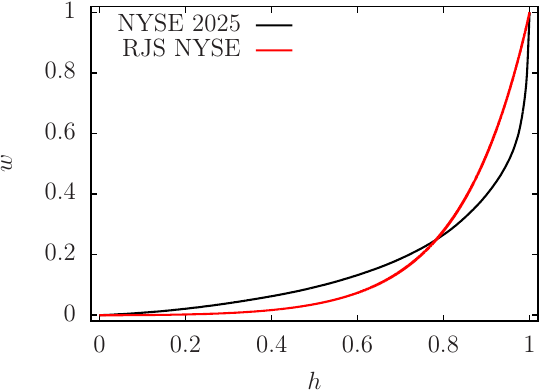}
\caption{\label{figA9}
Comparison of the Lorenz curve for the S\&P500 companies of NYSE 2025
(black; data from Ref.~\cite{sp500}) 
with the corresponding curve for the RJS model (red curve; $N=10000$) 
at same Gini coefficient $G=0.692$ obtained for $\eps=0.1582$. 
}
\end{center}
\end{figure}

Fig.~\ref{figA10}, compares the Lorenz curve for the London 
stock exchange (2024; data from Ref.~\cite{london}) with the RJS model. 
Here, the Gini coefficient 
$G=0.9126$ is higher than for the US and World cases and the corresponding 
value $\eps\approx 0.044$ for the RJS model is quite small. Due to 
the high value of $G$ 
the first probability $\rho_0=0.6545$ is very high indicating 
a strong RJ condensation with exactly $w=0$ for $h\in[0,\rho_0[$ 
in the RJS model. The chosen value $N=1637$ is identical to the number of 
considered companies but the RJS curve for $N=10000$ is identical on graphical 
precision (with a slightly modified value for $\eps$).

\begin{figure}[htbp]
\begin{center}
\includegraphics[width=0.9\textwidth]{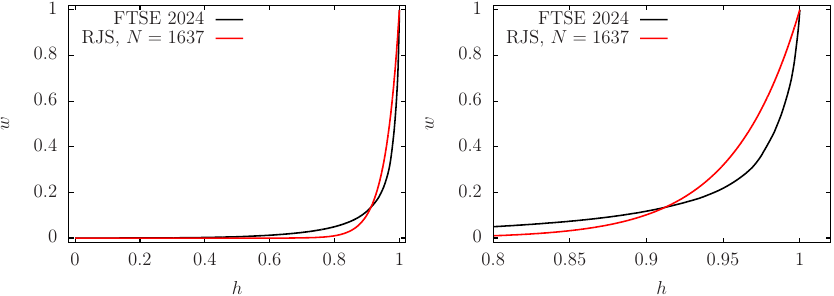}
\caption{\label{figA10}
Comparison of the Lorenz curve for the 1637 companies of the London 
stock exchange FTSE at 31 December 2024 
(black; data from Ref.~\cite{london}) 
with the corresponding curve for the RJS model (red curve; $N=1637$) 
at same Gini coefficient $G=0.9126$ obtained for $\eps=0.04387$. 
The left (right) panel shows the full range $h\in[0,1]$ (zoomed range 
$h\in[0.8,1]$). 
}
\end{center}
\end{figure}

Fig.~\ref{figA11}, compares the Lorenz curve for the Hong Kong
stock exchange (2025; data from Ref.~\cite{hk}) 
with the RJS model. Here, the Gini coefficient 
$G=0.9471$ is even higher than for the London stock exchange and the 
corresponding 
value $\eps\approx 0.027$ for the RJS model is even smaller. Due to 
the very high value of $G$ 
the first probability $\rho_0=0.7768$ is even higher (than $\rho_0$ 
for the London stock exchange) indicating 
a strong RJ condensation with exactly $w=0$ for the larger interval 
$h\in[0,\rho_0[$ in the RJS model. The chosen value 
$N=2683$ is identical to the number of 
considered companies but the RJS curve for $N=10000$ is identical on graphical 
precision (with a slightly modified value for $\eps$). 

\begin{figure}[htbp]
\begin{center}
\includegraphics[width=0.9\textwidth]{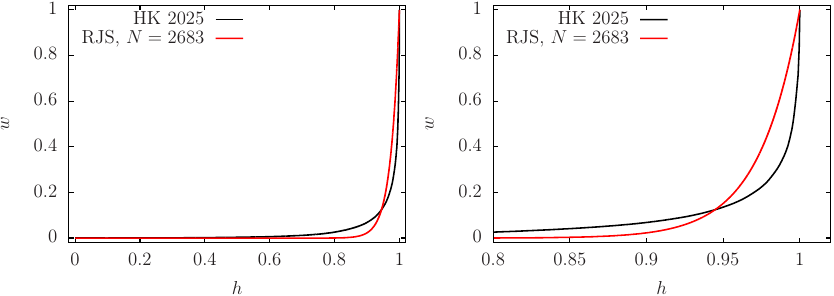}
\caption{\label{figA11}
Comparison of the Lorenz curve for the 2683 companies of the Hong Kong 
stock exchange at 19 June 2025
(black; data from Ref.~\cite{hk}) 
with the corresponding curve for the RJS model (red curve; $N=2683$) 
at same Gini coefficient $G=0.9471$ obtained for $\eps=0.02651$.  
The left (right) panel shows the full range $h\in[0,1]$ (zoomed range 
$h\in[0.8,1]$). 
}
\end{center}
\end{figure}

Fig.~\ref{figA12}, compares the Lorenz curve for the 30 Dow Jones companies 
(2025; data from Ref.~\cite{sp500}) with the RJS model. Here, the Gini coefficient 
$G=0.3096$ is very low and the corresponding 
value $\eps\approx 0.55$ for the RJS model is very high being in 
the region for $T<0$ with large $|T|$. The value of $G$ is even smaller
then $G=1/3$ for the curve $w=h^2$ corresponding to the RJS model with 
$\eps=0.5$ and $|T|\to\infty$. The chosen value 
$N=30$ is identical to the number of 
considered companies but despite the modest value of $N$ the RJS curve for 
$N=10000$ is identical on graphical 
precision (with a slightly modified value for $\eps$). 
We mention, that a comparison with the EQI model for a modest value of 
$E_0$ to fit approximately the finite initial slope in the data provides 
the energy value $\eps\approx 0.48<0.5$ corresponding to the regime of 
$T>0$ but still with large $|T|$. 
We note that this case is very special since these 30 companies are 
certainly not isolated and they constitute a subset of the 
504 companies of S\&P500 (which are not perfectly isolated as well). 

\begin{figure}[htbp]
\begin{center}
\includegraphics[width=0.65\textwidth]{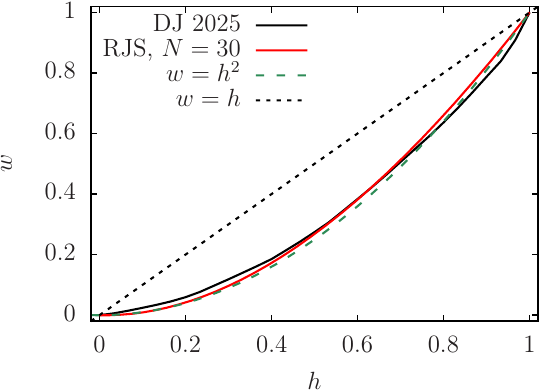}
\caption{\label{figA12}
Comparison of the Lorenz curve for the 30 Dow Jones companies of NYSE 2025
(black; data from the site of Ref. [33] taken at June 18, 2025) 
with the corresponding curve for the RJS model (red curve; $N=30$) 
at same Gini coefficient $G=0.3096$ obtained for $\eps=0.5528$. 
The dashed green (black) line represents the curve for $w=h^2$ ($w=h$) 
for the RJS model at $\eps=0.5, T\to\infty$ (perfect equipartition). 
}
\end{center}
\end{figure}

\subsection{Spectral reconstruction procedure} 

Let us briefly remind the construction of a Lorenz curve, 
already given in the main part, from a given 
model spectrum $E_m$ with appropriate values of $\mu$, $E$ and $T$ such that 
the two conditions (\ref{eqS1}) are verified. 
For this a set of points $(h_m,w_m)$ is determined with $h_0=w_0=0$, 
$h_{m+1}=h_m+\rho_m$ and $w_{m+1}=w_m+(E_m/E)\,\rho_m$ 
for $m=0,\ldots,N-1$ and where 
\begin{align}
\label{eqrho1}
\rho_m=\frac{E-\mu}{N(E_m-\mu_m)}
\end{align}
(see also (\ref{eqS2})). Then the conditions 
(\ref{eqS1}) assure that $h_{N}=w_{N}=1$ and the points $(h_m,w_m)$ provide 
for $0\le m\le N$ the associated Lorenz curve with both $h_m,w_m\in[0,1]$. 

The question arises if it is possible to invert this construction, i.e. 
to determine (``reconstruct'') for a given Lorenz curve $w(h)$ a 
certain effective spectrum $E_m$ with appropriate values of $\mu$ and 
$E$ such that its related Lorenz curve is very close to the Lorenz curve of real data 
with best possible precision (depending on the choice of $N$). 

This is indeed 
possible and to define an explicit reconstruction procedure 
let us assume we have some smooth Lorenz curve $w(h)$ with 
derivatives $w'(h)\ge 0$ and $w''(h)>0$ for all $h\in[0,1]$. In particular, 
we assume that we can compute numerically with high precision and in 
a reliable way the derivative 
$w'(h)$ which satisfies $w'(h_1)>w'(h_2)$ for all points with $h_1>h_2$ (this 
assumption may be problematic in practice; see below). 
For simplicity, we also assume 
that $w'(0)=0$ and we want to construct spectra with $E_0=0$ which is the 
most relevant case for the typical examples 
(it is not difficult to modify the method 
for the more general case $w'(0)\neq 0$ with $E_0\neq 0$). 
Furthermore, we choose $E=1$ to fix the global energy scale. 
Then the value of $E_m/E=E_m$ is (very close to) 
the derivative 
$w'(h_m)$ at the corresponding value of $h_m$. However, initially we do not 
know the value of $h_m$ for a given index $m$ (or the index $m$ value 
as a function of $h_m$). 

We choose some initial value for $\mu$ (close to zero), 
start with $m=N-1$ and want first to determine $E_m=E_{N-1}$. Here we know 
the last value $h_{m+1}=1$ at $m=N-1$. Therefore, we can at least 
approximately compute 
$E_m=w'(h_{m+1})$ and then the associated value of $\rho_m$ 
using (\ref{eqrho1}). This pair $(E_m,\rho_m)$ is yet not very 
precise since the 
derivative is taken at the right boundary of the interval $[h_m,h_{m+1}]$ 
and we can refine its value 
by recomputing $E_m=w'(h_{m+1}-\rho_m/2)$ using a small shift 
with the first approximate value 
of $\rho_m$ (which will then be updated with the more precise value of $E_m$ 
using (\ref{eqrho1})). In principle, one could iterate 
this refinement step until there is convergence of $(E_m,\rho_m)$. However, 
in our experience the method works best with precisely one refinement step
(to ensure later convergence for a good value of $\mu$). Once 
$\rho_m$ is known, we obtain $h_m=h_{m+1}-\rho_m$. Then we decrease 
$m$ by 1 and repeat this procedure to compute the next values 
of $E_m$, $\rho_m$ and $h_m$ at $m=N-2$. This provides a recursion 
for $m=N-1,N-2,\ldots,1,0$ and three sequences 
for $E_m$, $\rho_m$ and $h_m$ with decreasing $m$.

For the last value $E_0$ at $m=0$ we do not use the derivative but we simply 
fix it by $E_0=0$. Typically, in this regime the derivative is already very 
small. Ideally, the last value $h_0(\mu)$ should be $h_0(\mu)=0$ but this is 
only true for a specific value of $\mu$ which has to be found iteratively, 
e.g. to be determined numerically as the zero of the 
function $h_0(\mu)$ by some standard method (which is actually quite tricky 
for bad quality data with problematic convergence) and where this function 
is computed by a full reconstruction loop $m=N-1,\ldots,0$ for each value 
of $\mu$ as described above. 

Instead of searching numerically the zero of the function $h_0(\mu)$, 
one can also use another more practical method to determine the correct 
value of $\mu$. For this, one can at the last step $m=0$ manually fix the 
last density value 
and compute from (\ref{eqrho1}) a new modified value $\tilde\mu$ such that the 
condition $\rho_0(\tilde \mu)=h_1$ holds exactly and therefore $h_0=0$ 
is perfectly verified. The modified 
value $\tilde \mu$ can be reinjected in the procedure from the start 
resulting in a fixed-point iteration for $\mu$ which 
typically converges quite well and allows also to use the exact initial value 
$\mu=0$ at the first iteration (which is not a problem since this 
value is not used in the last step at $m=0$ with $E_0=0$). 
For this method the convergence is typically a bit slower, but 
more reliable, as compared to the secant 
method applied to $h_0(\mu)$ but the latter fails to converge in cases 
of bad quality data which influence the computation of 
$w'(h)$. In such a situation the fixed point iteration does not always provide 
a convergence with high precision as well but 
still the $\mu$ values stabilize in some small interval (with relative 
fluctuations $\sim 10^{-3}$ etc.) and any value in this interval can be used 
to have a nice reconstructed spectrum. 

Once the procedure is finished, we can use the obtained spectrum $E_m$ 
to recompute a new appropriate value of $\mu$ and the densities $\rho_m$ 
in the usual way by numerically solving 
(\ref{eqS3}) with the value $E=1$. In case of 
good convergence of the procedure this simply confirms the 
already obtained values of $\mu$ and $\rho_m$ but in case of a problematic 
convergence, this provide a final refinement of $\mu$ and $\rho_m$ which 
will match precisely the spectrum $E_m$ with $E=1$ according to (\ref{eqS1}) 
and (\ref{eqS2}). Using these refined densities, we can finally 
recompute the Lorenz curve associated to this spectrum in the usual way. 
This curve matches typically, also in the case of not perfect convergence, 
very well the original data with numerical errors below $10^{-3}$ (or less). 

The choice of the parameter $N$ for the size of the reconstructed spectrum 
is not very important, except it needs to be sufficiently large, e.g. $N=1000$.
The reconstructed spectrum $E_m$ provides, as a function of the rescaled 
level number $x=m/N$, essentially the same curves for different 
(sufficiently large) values of $N$ 
provided that the same (reliable) numerical implementation for the derivative 
function $w'(h)$ is used. 

The method depends in a very sensitive way on the quality of the numerical 
implementation of $w'(h)$, quality of input data and chosen interpolation 
procedure and this part is actually rather tricky. Usual linear interpolation 
for the initial data for the Lorenz curve $w(h)$ provides a piecewise 
constant derivative $w'(h)$ which works reasonably well in the above procedure 
concerning $\mu$-convergence and good matching of the initial Lorenz curve. 

If the raw data is of good quality, i.e. with support points that lie 
very accurately on a smooth function, one can also use a combination of 
rational interpolation (for the region where $w'(h)>1$) 
and polynomial interpolation (for the region with $w'(h)<1$) and 
in both cases with a small number of support points between 3 and 6 
which are closest to the value of $h$ for which we want to compute $w'(h)$. 
In both interpolation approaches, one can work out efficient formulas 
to exactly evaluate the derivative of the interpolation function.
However, if the data is of bad quality this procedure may be problematic 
for $\mu$-convergence and also violate the property that $w'(h_1)<w'(h_2)$ 
for $h_1<h_2$ which is crucial to obtain a correctly ordered spectrum 
(with $E_{m_1}\le E_{m_2}$ for $m_1<m_2$). In such a case, it may be 
necessary to clean the data by coarse-graining them (keeping only 15-20 
significant data points) and then recompute a new data set with 500 or 1000 
points using high quality interpolation (also rational/polynomial 
interpolation with 4-5 support points from the reduced set). 
In particular for the data of UK 2012-2014 from Ref. [31] with a lot of data 
points but with limited precision this was necessary.

In order, to keep things simple and reliable, we opted for a compromise 
between rational/polynomial interpolation and a piecewise constant 
derivative. Without going into too much details, we mention that 
we computed first discrete derivatives (from good quality data) and applied 
linear interpolation to obtain a piecewise linear numerical implementation 
of $w'(h)$ which respects that $w'(h_1)<w'(h_2)$ for $h_1<h_2$ and 
is still a continuous function. In this approach the support points for 
$h$ are now in the center of two former supports points 
(for which the discrete derivative was taken) and also a slight 
renormalization was applied to assure that the interpolated piecewise 
linear function satisfies numerically 
$\int_0^1 w'(h)dh=1$, a property which is very important for the 
reconstruction procedure.

Using this particular implementation of $w'(h)$, we have applied the 
above reconstruction procedure to all available data sets. Typically, 
the obtained reconstructed spectra initially increase slowly (linearly, 
with a possible quadratic correction) but at some critical value of 
$x_c=m_c/N\approx 0.7$-$0.9$ the increase becomes significantly stronger. 
Beyond 
this critical value the precise form of the obtained spectrum depends 
rather strongly on the chosen interpolation method and the obtained values of 
$E_m$ are not very reliable. This corresponds to the regime of the 
Lorenz curve with $h$ close to $1$ where both $w'(h)$ and $w''(h)$ may be very 
large and difficult to obtain with high precision by interpolation. 

\begin{figure}[htbp]
\begin{center}
\includegraphics[width=0.65\textwidth]{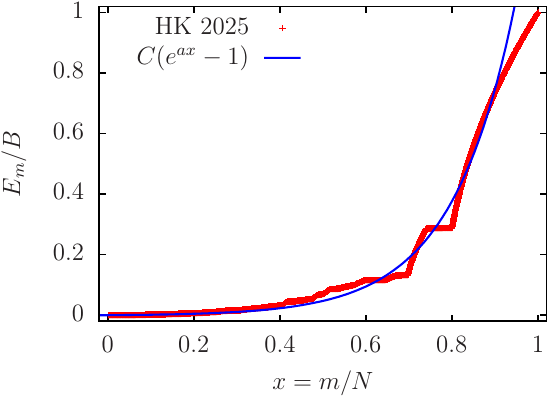}%
\end{center}
\caption{\label{figA13}
\label{fig_fits_HK_2025}
Rescaled reconstructed spectrum $E_m/B$ 
for the data of Hong Kong 2025 (red data points) 
versus rescaled level number $x=m/N$. The blue curve shows 
the curve $E_m/B=C\,(e^{ax}-1)$ with $C=0.00160\pm 0.00004$ 
and $a=6.83\pm 0.03$ obtained from a fit in the interval $x\in[0,0.9]$. 
The value of $a=6.83$ is used in Fig.~\ref{figI_11} for the blue curve 
of the RJE model. 
}
\end{figure}

Globally, the fit $E_m= C (e^{a (m/N)}-1)$ works rather well 
at least for 
some reasonable subinterval. For each data set, we perform two fits 
of this function for the intervals $x\in[0,0.7]$ and $x\in[0,0.9]$ that
provides two interesting values of $a$. We inject these two values 
of the parameter $a$ 
in the RJE model and determined which $a$ value gives a better agreement 
for the Lorenz curve (the value of $\eps$ is determined as usual by 
matching the Gini coefficient to be identical between initial Lorenz curve 
and the model curve). Figs.~\ref{figI_7}---\ref{figI_11} show the resulting 
RJE curves for 5 of our data sets, already  discussed in 
the previous section, with a very good agreement of the RJE model 
at the optimal fitted values of $a$. 

In Fig.~\ref{figA13}, we show as illustration one example of an 
reconstructed spectrum for the data set of the stock market Hong Kong 2025
using the value $N=2683$ and the piecewise linear derivative for $w'(h)$. 
The number $N=2683$ represents the number of companies used in the 
Hong Kong SE data but the precise choice of this value is not very important 
and the reconstruction procedure works also nicely for $N=1000$ or $N=10000$ 
for this example. In this case, the shown fit 
$E_m/B=C\,(e^{ax}-1)$ works quite well for the larger interval 
$x\in[0,0.9]$ and the resulting value of $a=6.83$ provides a nearly 
perfect Lorenz curve of the RJE model. 

We mention that the bandwidth $B$ of the reconstructed spectrum shown 
in Fig.~\ref{figA13} 
is $B=566.6$ and it corresponds to the initial choice $E=1$ to fix the 
global energy scale such that the 
rescaled energy of the reconstructed spectrum is 
$\eps_{rc}=1/B\approx 0.001765$.
The value of $\eps_{\rm RJE}$ for the fitted blue curve is slightly modified 
due to a modified bandwidth of the latter: 
$\eps_{\rm RJE}\approx \eps_{rc}/[C\,(e^a-1)]=0.001193$ 
which compares to the value $\eps=0.0008381$ given in the caption 
of Fig.~\ref{figI_11} obtained by matching the Gini coefficient. 
The slightly different value for $\eps$ is due to the matching of the 
Gini coefficient and the fact that the fit is far from perfect. Furthermore, 
also the data points for $x>0.9$ are not very reliable. 

However, here we do not want to enter deeply in such details and we use 
this reconstruction procedure more as a tool to determine and justify 
optimal values of $a$ for the RJE model. 
Globally this procedure is very sensitive to technical details and 
parameter choices which give potentially rather different spectra 
(for $x$ close to $1$ beyond a certain critical value $x_c$) but which all 
reproduce afterwards matching Lorenz curves to the initial data with good 
accuracy. 

Finally, we note that instead of the above reconstruction procedure
it is also possible simply to fit the value of parameter $a$ in the
spectrum $E_m$ of Eq.(\ref{RJEdef}) in such a way that the Lorenz curve
of the RJE model is closet to the real Lorenz curve which 
is done by minimizing 
a suitable metric to measure the distance between two curves 
(parameter $\varepsilon$ is fixed as usually to 
match Gini coefficient). We checked that the obtained $a$ values 
by minimizing different variants of such metrics 
are rather close to those obtained from the reconstruction procedure.
Thus both approaches allows to obtain very
good agreement between the model and real Lorenz curves.
The advantage of the reconstruction
procedure is related to 
a more physical understanding of the origins of spectrum $E_m$.

\section{Additional material for Part II}
\renewcommand{\theequation}{B.\arabic{equation}}
\renewcommand\thefigure{B.\arabic{figure}}
\renewcommand\thesubsection{B.\arabic{subsection}}
\setcounter{equation}{0}

\subsection{Netscience network}

\begin{figure}[htbp]
\begin{center}
\includegraphics[width=0.8\columnwidth]{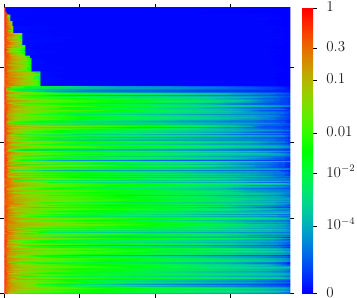}%
\end{center}
\caption{\label{figB1}
\label{fig_vecNet}
Color plot of the eigenvectors $\phi_n^{(m)}$ 
of the matrix $H$ (size $379\times 379$) for 
the \SN\ at $\kappa=0$. The values of the color bar correspond 
to $|\phi_n^{(m)}|^2$ at a certain position $(x=K_m(n),y=L(m))$ in 
the grid of the color plot. 
Each row shows a given eigenvector ordered with 
decreasing values of $|\phi_n^{(m)}|$ in the node index $n$
(from left to right) and therefore the $x$-axis corresponds to the 
ordering index $K_m(n)=1,\ldots, 379$ such that 
$|\phi_n^{(m)}|>|\phi_{\tilde n}^{(m)}|$
for $K_m(n)<K_m(\tilde n)$ (This index vector is different for each 
eigenmode $m$). 
For each such vector the support length $l(m)=$ number of nodes 
$n$ with $|\phi_n^{(m)}|>10^{-12}$ has been computed 
and the vectors have been ordred with decreasing values of $l(m)$ 
(bottom to top; ordering for identical $l(m)$ values is arbitrary) and 
the $y$-axis corresponds therefore to the ordering index $L(m)$ such 
$L(m)<L(\tilde m)$ for $l(m)>l(\tilde m)$ (lowest $L(m)$ values 
correspond to the bottom rows).
The tics indicate integer multiples of 100 for both index values 
$K_m$ and $L$. There are 104 out of 379 eigenvectors with a support 
length $l(m)$ significantly smaller than $N=379$ visible at the 
top 104 rows (other eigenvectors have either $l(m)=N$ or $l(m)\approx N$ due 
to the limited numerical precison). The eigenvalues $E_m$ of the 
104 modes with small $l(m)$ values correspond to nice fractional 
values (with denominator $\le 420$) and the majority of them 
correspond to the degenerate plateau values visible in the 
red curves in the bottom panels of Figure~\ref{fig_specNet}. 
The scale of the color bar is strongly non-linear to enhance small 
values of $|\phi_n^{(m)}|^2$. }
\end{figure}

In Fig.~\ref{figB1}, we show 
a color plot of the eigenvectors at $\kappa=0$. Each row represents 
an eigenvector with components in node space ordered horizontally 
according to the rank index $K_m$ of this vector (i.e. 
with decreasing values of $|\phi^{(m)}_n|$ with $n$ for fixed $m$) and 
the eigenvectors themselves are ordered vertically with decreasing values 
of their support length $l(m)$ (from bottom to top). 

The modes with limited support length $l(m)\le 48$ have 
energies $E_m=p/q$ with nice rational values (maximal $q=420$ 
and other $q\le 12$). We mention a few examples, such as 
the mode $E_{25}=-25/12$ with $l(25)=\xi_{\rm IPR}=2$ localized 
on the two nodes {\em Kim, D.} and {\em Goh} and with values 
$\phi^{(25)}_n=\pm 1/\sqrt{2}$ for these two nodes. 
Another example is the pair of two modes $E_{66}=E_{67}=-389/420$ with $l(66)=l(67)=3$ 
and $\xi_{\rm IPR}=2$ for both. These two modes are 
localized on the three nodes {\em Rajagopalan}, {\em Raghavan}, 
{\em Tomkins}. 

\begin{figure}[htbp]
\begin{center}
\includegraphics[width=0.8\columnwidth]{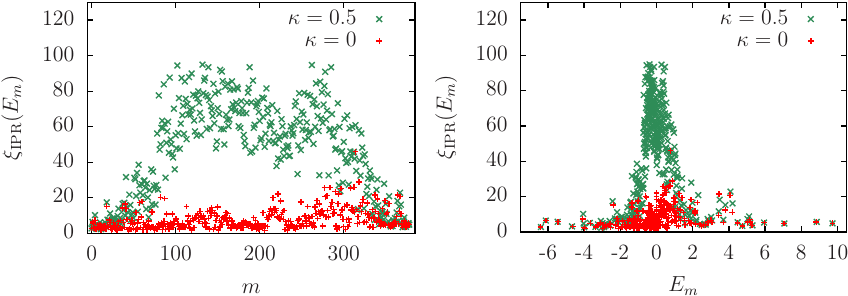}%
\end{center}
\caption{\label{figB2}
\label{fig_IPR_net}
$\xi_{\rm IPR}(E_m)$ of eigenvector $\phi^{(m)}$ 
of $H$ for the case of the \SN\ 
with $\kappa=0$ and $\kappa=0.5$ 
versus index $m$ (energy $E_m$) in left (right) panel.}
\end{figure}

Fig.~\ref{figB2} shows the IPR of eigenvectors for the \SN\ at 
$\kappa=0$ and $\kappa=0.5$ versus mode index $m$ and also mode energy 
$E_m$. The IPR values at $\kappa=0.5$ are roughly 
a factor 10 larger than the values at $\kappa=0$ and the maximal 
value of the IPR is 45.98 (94.94) for $\kappa=0$ ($\kappa=0.5$). 
However, the boundary modes have essentially the same (small) IPR values 
since due to the large energy gaps these modes are in the quantum perturbative 
regime even at $\kappa=0.5$ with coupling matrix elements 
$\sim \kappa/\sqrt{N}=0.5/\sqrt{379}\approx 1/40$ which is much smaller than 
the boundary energy gaps. 

\begin{figure}[htbp]
\begin{center}
\includegraphics[width=0.8\columnwidth]{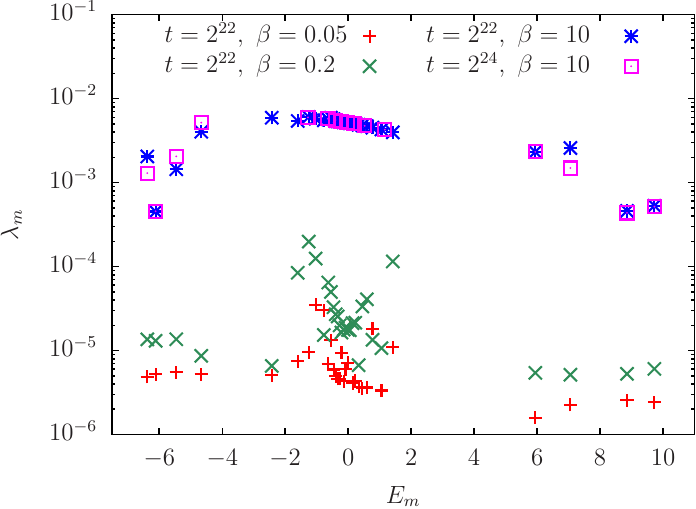}%
\end{center}
\caption{\label{figB3}
\label{fig_lyap}
Lyapunov exponent $\lambda_{m}$ dependence on $E_{m}$ with 
$m$ being the index of the initial state for the \SN\ with $\kappa=0.5$ 
and $N=379$. $\lambda_{m}$ 
has been determined from the fit $\ln\|\Delta\psi(t)\|
=a+b\ln(t)+\lambda_{m}\,t$. 
Shown are data from 4 data sets: with 32 selected modes 
at $t=2^{22}$ for $\beta=0.05$ (red $+$ symbol), 
$\beta=0.2$ (green $\times$ symbol), $\beta=10$ (blue $*$ symbol) 
and also with 16 selected modes at $t=2^{24}$ 
for $\beta=10$ (pink $\square$ symbol) obtained from a different 
computation. }
\end{figure}

Fig.~\ref{figB3} shows the Lyapunov exponent $\lambda_m$ 
for selected initial modes $m$ of the \SN\ at $\kappa=0.5$ and 
$\beta=0.05,0.2,10$. The quantity $\|\Delta\psi(t)\|$ mentionned in the 
figure caption is the vector norm of the difference 
$\Delta\psi=\psi_2(t)-\psi_1(t)$ where $\psi_1(t)$ and $\psi_2(t)$ 
are both solutions of (\ref{eqNLeq1}) with 
different but close initial conditions at the initial mode 
$\psi_1(t=0)\approx \psi_2(t=0)\approx \phi^{(m)}$. In this case 
the Lyapunov exponent is defined by 
$\lambda_m=\lim_{t\to\infty}(\ln\|\Delta\psi(t)\|)/t$ and 
the data in Fig.~\ref{figB3} has been extracted by the fit 
$\ln\|\Delta\psi(t)\|=a+b\ln(t)+\lambda_{m}\,t$. 

The modes at $\beta=10$ are clearly in the chaotic 
regime while boundary modes at $\beta=0.2$ and (most) modes at $\beta=0.05$ 
are in the KAM regime with much smaller (numerical) Lyapunov exponent 
(we expect that
for these states Lyapunov exponents become zero
in the limit $t \rightarrow \infty$).

\begin{figure}[htbp]
\begin{center}
\includegraphics[width=0.95\columnwidth]{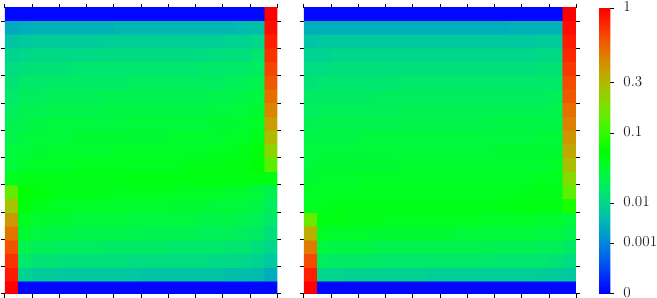}%
\end{center}
\caption{\label{figB4}
\label{fig_rhodensbarBoth}
Color plot of the coarse-grained thermalized occupation 
probabilities $\rho_m=T/(E_m-\mu)=(E-\mu)/[N(E_m-\mu)]$
for the \SN\ (\PN) at $\kappa=0.5$ in left (right) panel. 
The $x$-axis corresponds 
to the fraction $m/N\in[0,1]$ with $m=1,\ldots,N$ being the 
index of energies $E_m$ (left to right)
and the $y$-axis corresponds to the rescaled 
energy $\varepsilon=(E-E_1)/(E_N-E_1)$ (top to bottom 
for increasing values). The tics indicate integer multiples of 0.1 
for both quantities. 
The color values shown in the color bar 
correspond to the value of $\rho_m$ averaged over intervals of size 
$1/20$ (for $m/N$ on the $x$-axis) and computed for 
21 values $\varepsilon=i/20$, 
$i=0,1,\ldots, 20$ (for the $y$-axis; the 
minimal value $\varepsilon=0$ has been 
slightly enhanced and the maximal value $\varepsilon=1$ has been slightly 
reduced to have a stable computation of the thermalized $\mu$-value). 
To increase visibility of small values a non-linear color bar scale 
has been chosen (e.g. green color corresponds to $1/16$). }
\end{figure}

Fig.~\ref{figB4} presents coarse-grained color plots in 
the $m$-$E$ plane of the theoretical RJ-values $\rho_m=T/(E_m-\mu)$ 
where $T$ and $\mu$ are determined from the implicit equations 
(\ref{eqConstraints1}). The horizontal $x$-axis corresponds 
to $m/N\in]0,1]$ and the vertical $y$-axis to 
the rescaled energy $\varepsilon=(E-E_1)/(E_N-E_1)\in [0,1]$. The left (right) 
panel corresponds to the energy spectrum of $H$ at $\kappa=0.5$ 
for the \SN\ with $N=379$ (\PN\ with $N=5908$). 
One clearly sees a condensation 
on the modes with minimal $m\sim 1$ (maximal $m\approx N$) at 
$E\approx E_1$ ($E\approx E_N$) while for intermediate energies the 
distribution of $\rho_m$ is more uniform in $m$. 

\begin{figure}[htbp]
\begin{center}
\includegraphics[width=0.95\columnwidth]{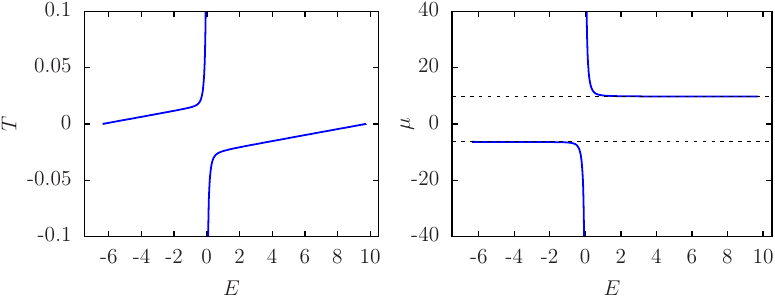}%
\end{center}
\caption{\label{figB5}
\label{fig_Tmu}
The left (right) panel shows the temperature $T$ 
(the chemical potential 
$\mu$) versus the energy 
$E$ for the \SN\ at $\kappa=0.5$. 
The dashed black lines in the right panel correspond to the values of 
$E_1=-6.3829$ and $E_N=9.7256$ showing that either $\mu<E_1$ (for $T>0$) 
or $\mu>E_N$ (for $T<0$). }
\end{figure}

Fig.~\ref{figB5} shows the dependence of $T$ and $\mu$ on $E$ obtained 
by solving the implicit equations (\ref{eqConstraints1}) with 
$\rho_m=T/(E_m-\mu)$. Both curves show the usual behavior with 
$T>0$ ($T<0$) and $\mu<E_1$ ($\mu>E_N$) for $E<0$ ($E>0$) 
with $T\to 0$ for $E\to E_1$ or $E\to E_N$. Furthermore, for 
$E\to 0$ we have $|T|\to\infty$ and $|\mu|\sim NT\to\infty$ corresponding 
to uniform $\rho_m\to 1/N$. 

It is not very difficult to show that in the limit 
of large $|\mu|$ the chemical potential is given by 
the equation 
\begin{align}
\label{eqlargemu}
\mu&\approx E+\frac{\sum_{m}(E-E_m)^2}{
\sum_{m}(E-E_m)}\ .
\end{align}
Here the denominator has the same sign as $E-E_c$ where 
$E_c=(\sum_m E_m)/N=\mbox{Tr}(H)/N\sim \kappa/N\approx 0$ 
is the critical energy (``energy center of mass'') 
at which the chemical potential and the 
temperature switch their sign. Here the trace of $H$ is entirely 
given by the trace of the GOE perturbation 
$\kappa H_{\rm GOE}$ since Tr$(A)=0$ (the 
absence of self links implies $A_{nn}=0$). 

Note that Figs.~\ref{figB4} and \ref{figB5}, which simply provide 
a generic illustration of the basic properties of RJ-thermalization, 
are rather similar to Figs.~\ref{figA2} and \ref{figA1} of 
Appendix A obtained from a different simple uniform 
spectrum $E_m=m/N$.

\begin{figure}[htbp]
\begin{center}
\includegraphics[width=0.8\columnwidth]{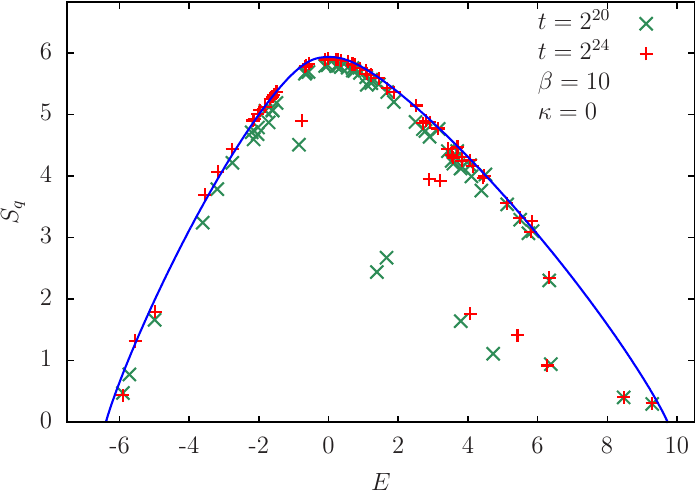}%
\end{center}
\caption{\label{figB6}
\label{fig_SE_q_beta10_kappa0}
Entropy $S_q$ versus energy $E$ for 
the \SN\ with $\kappa=0$, $\beta=10$, $N=379$. 
Shown are 64 selected modes at 
$t=2^{24}$ (red $+$ symbols) and 
$t=2^{20}$ (green $\times$ symbols). 
The blue line shows the energy dependence of the theoretial 
thermalized entropy. 
$S_q$ has been computed by Equation~(\ref{eqSq}) using $\rho_m$ values 
obtained as the time average $\rho_m=\langle |C_m(\tilde t)|^2\rangle$ 
for $t/2<\tilde t\le t$ (for $t=2^{24}$ or $t=2^{20}$ according 
to the selected data in this figure). }
\end{figure}

For the case of the \SN\ with $\kappa=0$ and $\beta=10$, we show in 
Fig.~\ref{figB6} the energy dependence of $S_q$. 
Here the agreement with the theoretical curve is similar 
as for the case with $\kappa=0.5$ but for this longer iteration 
time scales $t=2^{24}$ are required. 

\begin{figure}[htbp]
\begin{center}
\includegraphics[width=0.95\columnwidth]{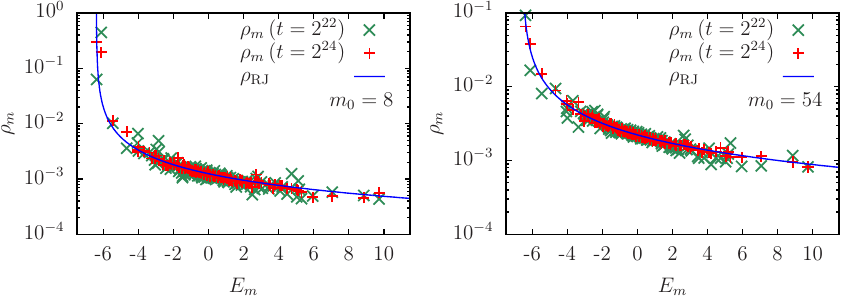}%
\end{center}
\caption{\label{figB7}
\label{fig_stateNET_beta4}
As Figure~\ref{fig_stateNET} 
for the \SN\ with $\kappa=0.5$, $\beta=4$, $N=379$ and 
the two initial modes $m_0=8, 54$. 
The blue curve shows the RJ theoretical curve 
$\rho_{\rm RJ}(E)=T/(E-\mu)$. 
The values of $T$, $\mu$ and $\langle E\rangle$ for the 
2 initial modes $m_0=8, 54$ are 
$T=0.008006, 0.01461$, $\mu=-6.4, -6.574$ and 
$\langle E\rangle =-3.366, -1.038$. }
\end{figure}

In Fig.~\ref{figB7}, we present two examples of nicely thermalized states
for the \SN\ with $\kappa=0.5$ and $\beta=4$, Due to the smaller 
coupling value of $\beta$ the fluctuations at the shorther time $t=2^{22}$ 
are stronger than for similar states at $\beta=10$. 

\begin{figure}[htbp]
\begin{center}
\includegraphics[width=0.9\columnwidth]{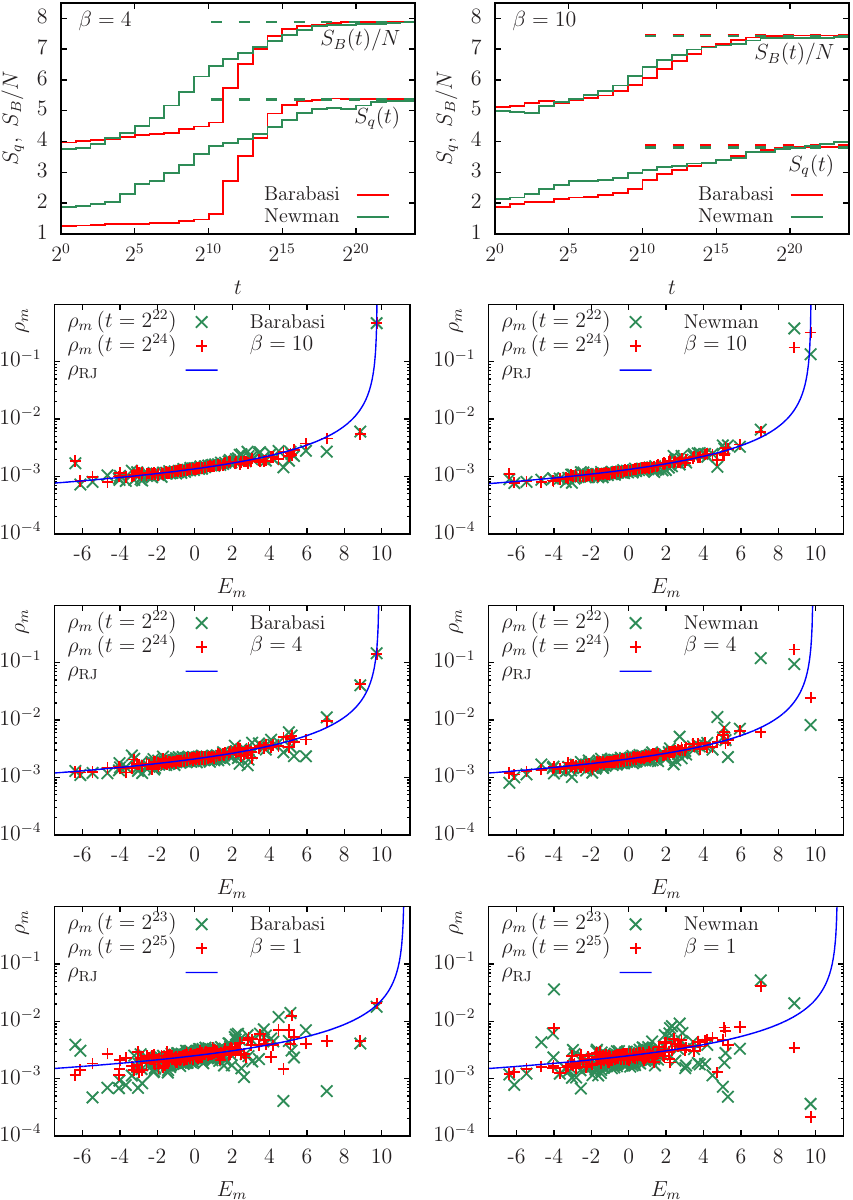}%
\end{center}
\caption{\label{figB8}
\label{fig_BN}
Results for the \SN\ with $\kappa=0.5$ and two initial 
states localized either on the node of {\it Barbasi} or {\it Newman}. 
Top panels show the time dependence of $S_q$ and $S_B/N$ 
for $\beta=4$ (left) and $\beta=10$ (right). The two top (bottom) 
curves correspond to $S_B(t)/N$ ($S_q(t)$) and dashed lines 
indicate the theoretical thermalized values for each case. 
The 6 bottom panels show $\rho_m$ versus mode energy $E_m$ 
for both initial conditions 
and $\beta=10,4$ ($t=2^{22},2^{24}$) and $\beta=1$ 
($t=2^{23},2^{25}$). See text and captions of Figs. 
\ref{fig_stateNET}-\ref{fig_Stime_B_Raw} for more technical details.}
\end{figure}

Fig.~\ref{figB8} shows some results for the intial condition 
localized on the nodes of {\em Barabasi} or {\em Newman} (top two PageRank  
nodes for the \SN). 
In its first two top panels we see the time dependence of $S_q(t)$ and 
$S_B(t)/N$ for both cases and the two values $\beta=4$ and $\beta=10$. 
All eight curves saturate to the corresponding theoretical thermalized 
entropy values which are obtained from the usual value for 
$E=\langle E\rangle=\sum_m E_m\rho_m(t=2^{24})$ to determine $T$ and $\mu$ 
and with the numerical final values of $\rho_m(t=2^{24})$. 
For the case $\beta=4$ the onset of thermalization is somewhat delayed 
in the initial phase with smaller entropy values for $t<2^{14}$
(in comparison to the case $\beta=10$). 
For such states the conserved energy in (\ref{eqIntegrals}) 
is given by the simple formula 
$E=\mathcal{H}=H_{n_0,n_0}+\beta/2\approx  \beta/2$ since 
$H_{n_0,n_0}\sim 1/\sqrt{N}$ (absence of self links with $A_{n_0,n_0}=0$ 
and small diagonal matrix element from the GOE perturbation). 

The 6 panels in the bottom three rows of Fig.~\ref{figB8} show, 
in the same style as Fig.~\ref{figII_3}, for 
different states the dependence of $\rho_m$ on $E_m$ for the 
largest iterations times $t=2^{24}$ ($\beta=10,4$) 
or $t=2^{25}$ ($\beta=1$) (red $+$ symbols) 
and at a shorter time $t/100$ (green $\times$ symbols). 
Furthermore, each panel also presents the theoretical curve  
$\rho_{\rm RJ}(E_m)=T/(E_m-\mu)$ (blue line) where 
$T$ and $\mu$ are obtained by solving the constraints (\ref{eqConstraints1}) 
at $E=\langle E\rangle =\sum_m E_m\rho_m(t)$ for the numerical data of 
$\rho_m(t)$ at the largest available value of $t$. 
The obtained values of $T$, $\mu$ and $\langle E\rangle$ for the 
3 cases $\beta=10,4,1$ of {\em Barabasi} are 
$T=-0.01322,-0.02075,-0.02822$, $\mu=9.754,9.853,11.2$
and $\langle E\rangle= 4.744, 1.99, 0.5012$.
For {\em Newman} these values (at same $\beta$-order) are 
$T=-0.01292,-0.02074,-0.02811$, $\mu=9.753,9.853,11.16$ 
and $\langle E\rangle=4.855,1.991,0.5069$. 
The negative values of $T$ are coherent with 
the positive energy values above the critical value $E_c\approx 0$ 
for the transition from positive to negative $T$ (see 
also Eq. (\ref{eqlargemu}) and the discussion of Fig.~\ref{figB4} above). 

From the physical point of view, we see that globally for all cases 
the states thermalize well to the theoretical curve for (nearly) 
all values $\rho_m$. The data points for 
the shorter time value $t/100$ show stronger fluctuations as expected 
and for $\beta=1$ the quality of convergence is also a bit reduced, probably 
larger iterations times are still needed here. 

\subsection{Politician network}

\begin{figure}[htbp]
\begin{center}
\includegraphics[width=0.95\columnwidth]{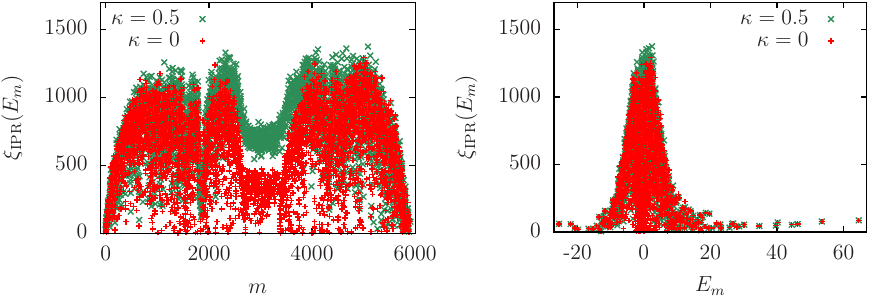}%
\end{center}
\caption{\label{figB9}
\label{fig_IPR_polit}
$\xi_{\rm IPR}(E_m)$ of eigenvector $\phi^{(m)}$ 
of $H$ for the case of the \PN\ 
with $\kappa=0$ and $\kappa=0.5$ 
versus index $m$ (energy $E_m$) in left (right) panel.}
\end{figure}

Fig.~\ref{figB9}, shows the IPR values of eigenmodes for the \PN\ 
at $\kappa=0$ and $\kappa=0.5$ 
(in the same style as Fig.~\ref{figB2}). 
For a significant number of boundary modes the IPR is 
very small with sames values between $\kappa=0$ and $\kappa=0.5$. 
Due to the large boundary energy gaps the states are essentially the 
same for both $\kappa$ values. Modes in the bulk have a quite large 
distribution of IPR values between localized states (IPR $\sim 10$) 
and maximal values 
IRP $\sim 1200$ (for both $\kappa$ values). 
There is also a signifiant reduction of typical IPR values for energies 
$E_m\approx 0$ by a factor 3 (2) for $\kappa=0$ ($\kappa=0.5$). 
Furthermore, the probability (density of data points) to find very 
small IPR values (for center modes) is strongly reduced for $\kappa=0.5$ 
in comparison to $\kappa=0$. 

\begin{figure}[htbp]
\begin{center}
\includegraphics[width=0.95\columnwidth]{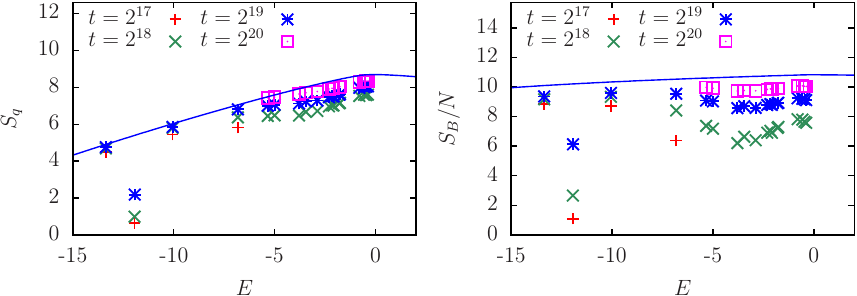}%
\end{center}
\caption{\label{figB10}
\label{figSEPolit}
Entropy $S_q$ ($S_B/N$) versus energy $E$ in left (right) panel for 
some initial modes of the \PN\ with $\kappa=0.5$, $N=5908$, 
$\beta=10$. For 
each case three data points for the three longest time 
values are shown to illustrate the thermalization. 
The first four modes ($m_0=8,16,32, 128$) were computed 
with the time step $dt=1/32$ for the symplectic integrator 
up to $t=2^{19}$. For other modes $S_q$ and $S_B$ were computed with $dt=1/16$ 
(for $128<m_0\le 600$) or $dt=0.1$ (for $600<m_0$) both up to $t=2^{20}$. 
The blue line shows the energy dependence of the theoretical 
thermalized entropy for both entropy quantities. }
\end{figure}

In Fig.~\ref{figB10}, we show for a selected number of modes 
with initial energies in the interval $-14<E_{m_0}<0$ the 
dependence of both entropy values $S_q$ and $S_B/N$ on the 
final (linear) energy $E=\langle E\rangle=\sum_m E_m\rho_m$ 
(which is typically very close to $E_{m_0}$). 
For each data point the values for three successive 
time values $t=2^{l-2},2^{l-1},2^{l}$ (with either $l=19$ for $m_0\le 128$ 
or $l=20$ for $m_0>128$) are shown indicating a clear tendency for 
convergence to the theoretical curve. For three of the four modes with 
$m_0\le 128$ and $S_q$ the last data points are actually already 
very close to the theoretical curve and the remaining mode 
at $E\approx -12$ shows a strong increase of $S_q$ between 
the last two time values indicating a potential convergence at later times.

\subsection{Entropy in the RMT model}

\begin{figure}[htbp]
\begin{center}
\includegraphics[width=0.95\columnwidth]{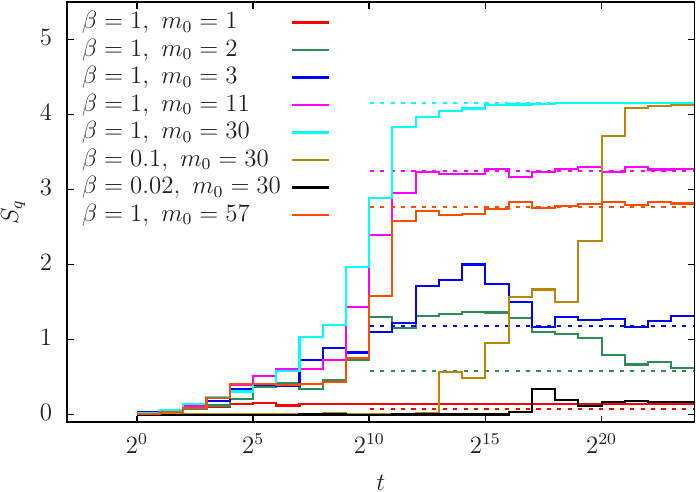}%
\end{center}
\caption{\label{figB11}
\label{fig_Stime_q_RawRMT}
Time dependence of $S_q$ for the case of a GOE matrix ($N=64$, 
semicircle radius 1) 
using the data of \cite{rmtprl}. The full lines correspond to 
$S_q(t)$ for the cases of Figure~S1 of \cite{rmtprl} and the 
dotted lines indicate the theoretical thermalized RJ values 
(using the color for the cases with $\beta=1$). 
This figures is very similar 
to Figure~S1 of \cite{rmtprl}. It is shown here for convenience. }
\end{figure}

\begin{figure}[htbp]
\begin{center}
\includegraphics[width=0.95\columnwidth]{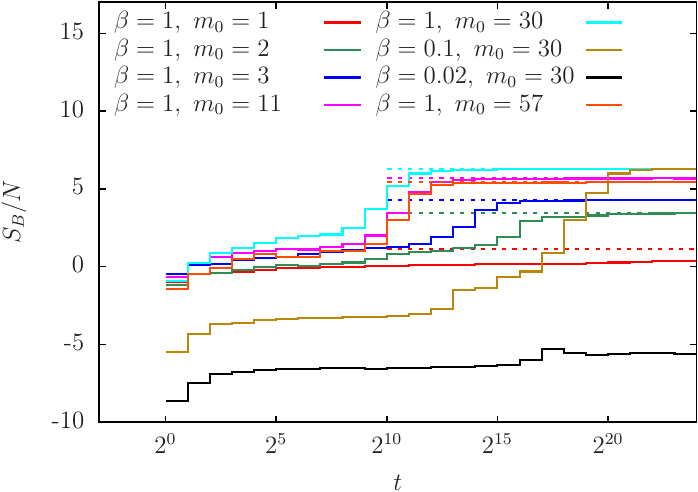}%
\end{center}
\caption{\label{figB12}
\label{fig_Stime_B_RawRMT}
Time dependence of $S_B/N$ for the case of a GOE matrix ($N=64$, 
semicircle radius 1) 
using the data of \cite{rmtprl}. The full lines correspond to 
$S_B(t)/N$ for the cases of Figure~S1 of \cite{rmtprl} and the 
dotted lines indicate the theoretical thermalized RJ values 
(using the color for the cases with $\beta=1$). 
$S_B$ has been computed by (\ref{eqSB}) using $h_B=1/N^2$. }
\end{figure}

Figs.~\ref{figB11} and \ref{figB12} show the time dependence 
of $S_q(t)$ and $S_B(t)/N$ respectively for a pure GOE matrix 
(corresponding to $A=1$ and $\kappa=1$ for $N=64$) for the 
cases already shown in Fig. S1 of \cite{rmtprl} which is actually 
very similar to Fig.~\ref{figB11} (the latter shows more theoretical 
values as well). For $S_q$ we have a nonmonotonic time dependence for
the initial states at $m_0 = 2,3$ at intermediate times. However, 
at large times these modes are well thermalized and close to
the RJ condensate phase. For $S_B$ the time dependence is always monotonic
in the chaotic regime.

\subsection{Wealth inequality and Lorenz curves}

\begin{figure}[htbp]
\begin{center}
\includegraphics[width=0.95\columnwidth]{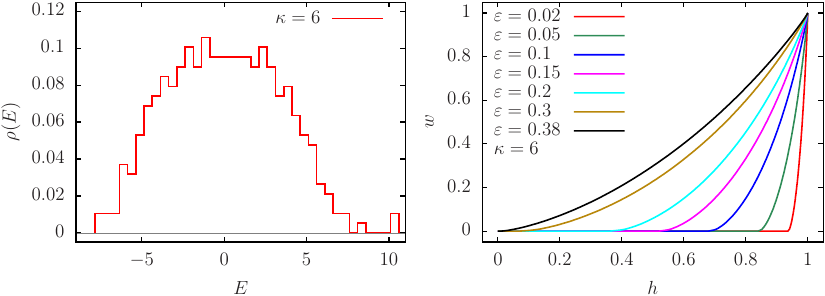}%
\end{center}
\caption{\label{figB13}
\label{fig_DensLor_kappa6}
Density of states (Lorenz curves) for the \SN\ at $\kappa=6$ 
in left (right) panel.}
\end{figure}

Fig.~\ref{figB13} shows the density of state for the \SN\ and a set 
of Lorenz curves for the case of a very strong value of $\kappa$. 
Now, the density of states is rather close to the semicercle law 
with radius $\kappa=6$ and the behavior for $h>h_0$ ($h_0=$ maximal value at 
which $w(h)=0$) corresponds more to a smooth curve instead of a straight line 
as for the case $\kappa=0.5$ shown Fig.~\ref{figII_8}.

\end{appendices}

%% The style puts already the ``Reference'' title,
%%\section*{References} %% => can be commented

\end{document}